\documentclass[aps,prd,twocolumn,amsmath,amssymb,superscriptaddress,showpacs]{revtex4}
\usepackage{epsfig}
\usepackage{graphicx}
\usepackage{dcolumn}
\usepackage{amsmath}
\usepackage{xspace}

\newcommand{\rphi}{\ensuremath{r}-\ensuremath{\phi}}
\newcommand{\mett}{\mbox{$E\!\!\!\!/_{T}$}}

\begin{document}
\title{Search for Anomalous Production of Events with Two Photons and Additional Energetic 
Objects at CDF}
\affiliation{Institute of Physics, Academia Sinica, Taipei, Taiwan 11529, Republic of China} 
\affiliation{Argonne National Laboratory, Argonne, Illinois 60439} 
\affiliation{University of Athens, 157 71 Athens, Greece} 
\affiliation{Institut de Fisica d'Altes Energies, Universitat Autonoma de Barcelona, E-08193, Bellaterra (Barcelona), Spain} 
\affiliation{Baylor University, Waco, Texas  76798} 
\affiliation{Istituto Nazionale di Fisica Nucleare Bologna, $^{cc}$University of Bologna, I-40127 Bologna, Italy} 
\affiliation{Brandeis University, Waltham, Massachusetts 02254} 
\affiliation{University of California, Davis, Davis, California  95616} 
\affiliation{University of California, Los Angeles, Los Angeles, California  90024} 
\affiliation{University of California, San Diego, La Jolla, California  92093} 
\affiliation{University of California, Santa Barbara, Santa Barbara, California 93106} 
\affiliation{Instituto de Fisica de Cantabria, CSIC-University of Cantabria, 39005 Santander, Spain} 
\affiliation{Carnegie Mellon University, Pittsburgh, PA  15213} 
\affiliation{Enrico Fermi Institute, University of Chicago, Chicago, Illinois 60637}
\affiliation{Comenius University, 842 48 Bratislava, Slovakia; Institute of Experimental Physics, 040 01 Kosice, Slovakia} 
\affiliation{Joint Institute for Nuclear Research, RU-141980 Dubna, Russia} 
\affiliation{Duke University, Durham, North Carolina  27708} 
\affiliation{Fermi National Accelerator Laboratory, Batavia, Illinois 60510} 
\affiliation{University of Florida, Gainesville, Florida  32611} 
\affiliation{Laboratori Nazionali di Frascati, Istituto Nazionale di Fisica Nucleare, I-00044 Frascati, Italy} 
\affiliation{University of Geneva, CH-1211 Geneva 4, Switzerland} 
\affiliation{Glasgow University, Glasgow G12 8QQ, United Kingdom} 
\affiliation{Harvard University, Cambridge, Massachusetts 02138} 
\affiliation{Division of High Energy Physics, Department of Physics, University of Helsinki and Helsinki Institute of Physics, FIN-00014, Helsinki, Finland} 
\affiliation{University of Illinois, Urbana, Illinois 61801} 
\affiliation{The Johns Hopkins University, Baltimore, Maryland 21218} 
\affiliation{Institut f\"{u}r Experimentelle Kernphysik, Karlsruhe Institute of Technology, D-76131 Karlsruhe, Germany} 
\affiliation{Center for High Energy Physics: Kyungpook National University, Daegu 702-701, Korea; Seoul National University, Seoul 151-742, Korea; Sungkyunkwan University, Suwon 440-746, Korea; Korea Institute of Science and Technology Information, Daejeon 305-806, Korea; Chonnam National University, Gwangju 500-757, Korea; Chonbuk National University, Jeonju 561-756, Korea} 
\affiliation{Ernest Orlando Lawrence Berkeley National Laboratory, Berkeley, California 94720} 
\affiliation{University of Liverpool, Liverpool L69 7ZE, United Kingdom} 
\affiliation{University College London, London WC1E 6BT, United Kingdom} 
\affiliation{Centro de Investigaciones Energeticas Medioambientales y Tecnologicas, E-28040 Madrid, Spain} 
\affiliation{Massachusetts Institute of Technology, Cambridge, Massachusetts  02139} 
\affiliation{Institute of Particle Physics: McGill University, Montr\'{e}al, Qu\'{e}bec, Canada H3A~2T8; Simon Fraser University, Burnaby, British Columbia, Canada V5A~1S6; University of Toronto, Toronto, Ontario, Canada M5S~1A7; and TRIUMF, Vancouver, British Columbia, Canada V6T~2A3} 
\affiliation{University of Michigan, Ann Arbor, Michigan 48109} 
\affiliation{Michigan State University, East Lansing, Michigan  48824}
\affiliation{Institution for Theoretical and Experimental Physics, ITEP, Moscow 117259, Russia} 
\affiliation{University of New Mexico, Albuquerque, New Mexico 87131} 
\affiliation{Northwestern University, Evanston, Illinois  60208} 
\affiliation{The Ohio State University, Columbus, Ohio  43210} 
\affiliation{Okayama University, Okayama 700-8530, Japan} 
\affiliation{Osaka City University, Osaka 588, Japan} 
\affiliation{University of Oxford, Oxford OX1 3RH, United Kingdom} 
\affiliation{Istituto Nazionale di Fisica Nucleare, Sezione di Padova-Trento, $^{dd}$University of Padova, I-35131 Padova, Italy} 
\affiliation{LPNHE, Universite Pierre et Marie Curie/IN2P3-CNRS, UMR7585, Paris, F-75252 France} 
\affiliation{University of Pennsylvania, Philadelphia, Pennsylvania 19104}
\affiliation{Istituto Nazionale di Fisica Nucleare Pisa, $^{ee}$University of Pisa, $^{ff}$University of Siena and $^{gg}$Scuola Normale Superiore, I-56127 Pisa, Italy} 
\affiliation{University of Pittsburgh, Pittsburgh, Pennsylvania 15260} 
\affiliation{Purdue University, West Lafayette, Indiana 47907} 
\affiliation{University of Rochester, Rochester, New York 14627} 
\affiliation{The Rockefeller University, New York, New York 10021} 
\affiliation{Istituto Nazionale di Fisica Nucleare, Sezione di Roma 1, $^{hh}$Sapienza Universit\`{a} di Roma, I-00185 Roma, Italy} 

\affiliation{Rutgers University, Piscataway, New Jersey 08855} 
\affiliation{Texas A\&M University, College Station, Texas 77843} 
\affiliation{Istituto Nazionale di Fisica Nucleare Trieste/Udine, I-34100 Trieste, $^{ii}$University of Trieste/Udine, I-33100 Udine, Italy} 
\affiliation{University of Tsukuba, Tsukuba, Ibaraki 305, Japan} 
\affiliation{Tufts University, Medford, Massachusetts 02155} 
\affiliation{Waseda University, Tokyo 169, Japan} 
\affiliation{Wayne State University, Detroit, Michigan  48201} 
\affiliation{University of Wisconsin, Madison, Wisconsin 53706} 
\affiliation{Yale University, New Haven, Connecticut 06520} 
\author{T.~Aaltonen}
\affiliation{Division of High Energy Physics, Department of Physics, University of Helsinki and Helsinki Institute of Physics, FIN-00014, Helsinki, Finland}
\author{J.~Adelman}
\affiliation{Enrico Fermi Institute, University of Chicago, Chicago, Illinois 60637}
\author{B.~\'{A}lvarez~Gonz\'{a}lez$^v$}
\affiliation{Instituto de Fisica de Cantabria, CSIC-University of Cantabria, 39005 Santander, Spain}
\author{S.~Amerio$^{dd}$}
\affiliation{Istituto Nazionale di Fisica Nucleare, Sezione di Padova-Trento, $^{dd}$University of Padova, I-35131 Padova, Italy} 

\author{D.~Amidei}
\affiliation{University of Michigan, Ann Arbor, Michigan 48109}
\author{A.~Anastassov}
\affiliation{Northwestern University, Evanston, Illinois  60208}
\author{A.~Annovi}
\affiliation{Laboratori Nazionali di Frascati, Istituto Nazionale di Fisica Nucleare, I-00044 Frascati, Italy}
\author{J.~Antos}
\affiliation{Comenius University, 842 48 Bratislava, Slovakia; Institute of Experimental Physics, 040 01 Kosice, Slovakia}
\author{G.~Apollinari}
\affiliation{Fermi National Accelerator Laboratory, Batavia, Illinois 60510}
\author{A.~Apresyan}
\affiliation{Purdue University, West Lafayette, Indiana 47907}
% ...

\collaboration{CDF Collaboration\footnote{With visitors from $^a$University of Massachusetts Amherst, Amherst, Massachusetts 01003,
$^b$Universiteit Antwerpen, B-2610 Antwerp, Belgium, 
$^c$University of Bristol, Bristol BS8 1TL, United Kingdom,
$^d$Chinese Academy of Sciences, Beijing 100864, China, 
$^e$Istituto Nazionale di Fisica Nucleare, Sezione di Cagliari, 09042 Monserrato (Cagliari), Italy,
$^f$University of California Irvine, Irvine, CA  92697, 
$^g$University of California Santa Cruz, Santa Cruz, CA  95064, 
$^h$Cornell University, Ithaca, NY  14853, 
$^i$University of Cyprus, Nicosia CY-1678, Cyprus, 
$^j$University College Dublin, Dublin 4, Ireland,
$^k$University of Edinburgh, Edinburgh EH9 3JZ, United Kingdom, 
$^l$University of Fukui, Fukui City, Fukui Prefecture, Japan 910-0017
$^m$Kinki University, Higashi-Osaka City, Japan 577-8502
$^n$Universidad Iberoamericana, Mexico D.F., Mexico,
$^o$University of Iowa, Iowa City, IA  52242,
$^p$Kansas State University, Manhattan, KS 66506
$^q$Queen Mary, University of London, London, E1 4NS, England,
$^r$University of Manchester, Manchester M13 9PL, England,
$^s$Muons, Inc., Batavia, IL 60510, 
$^t$Nagasaki Institute of Applied Science, Nagasaki, Japan, 
$^u$University of Notre Dame, Notre Dame, IN 46556,
$^v$University de Oviedo, E-33007 Oviedo, Spain, 
$^w$Texas Tech University, Lubbock, TX  79609, 
$^x$IFIC(CSIC-Universitat de Valencia), 56071 Valencia, Spain,
$^y$Universidad Tecnica Federico Santa Maria, 110v Valparaiso, Chile,
$^z$University of Virginia, Charlottesville, VA  22906
$^{aa}$Bergische Universit\"at Wuppertal, 42097 Wuppertal, Germany,
$^{bb}$Yarmouk University, Irbid 211-63, Jordan
$^{jj}$On leave from J.~Stefan Institute, Ljubljana, Slovenia, 
}}
\noaffiliation

\date{\today}  % *** ALWAYS *** use the current date.
%%
%%  Abstract
\begin{abstract}
We present results of a search for anomalous production of two photons together with an electron, muon, 
$\tau$ lepton, missing transverse energy, or jets using $p\bar{p}$ collision data from 1.1-2.0~fb$^{-1}$ 
of integrated luminosity collected by the Collider Detector at Fermilab (CDF). The event yields and 
kinematic distributions are examined for signs of new physics without favoring a specific model of new 
physics. The results are consistent with the standard model expectations. The search employs several new 
analysis techniques that significantly reduce instrumental backgrounds in channels with an electron and 
missing transverse energy. 
\end{abstract}

%%
%% PACS

%%
%%  Now make the title page...
\maketitle

%%____________________________________________________________________________________
%%------------------------ Introduction ----------------------------------------------
%%____________________________________________________________________________________
\section{Introduction}
\label{sec:introduction}

Over the last twenty years, the rapid pace of developments in phenomenology and model-building has left 
experimentalists at the collider energy frontier with a wide array of new physics scenarios to investigate~\cite{models}.
We are also assured that the number of models which have not yet been described is large. Since each 
search requires substantial resources, only a few new physics scenarios can be the focus of dedicated 
efforts. We address this problem by performing broad searches in available data samples for any discrepancy 
with the standard model (SM)~\cite{SM} in event yields or kinematic distributions.  While this approach is not 
optimized for any particular scenario, it could possibly increase the chance of an unpredicted discovery.

In this article we investigate a sample of data collected by the CDF II detector in $p\bar{p}$ collisions 
at $\sqrt{s}=1.96$~TeV at the Fermilab Tevatron. We restrict ourselves to a ``baseline'' sample with two 
isolated, central (0.05$<$$|\eta|$$<$1.05) photons ($\gamma$) with $E_{T}$$>$13~GeV~\cite{eta}. We then 
select subsamples which also contain at least one more energetic, isolated and well-identified object or where 
two photons are accompanied by large missing transverse energy ($\mett$).  The additional object may be an 
electron ($e$), muon ($\mu$), or $\tau$ lepton ($\tau$). The $\mett$ is calculated from the imbalance 
in the energy of visible particles projected to the plane transverse to the beams. The integrated luminosity 
for each subsample varies from 1.1 to 2.0~fb$^{-1}$. 

The $\gamma\gamma$$+$$X$ ($X$=$e/\mu$, $\tau$, and $\mett$) signatures are present in many new physics 
scenarios beyond the SM. Examples include models with the gauge-mediated supersymmetry breaking (GMSB)~\cite{ggMETgmsb}, 
extended Higgs sector~\cite{Hsector}, technicolor models~\cite{tRho}, 4$^{th}$ generation fermions~\cite{gen4}, 
and theories with large extra dimensions~\cite{ledH}.  
 
The CDF collaboration has previously performed a search for anomalous production of two photons and 
an additional energetic object ($\mett$, $e$, $\mu$, $\tau$, $\gamma$, jets, and $b$-quarks) in 85~pb$^{-1}$ 
of the Tevatron Run~I data~\cite{run1evnt}. Apart from the observation of a single $ee\gamma\gamma\mett$ 
candidate event, the results were consistent with the SM predictions. The $ee\gamma\gamma\mett$ event sparked 
considerable theoretical interest, because this signature is very rare in the SM and the event's topology is 
consistent with that of a decay of a pair of new heavy particles. In Run~II, both the CDF~\cite{run2cdf} 
and D0~\cite{run2d0} collaborations searched for production of $\gamma\gamma+\mett$ events in the context of 
GMSB models using data corresponding to 0.20~fb$^{-1}$ and 1.1~fb$^{-1}$ of integrated luminosity, 
respectively. A search for anomalous production of $\gamma\gamma+e/\mu$ events with energetic central photons 
and leptons ($E_T^{\gamma,e}$$>$25 GeV and $p_T^{\mu}$$>$25 GeV/$c$) in 0.93~fb$^{-1}$ of data was performed 
by the CDF collaboration as a part of a broader signature-based search for new physics in $l+\gamma+X$ 
($l=e,\mu$ and $X=\gamma,l,\mett$) events~\cite{lgmetPRD}. Other signatures involving two photons were studied 
in CDF searches reported in Ref.~\cite{RSdipho,Hdipho,vista}. The current model-independent analysis is improved upon 
previous diphoton searches both in terms of refined experimental techniques and of amount of data analyzed. 
It also probes a wider kinematic range compared to the CDF analyses reported in Ref.~\cite{lgmetPRD} and Ref.~\cite{vista}.

This paper is organized as follows. It begins with a description of the CDF II detector and the baseline diphoton 
sample.  Then, each $\gamma\gamma$$+$$X$ ($X$=$e/\mu$, $\tau$, and $\mett$) subsample is discussed in separate 
sections where we describe the definition of the subsamples, the calculation of the SM predictions, and the 
comparison of the data and the predictions. The details of several techniques are postponed to appendices.

%%____________________________________________________________________________________
%%------------------------ Detector ----------------------------------------------
%%____________________________________________________________________________________
\section{Detector Overview}
\label{sec:detector}

The CDF II detector is a cylindrically symmetric apparatus designed to study $p\bar{p}$ collisions at 
the Fermilab Tevatron. The detector has been described in detail elsewhere~\cite{CDF}; only the detector 
components that are relevant to this analysis are briefly discussed below. The magnetic spectrometer 
consists of tracking devices inside a 3-m diameter, 5-m long superconducting solenoid magnet which provides an 
axial magnetic field of 1.4 T. A set of silicon microstrip detectors (L00, SVX, and ISL)~\cite{L00,ISL,SVX} 
and a 3.1-m long drift chamber (COT)~\cite{COT} with 96 layers of sense wires measure momenta and trajectories (tracks) 
of charged particles in the pseudorapidity regions of $\left|\eta\right|$$<$2 and $\left|\eta\right|$$<$1~\cite{eta}, 
respectively. Surrounding the magnet coil is the projective-tower-geometry sampling calorimeter, which is used 
to identify and measure the energy and position of photons, electrons, jets, and \mett. The calorimeter consists 
of lead-scintillator electromagnetic and iron-scintillator hadron compartments and it is divided into a central 
barrel ($\left|\eta\right|$$<$1.1) and a pair of ``end plugs'' that cover the region 1.1$<$$\left|\eta\right|$$<$3.6. 
The central calorimeter is composed of towers with a segmentation of $\Delta\eta\times\Delta\phi\simeq 0.1\times 15^o$. 
The energy resolution of the central electromagnetic calorimeter for electrons is $\sigma(E_T)/E_T=13.5\%/
\sqrt{E_T({\rm GeV})}\oplus 1.5\%$~\cite{cem}, while the energy resolution of the central hadron calorimeter for 
charged pions that do not interact in the electromagnetic section is $\sigma(E_T)/E_T=50\%/\sqrt{E_T({\rm GeV})}
\oplus 3\%$~\cite{cwha}. In the plug calorimeter, the segmentation varies from $\Delta\eta\times\Delta\phi\simeq 
0.1\times 7.5^o$ for 1.1$<$$|\eta|$$<$1.8 to $\Delta\eta\times\Delta\phi\simeq 0.6\times 15^o$ for $|\eta|$$=$3.6. 
The corresponding plug electromagnetic and hadron calorimeter energy resolutions are $\sigma(E)/E=14.4\%/
\sqrt{E({\rm GeV})}\oplus 0.7\%$ and $\sigma(E)/E=74\%/\sqrt{E({\rm GeV})}\oplus 4\%$, respectively~\cite{plug}.
The additional system in the central region is used for identification and precise position measurement of photons
and electrons. Multiwire proportional chambers with cathode-strip readout (the CES system) are located at the depth of 
six radiation lengths (near shower maximum) in the central electromagnetic calorimeter. Cathode strips and anode wires,
with a channel spacing between 1.5~cm and 2~cm, running along the azimuthal (strips) and the beam line (wires) 
direction give location and two-dimensional profiles of the electromagnetic showers. The position resolution of the 
CES is 2~mm for a 50~GeV photon. The electromagnetic compartments of the calorimeter are also used to measure the 
arrival time of particles depositing energy in each tower~\cite{EMtiming}. Muons from collisions as well as cosmic 
rays are identified using systems which are located outside the calorimeters: the central muon detector (CMU) and the 
central muon upgrade detector (CMP) in the pseudorapidity region of $\left|\eta\right|$$<$0.6, and the central muon 
extension (CMX) for the pseudorapidity region of 0.6$<$$\left|\eta\right|$$<$1.0~\cite{muon_systems}. The CMU system 
uses four layers of planar drift chambers and detects muons with $p_T$$>$1.4~GeV/$c$. The CMP system, located behind 
a 0.6~m thick steel absorber outside the magnetic return yoke, consists of an additional four layers of planar drift 
chambers and detects muons with $p_T$$>$2.2~GeV/$c$. The CMX detects muons with $p_T$$>$1.4~GeV/$c$ using four to 
eight layers of drift chambers, depending on the polar angle. A system of Cherenkov luminosity counters (CLC)~\cite{CLC}, 
located around the beam pipe and inside the plug calorimeters, is used to measure a number of inelastic $p\bar p$ 
collisions per bunch crossing, and thereby the luminosity.  

The online event selection at CDF is done by a three-level trigger~\cite{trigger} system with each level providing a 
rate reduction sufficient to allow for processing at the next level with minimal deadtime. The Level-1 uses custom 
designed hardware to find physics objects based on a subset of the detector information. The Level-2 trigger consists
of custom hardware to do a limited event reconstruction which can be processed in programmable processors. The
Level-3 trigger uses the full detector information and consists of a farm of computers that reconstruct the data and 
apply selection criteria similar to the offline requirements.       

%%____________________________________________________________________________________
%%------------------------ Data Selection and Reconstruction -------------------------
%%____________________________________________________________________________________
\section{Data Selection and Event Reconstruction}
\label{sec:data}

The search for anomalous production of $\gamma\gamma+\mett$ and $\gamma\gamma+\tau$ events is performed 
with data corresponding to 2.0$\pm$0.1~fb$^{-1}$ of luminosity integrated from the beginning of Run~II. 
The search for anomalous 
$\gamma\gamma+e/\mu$ events utilizes a smaller dataset of 1.1$\pm$0.1~fb$^{-1}$ of integrated luminosity. 
The online (trigger) requirements and offline selection criteria that are common to all three final 
states are discussed below. Additional requirements of each analysis are explained separately.

The inclusive $\gamma\gamma$ events are selected online by a three-level trigger that requires two 
isolated electromagnetic (EM) clusters with $E_{T}^{\gamma}$$>$12 GeV (diphoton-12 trigger) or two 
electromagnetic clusters with $E_{T}^{\gamma}$$>$18 GeV and no isolation requirement (diphoton-18 trigger). 
A detailed description of diphoton triggers can be found in Appendix~\ref{sec:trigger}. The triggered 
$\gamma\gamma$ candidate events are then subjected to the offline selection. Each event is required 
to have two central EM clusters (photon candidates) inside a well-instrumented region of the calorimeter 
(approximately 0.05$<$$|\eta|$$<$1.05) with $E_{T}$$>$13~GeV. For each photon candidate, the transverse 
shower profile in the CES and the amount of energy leaked into the hadron calorimeter must be consistent with 
those of a single electromagnetic shower. We distinguish photons from electrons by making sure that no 
high-$p_T$ charged track points to the EM cluster. Both photon candidates are also required to be isolated 
energy clusters in the calorimeter in order to suppress background due to jets. More details of the standard 
photon identification criteria can be found in Appendix~\ref{sec:photon}. To reduce contamination due to 
cosmic-ray, beam-related, and other non-collision backgrounds, the event must contain a well-reconstructed 
vertex, formed by tracks, with $|z|$$<$60 cm. If multiple vertices are reconstructed, the vertex with 
the largest $\sum p_T$ of the associated tracks is selected.  Unless noted otherwise, the transverse 
energy of all calorimeter objects is calculated with respect to this primary vertex. Finally, the arrival 
time of both photon candidates, corrected for average path length, has to be consistent with the $p\bar{p}$ 
collision time. It should be pointed out that due to the photon timing requirements, we are only sensitive 
to new physics processes where photons are produced in decays of new particles with small lifetime ($<$1~ns).  

Inclusive $\gamma\gamma$ events satisfying the above criteria form the baseline $\gamma\gamma$ signal 
sample used in all three analyses. This sample consists of real $\gamma\gamma$ events (approximately $30\%$), 
{\it jet-}$\gamma$ ($45\%$) and {\it jet-jet} ($25\%$)~\cite{run2diphoMET} events where one or both jets are 
misidentified as a photon. (An object misidentified as a photon is referred to as a ``fake'' photon.) 
The $\gamma\gamma+e/\mu$, $\gamma\gamma+\tau$, and $\gamma\gamma+\mett$ candidate events are then selected 
from the base signal sample by requiring additional objects of interest or significant $\mett$. 
We also select a control sample of $\gamma\gamma$ events by applying less stringent photon identification 
requirements as discussed in Appendix~\ref{sec:photon}. To avoid an overlap with the signal sample, at least 
one photon candidate from the control sample must fail the standard photon cuts. The control sample is ideal 
for testing our analysis techniques because it has a similar event topology, but is dominated by background 
events (the fraction of real $\gamma\gamma$ events in it is approximately 5$\%$).

Our baseline signal and control $\gamma\gamma$ samples consist of 31,116 and 42,708 events, respectively, 
in data corresponding to 2.0 fb$^{-1}$ of integrated luminosity.

%%____________________________________________________________________________________
%%------------------------ Analysis --------------------------------------------------
%%____________________________________________________________________________________
\section{Searches for Anomalous Production of $\gamma\gamma+X$ Events}
\label{sec:analysis}

In this Section, we describe in detail three separate searches for anomalous production of $\gamma\gamma+e/\mu$, 
$\gamma\gamma+\tau$, and $\gamma\gamma+\mett$ events. All analyses use the same baseline $\gamma\gamma$ samples 
and utilize the same definitions of the additional objects and kinematic variables: electrons, muons, $\tau$ 
leptons, jets, soft unclustered energy, $\mett$, and $H_T$. The $H_T$ is defined as a scalar sum of $\mett$ and 
$E_T$ of all identified photons, leptons, and jets. The detailed descriptions of these objects can be found in 
Appendices~\ref{sec:electron}-\ref{sec:soft}.

%%__________ gg+e/mu __________________________________________________________________
\subsection{The $\gamma\gamma+e/\mu$ Final State}
\label{sec:dipho+e/mu}

We search for anomalous production of events containing two photons and at least one additional electron or 
muon in data corresponding to 1.1$\pm$0.1~fb$^{-1}$ of integrated luminosity. The events of interest are 
derived from the $\gamma\gamma$ baseline sample described in Section~\ref{sec:data}. The electron identification 
criteria are similar to those for the photon except that an electron candidate must have an energetic track 
pointing to the EM cluster. The momentum, $p$, of this track has to be consistent with the energy deposited 
in the EM calorimeter. The electron identification requirements are described in detail in Appendix~\ref{sec:electron}.

A well-reconstructed COT track is identified as a muon candidate if it is matched to hit segments (stubs) in 
the central muon detectors, and its energy deposition pattern in the EM and HAD calorimeters is consistent with 
that left by a minimum ionizing particle. Details on the muon identification requirements can be found in 
Appendix~\ref{sec:muon}. 

The selected $\gamma\gamma e$ and $\gamma\gamma\mu$ events must have at least one electron or muon candidate 
with $E_T^e$$>$20~GeV and $p_T^{\mu}$$>$20~GeV/$c$, respectively. We compare the observed number of events and 
kinematic distributions in the data with those from our SM background predictions. Backgrounds for the 
$\gamma\gamma e$ and $\gamma\gamma\mu$ signatures of new physics include: 
\begin{enumerate} 
\item The SM production of $Z$$\rightarrow$$l^+l^-$ and $W$$\rightarrow$$l\nu$ in association with two 
photons ($Z\gamma\gamma$, $W\gamma\gamma$), where photons are radiated from either the initial state quarks, 
charged electroweak boson ($W$), or the final--state leptons. 
\item Backgrounds due to misidentified particles (fake photons or leptons)
  \begin{enumerate}
  \item electrons misidentified as photons (e.g., $Z\gamma$ events),
  \item jets misidentified as photons (e.g, $W\gamma+jet$ or $Z\gamma+jet$ events),
  \item jets misidentified as leptons (mostly $\gamma\gamma$ candidate events with an additional jet).
 \end{enumerate}
\end{enumerate} We describe below how these background contributions are estimated.
 
% SM
The SM $Z\gamma\gamma$ and $W\gamma\gamma$ contributions are estimated from Monte Carlo (MC) simulation. The 
MC samples are generated using the leading-order (LO) matrix-element generator {\sc madgraph}~\cite{madgraph}. 
The output of {\sc madgraph} is fed into {\sc pythia}~\cite{pythia} to carry out parton fragmentation, 
simulation of the underlying event and additional $p\bar p$ interactions in the same bunch crossing, as well 
as initial- and final-state radiation. The output of {\sc pythia} is then processed through the {\sc geant}-based 
detector simulation~\cite{geant} followed by the same reconstruction program as that for the data. To account 
for an imperfect modeling of the CDF-II detector, the MC predictions are corrected for small differences 
(1-10$\%$) in photon and lepton identification and trigger efficiences between data and MC. In addition, 
the LO cross sections predicted by {\sc madgraph} are scaled to the next-to-leading-order (NLO) cross sections 
according to the $K$-factors in Ref.~\cite{cdf6601}. These $K$-factors are functions of the dilepton mass and 
$E_T$ of the highest-$E_T$ photon, and their values range from 1.36 to 1.62 with an average of $\sim$1.4 for the 
kinematic range of $Z\gamma\gamma$ and $W\gamma\gamma$ production. The uncertainty for this background prediction 
includes statistical uncertainty due to the finite size of MC samples, 6$\%$ systematic uncertainty on the measured 
integrated luminosity, and 7$\%$ systematic uncertainty on the $Z\gamma\gamma$ and $W\gamma\gamma$ cross sections 
due to uncertainties in the factorization and renormalization scales and parton distribution functions 
(PDF)~\cite{pdf}.

% fake photons
The rest of the background contains at least one misidentified object. Fake photons can arise from the hard 
bremsstrahlung of electrons in the detector material, inefficient electron-track reconstruction, or decays of 
$\pi^0$, $\eta^0$, or $K_s^0$ in jets. Although these background sources yield real photons in the final state, 
they are referred to as "fake'' in this analysis, to be distinguished from the photons possibly produced by new 
exotic particles. The number of $\gamma\gamma+l$ ($l=e,\mu$) events where at least one of the photons is faked 
by an electron is estimated from the $l\gamma+e$ data (collected with the diphoton triggers). We obtain the 
prediction by applying a $e$$\to$$\gamma$ misidentification probability as a function of the $E_T$ of electron 
(about 2.7$\%$ and 1.5$\%$ for electrons with $E_T=$20~GeV and 40~GeV, respectively) to the selected $l\gamma+e$ 
events. More details about this misidentification probability and its uncertainty are included in 
Appendix~\ref{sec:e2phoFkRt}. To estimate the number of $\gamma\gamma+l$ events where at least one of the photons 
is a misidentified jet, we select the $l\gamma+jet$ data collected with inclusive lepton triggers and multiply 
them by a $jet$$\to$$\gamma$ misidentification probability as a function of the jet's $E_T$ (about 0.2$\%$ and 
0.04$\%$ for jets with $E_T=$13~GeV and $>$50~GeV, respectively). The description of the $jet$$\to$$\gamma$ 
misidentification probability and its associated uncertainty can be found in Appendix~\ref{sec:Jet2phoFkRt}.    
Also note that both $l\gamma+e$ and $l\gamma+jet$ samples may contain events with fake leptons.  

% fake leptons
The last source of background is events with two real photons and a fake lepton from the direct diphoton 
production with additional jets. The number of ``diphoton + fake lepton'' events is obtained by applying 
$E_T$-dependent misidentification probabilities from Ref.~\cite{canepa} to the events with two photon 
candidates and an object which may fake a lepton. These objects are jets for electrons and isolated tracks 
for muons. The probability for a jet (isolated track) with $E_T$($p_T$)=50~GeV to fake a central electron 
(muon) is $\sim$0.01$\%$ ($\sim$1$\%$). Details of the misidentification probabilities and their uncertainties 
are discussed in Appendix~\ref{sec:Jet2eleFkRt}. According to earlier studies~\cite{run2diphoMET}, only 
29$\%$$\pm$4$\%$ of observed diphotons are real diphoton events. In order to avoid duplication with the fake 
photon contribution estimated above, the number of ``diphoton + fake lepton'' events is multiplied by the real 
diphoton fraction (29$\%$$\pm$4$\%$), which gives the number of ``real $\gamma\gamma$ + fake lepton'' events. 

% cosmics
The fake photon signature can also be produced as a result of the bremsstrahlung of cosmic muons as they pass 
through the calorimeters. However, the probability for a real photon event to overlap with such a cosmic event 
is found to be very small: 1.5$\times$10$^{-8}$ (see Ref.~\cite{run2diphoMET}). Therefore, the cosmic backgrounds 
are negligible in the $\gamma\gamma e$ and $\gamma\gamma \mu$ searches.

\begin{figure*}[tbp]
\begin{center}
   \includegraphics[width=0.3\linewidth]
   {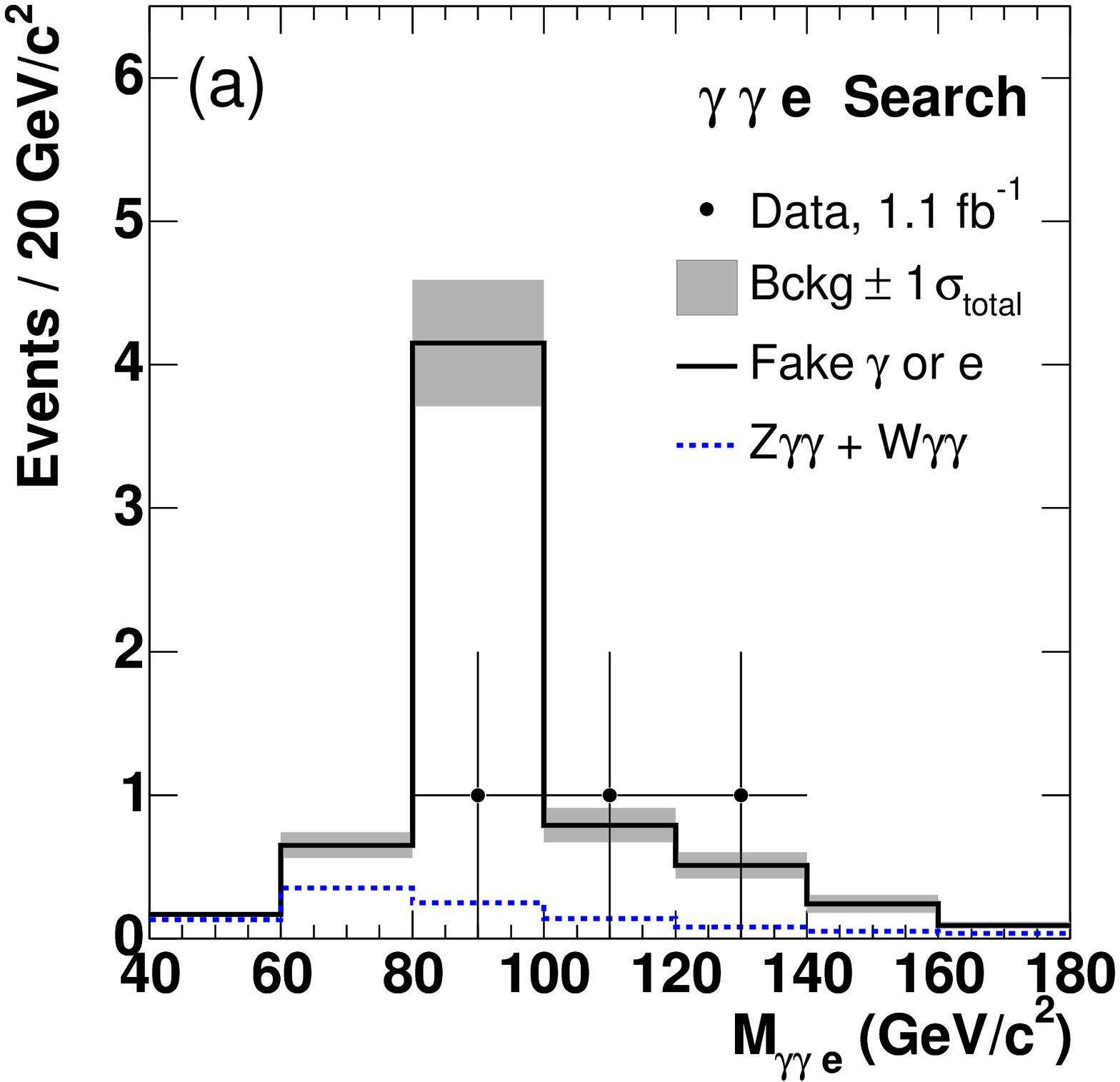}
   \includegraphics[width=0.3\linewidth]
   {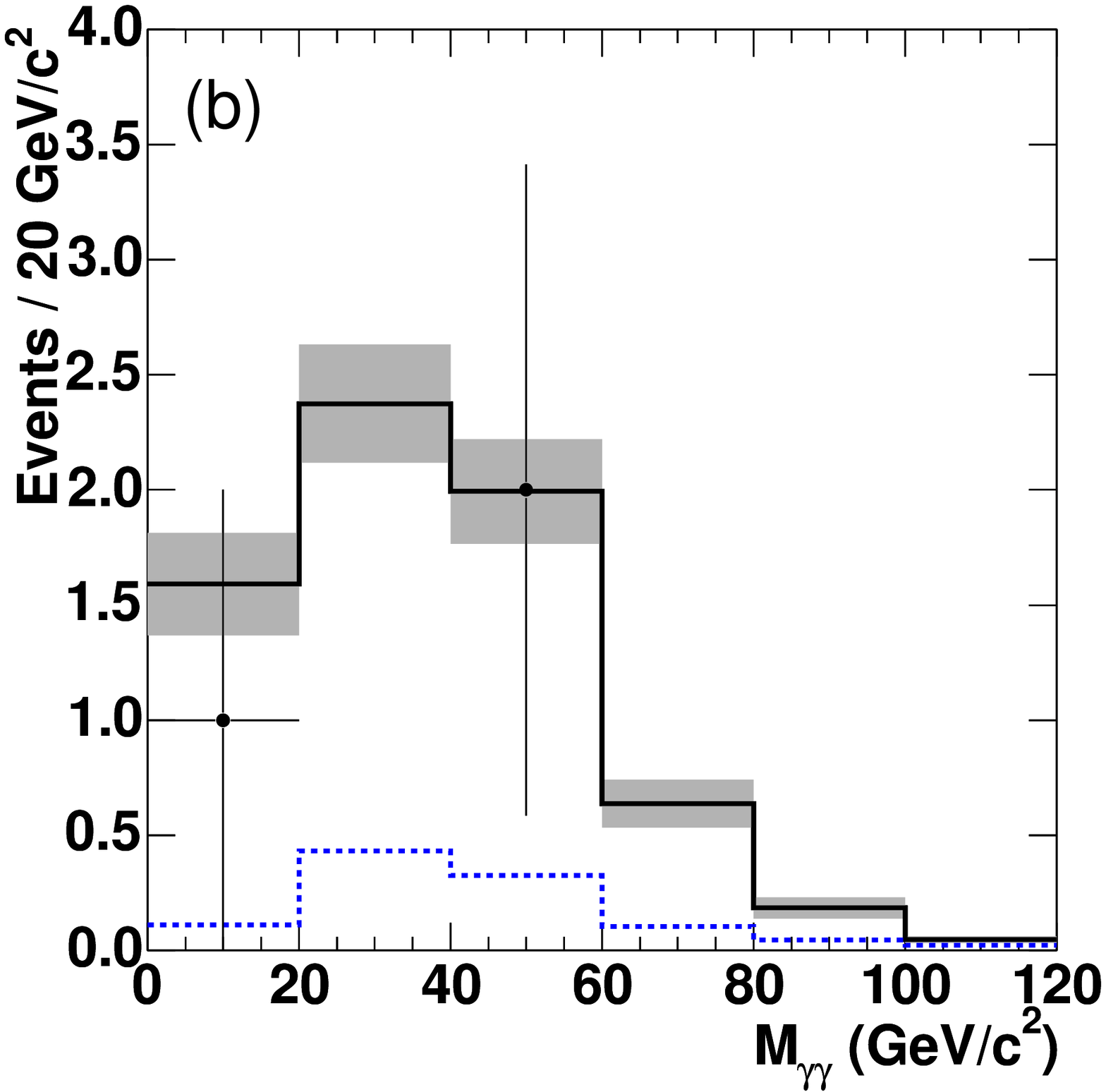}
   \includegraphics[width=0.3\linewidth]
   {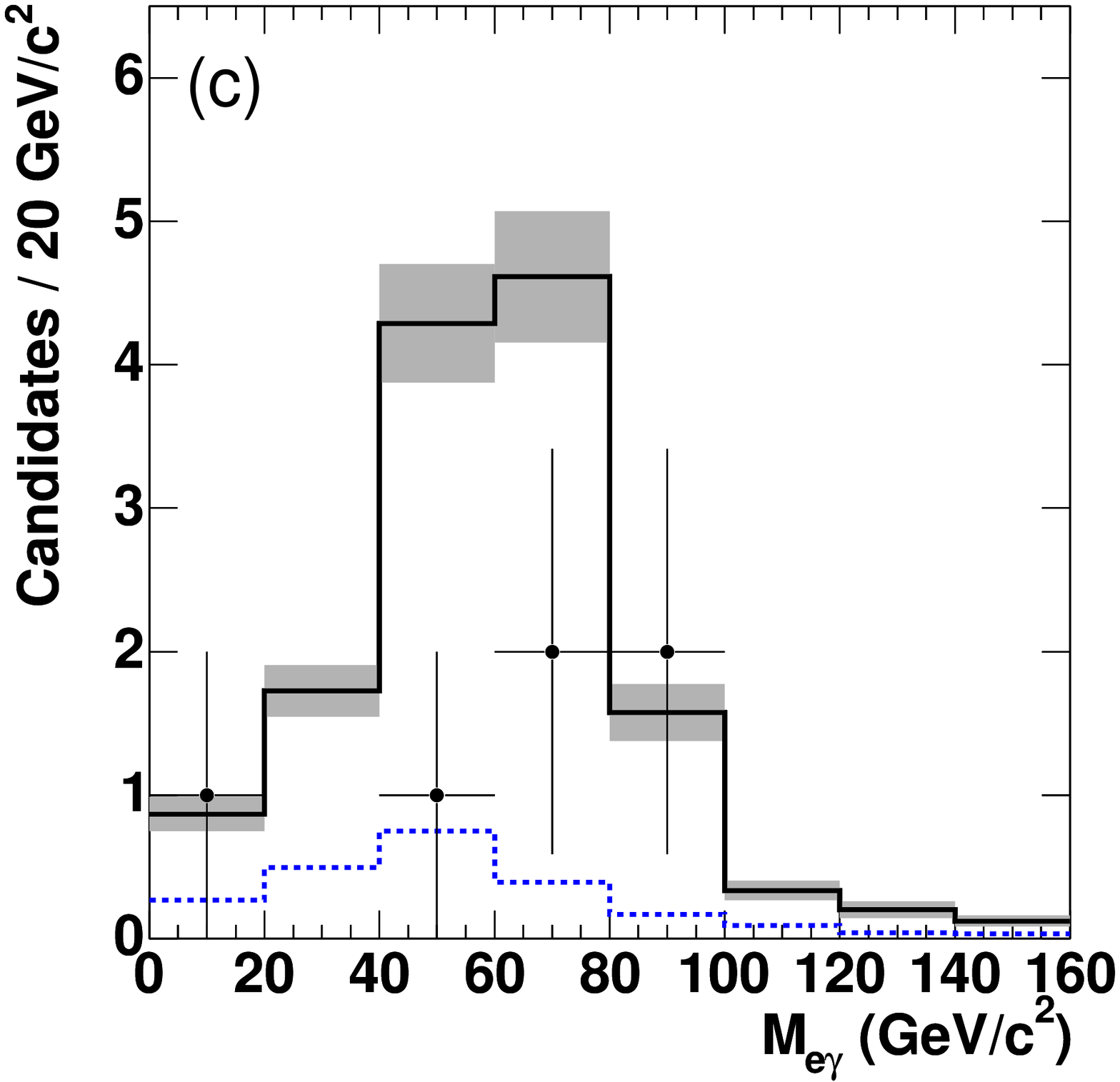}
   \includegraphics[width=0.3\linewidth]
   {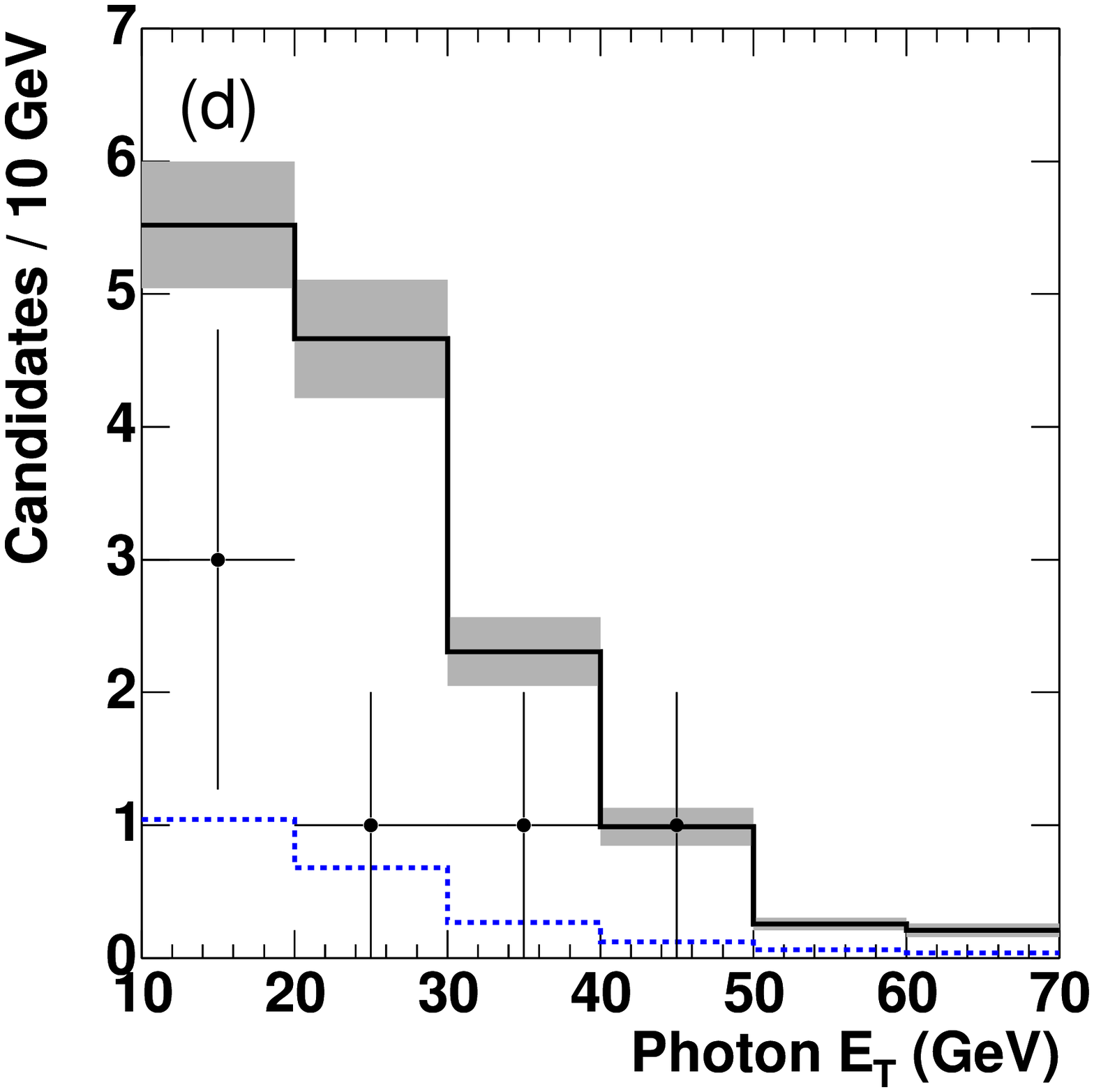}
   \includegraphics[width=0.3\linewidth]
   {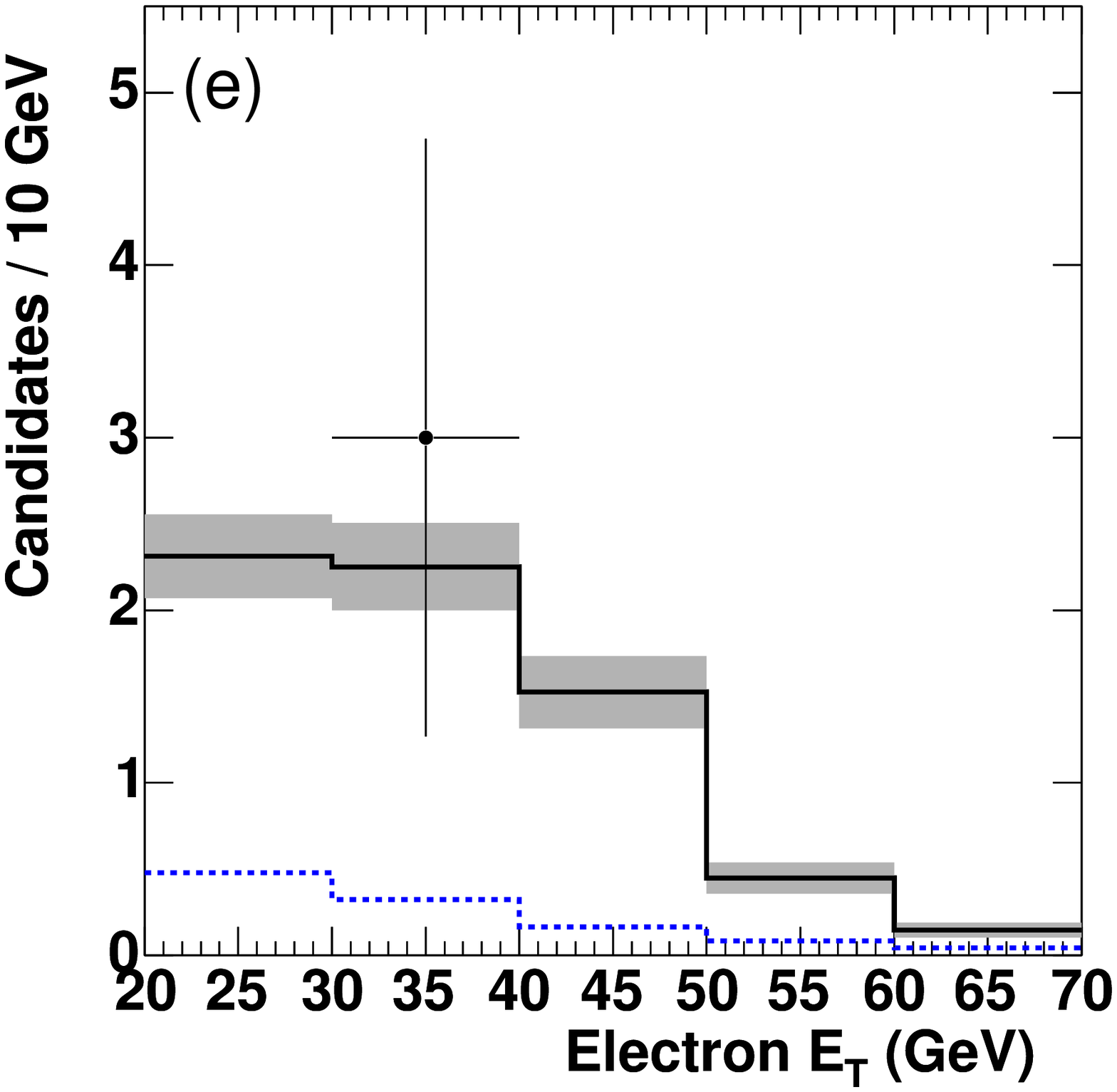}
   \includegraphics[width=0.3\linewidth]
   {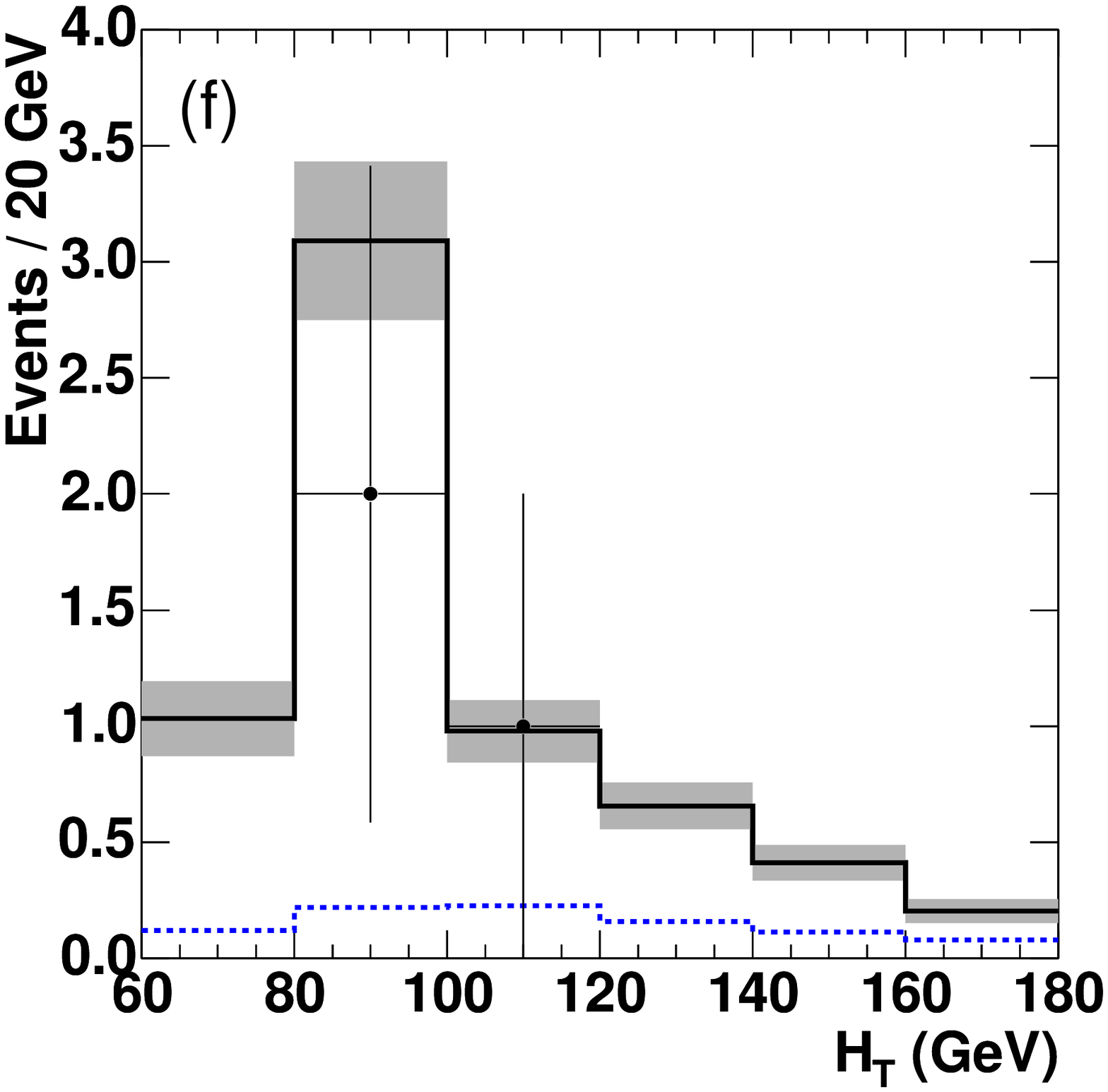}
   \includegraphics[width=0.3\linewidth]
   {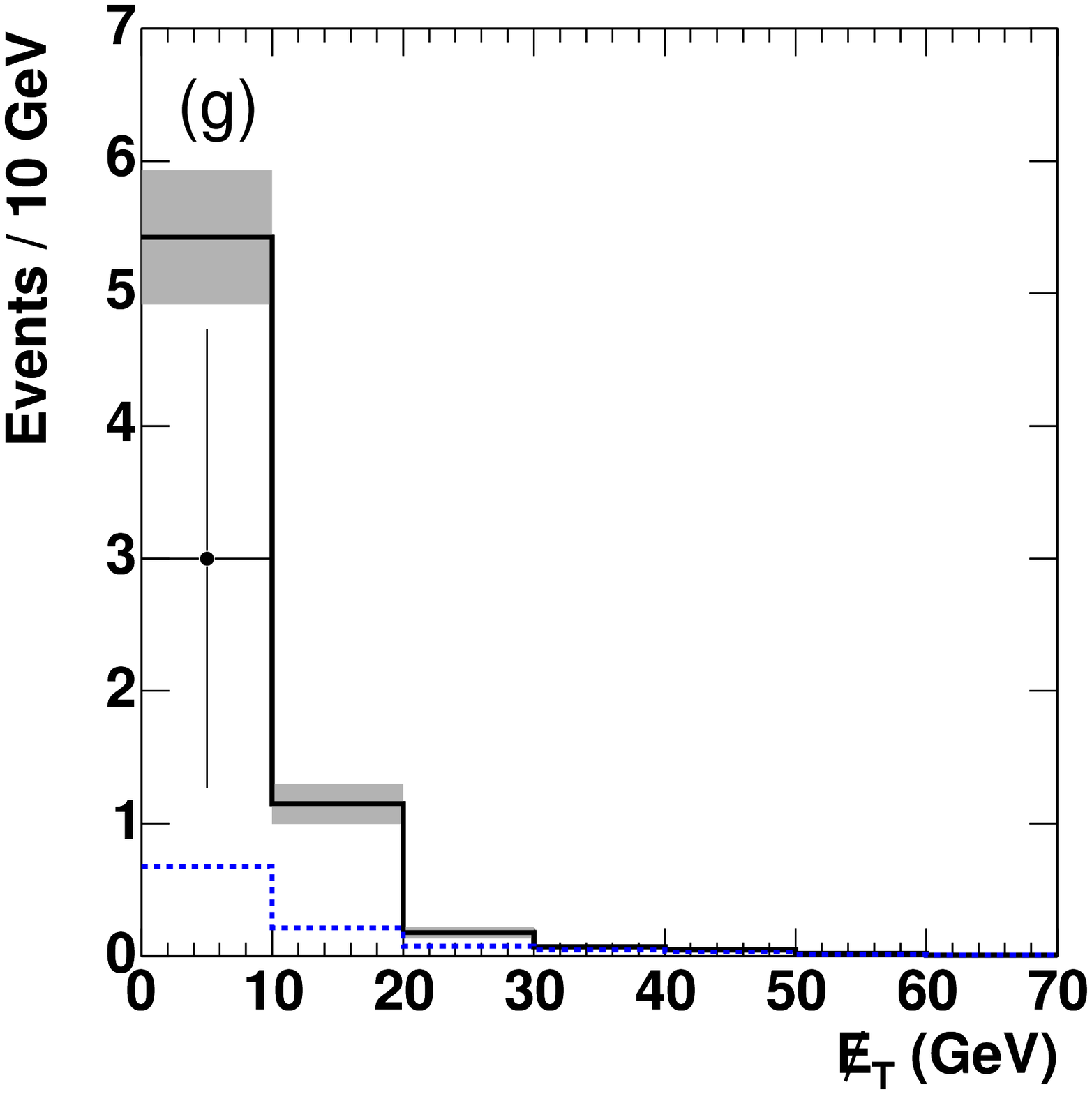}
   \includegraphics[width=0.3\linewidth]
   {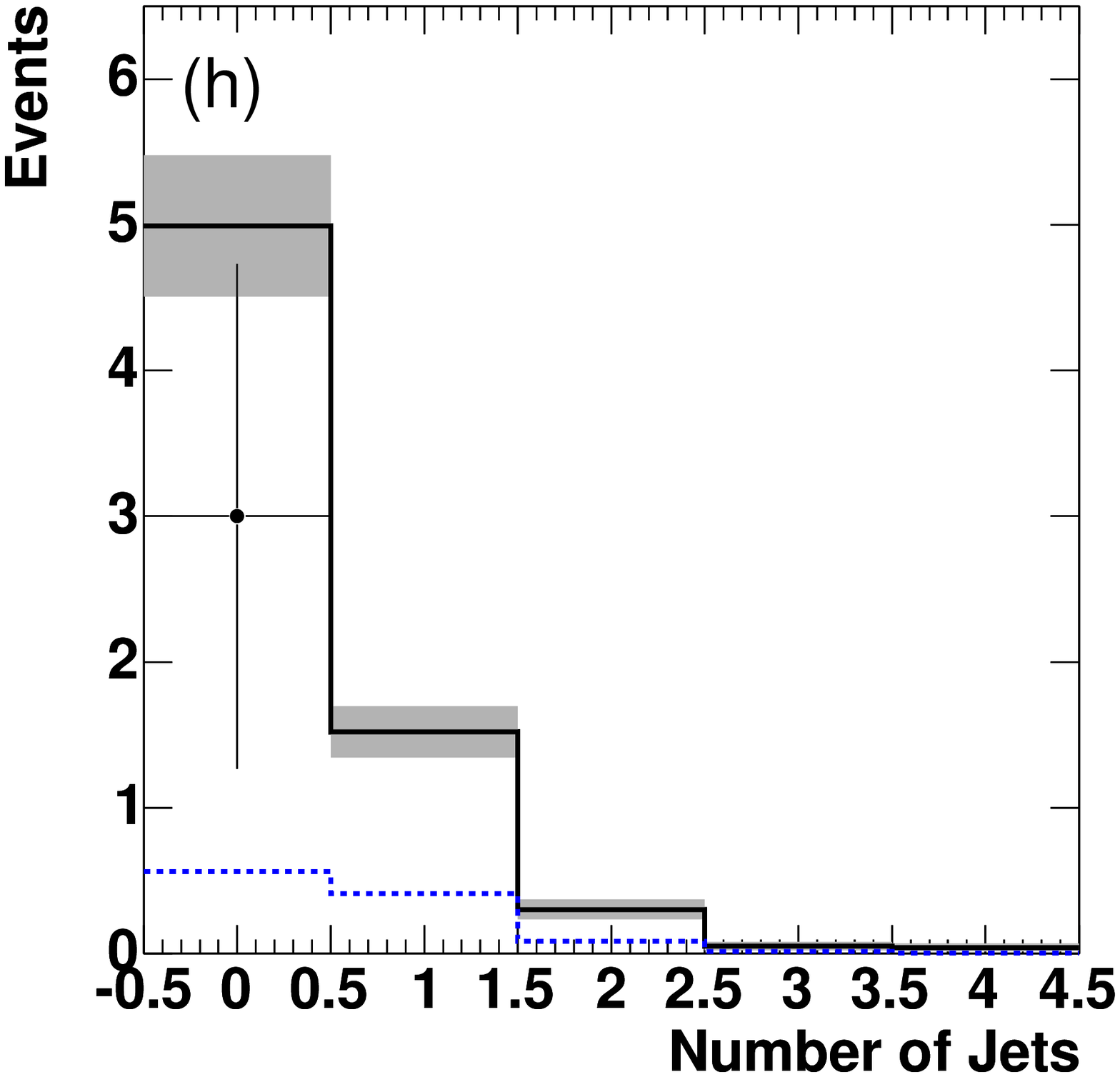}
\end{center}
\caption  		 
{\label{fig:combinee}
Kinematic distributions of the $\gamma\gamma e$ events from the SM (dashed line) and total (solid line) background 
predictions as well as the three events observed in the data (marker). The total backgound includes SM and fake 
contributions. The gray boxes indicate the uncertainty in background determination. Each photon is required to 
have an $E_T$$>$13~GeV. Distributions from the top left to the bottom right are: a) three-body invariant mass; 
b) invariant mass of two photons; c) invariant mass of each electron-photon pair; d) $E_T$ of each photon; 
e) $E_T$ of the electron; f) $H_T$, scalar sum of $\mett$ and $E_T$ of all identified photons, electrons, and 
jets; g) $\mett$; and h) number of jets with $E_T$$>$15~GeV.  
}
\end{figure*}

\begin{figure*}[tbp]
\begin{center}
\includegraphics[width=0.3\linewidth]
{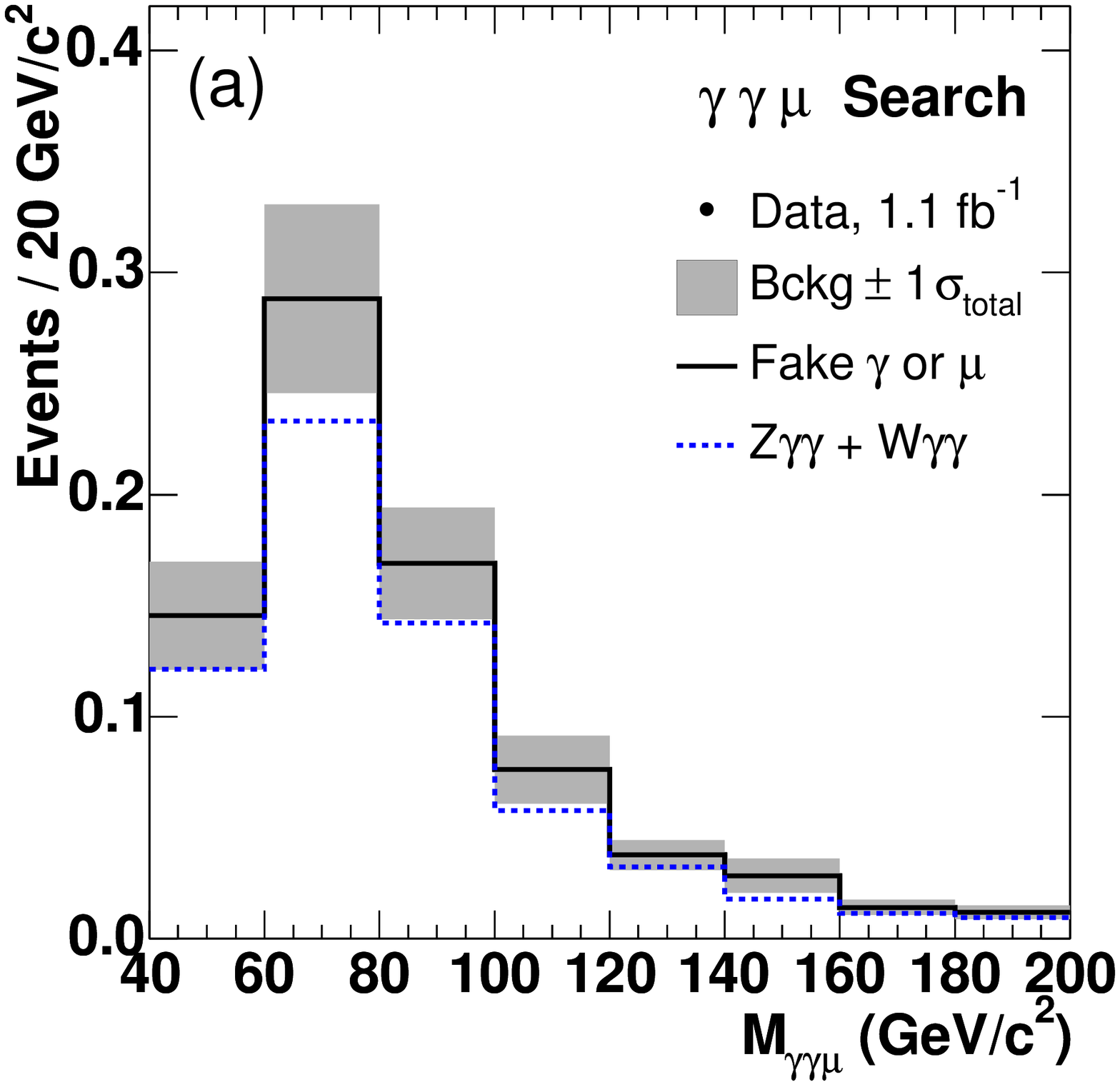}
\includegraphics[width=0.3\linewidth]
{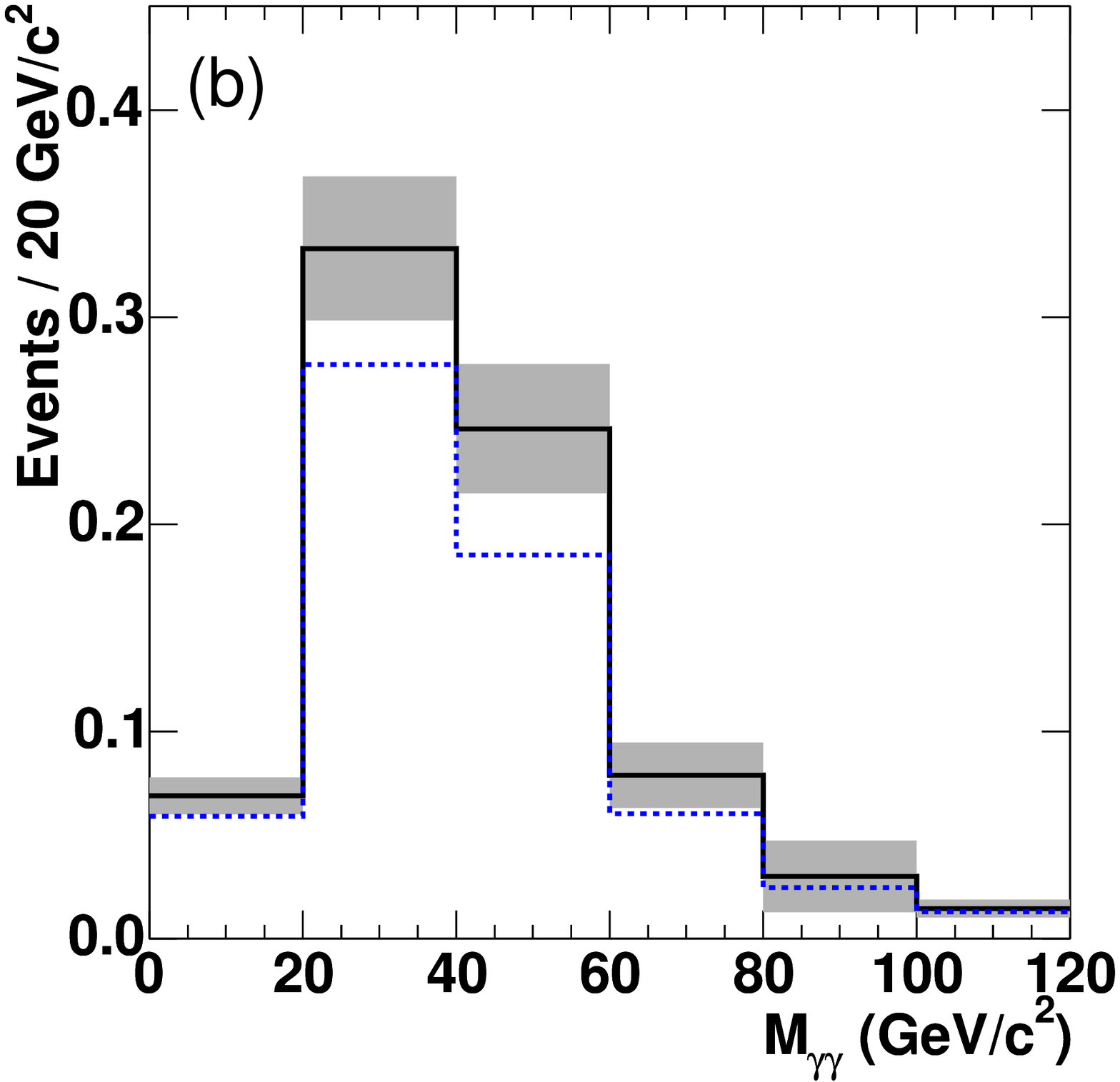}
\includegraphics[width=0.3\linewidth]
{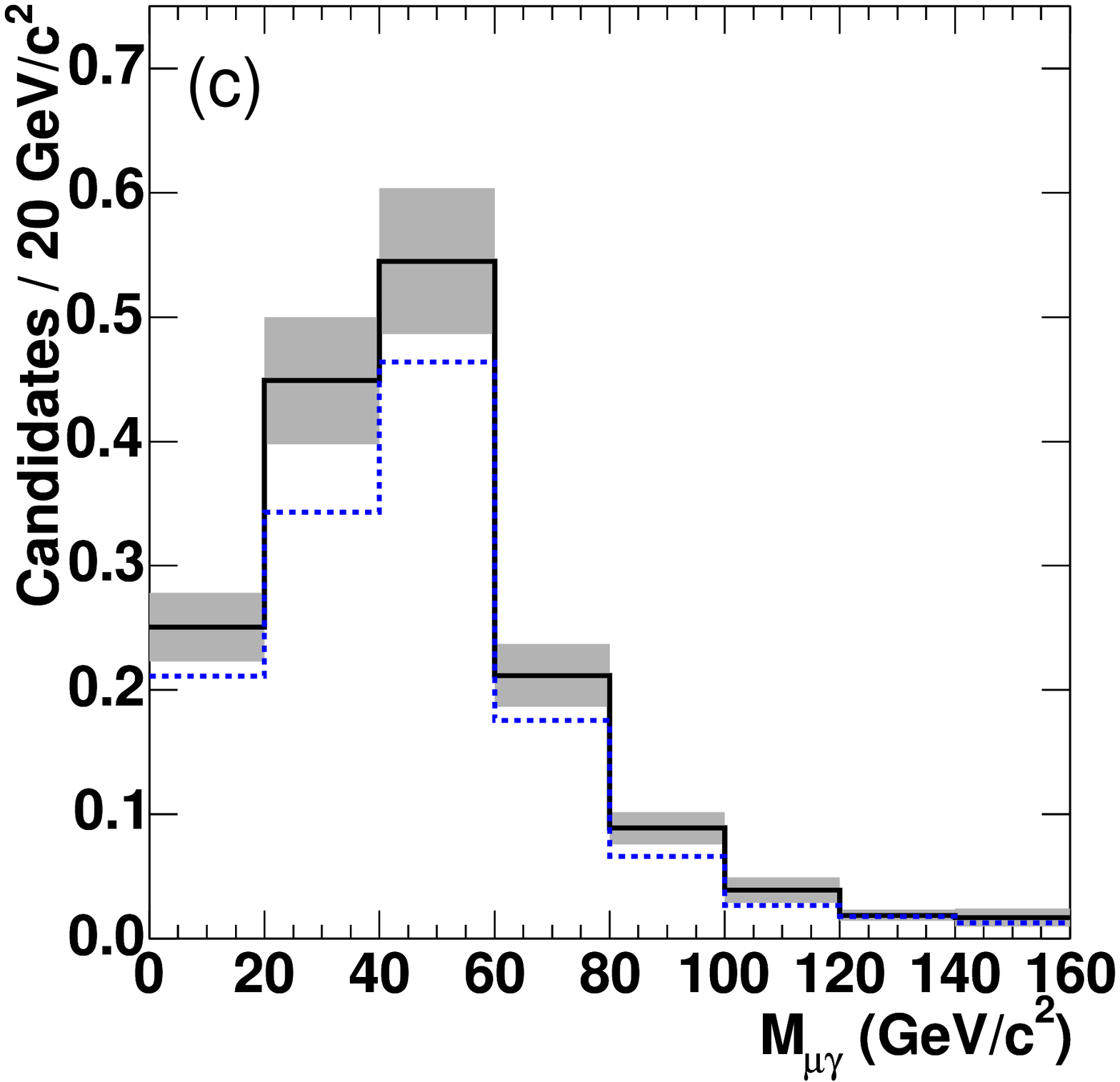}
\includegraphics[width=0.3\linewidth]
{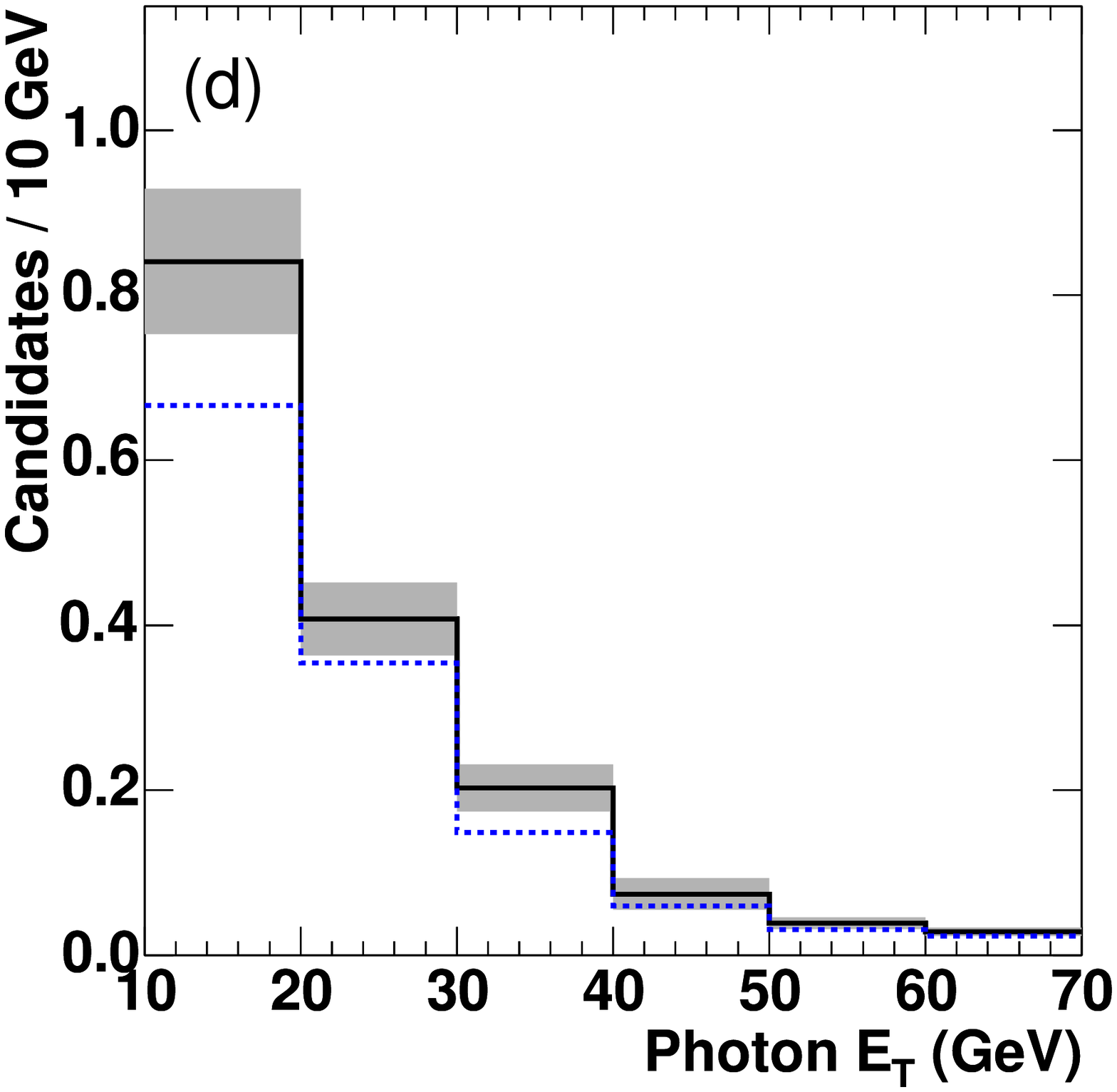}
\includegraphics[width=0.3\linewidth]
{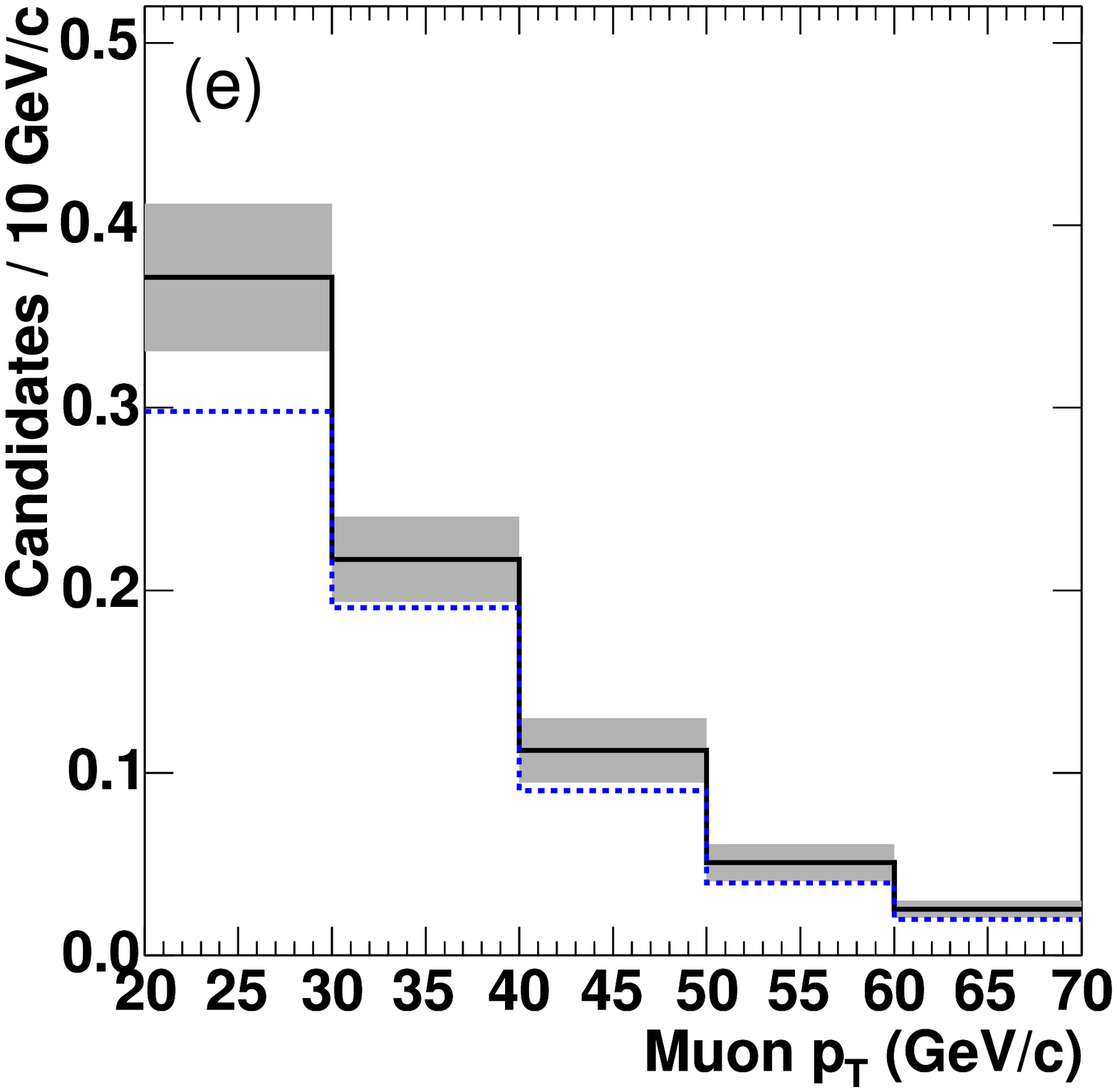}
\includegraphics[width=0.3\linewidth]
{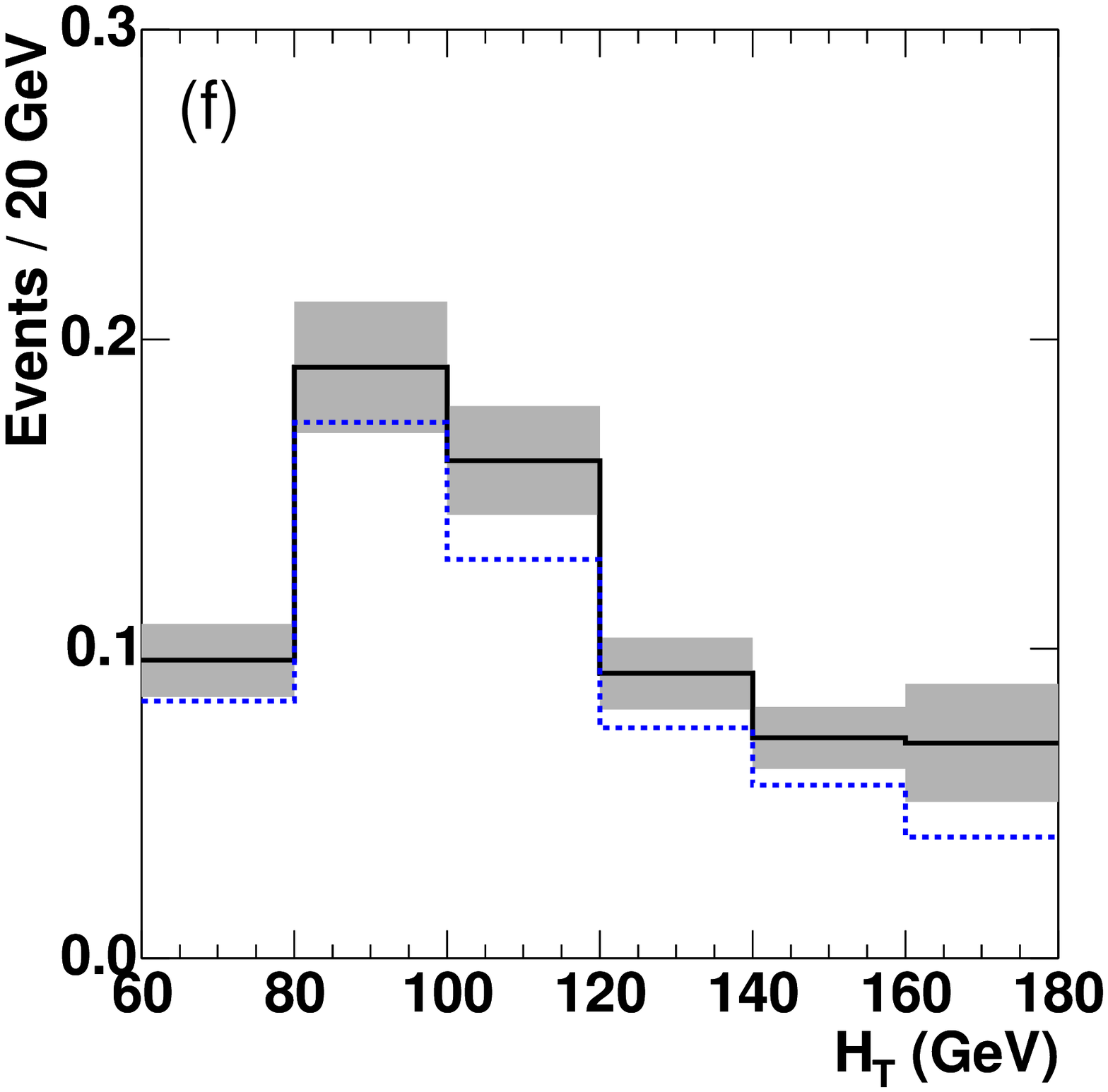}
\includegraphics[width=0.3\linewidth]
{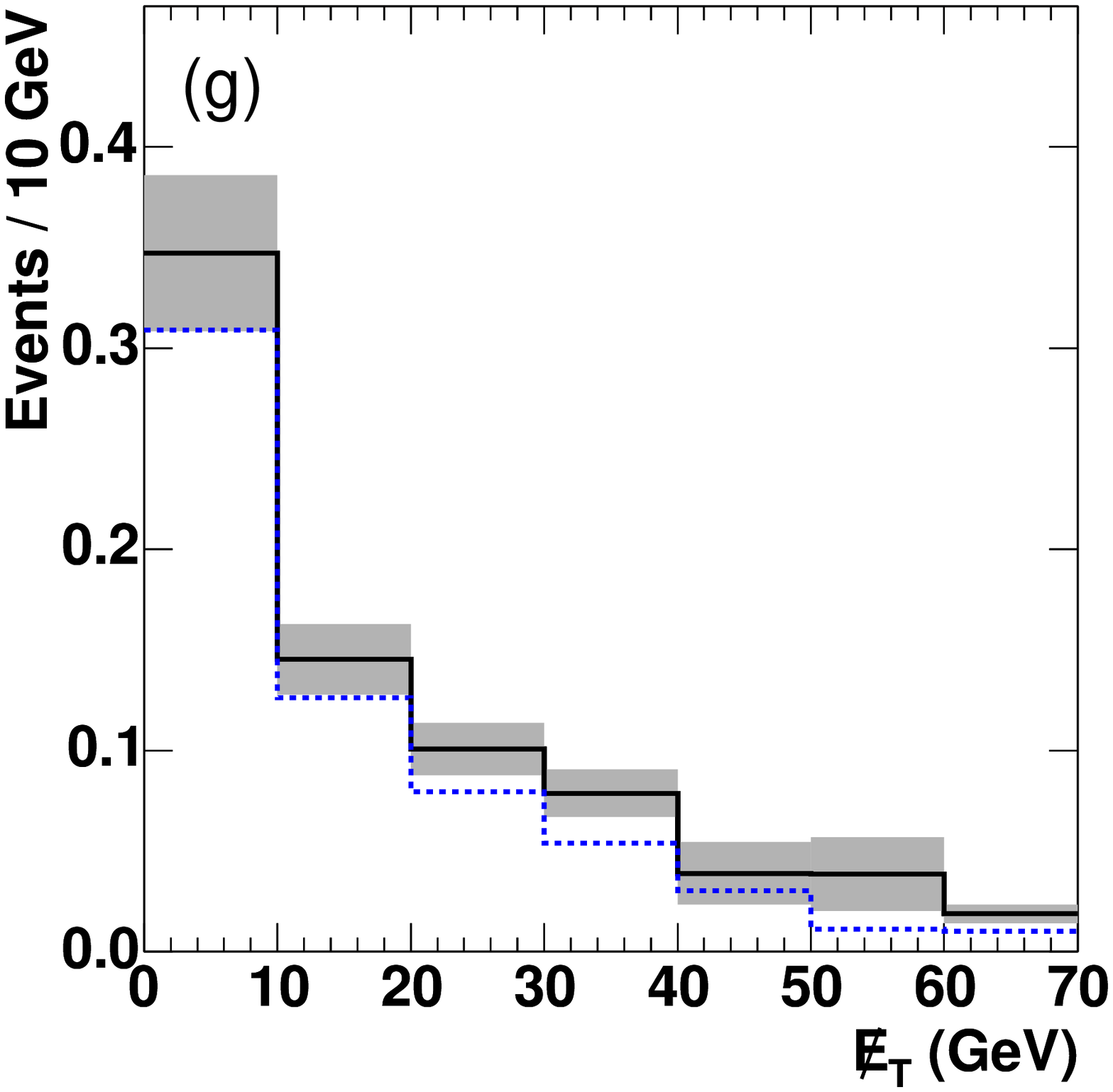}
\includegraphics[width=0.3\linewidth]
{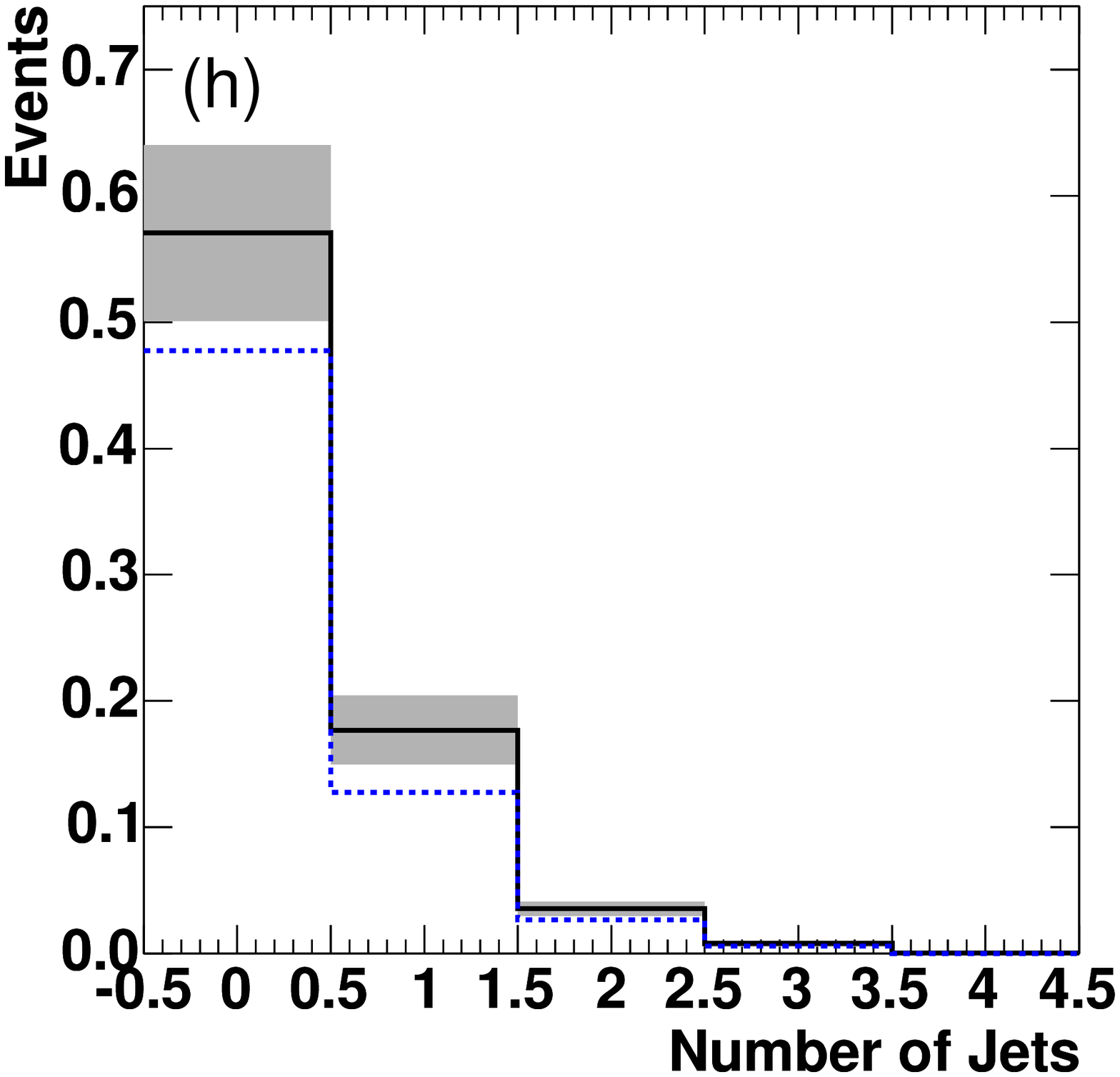}
\end{center}
\caption  		 
{\label{fig:combinem}
Kinematic distributions of the $\gamma\gamma e$ events from the SM (dashed line) and total (solid line) background 
predictions. The total backgound includes SM and fake contributions. The gray boxes indicate the uncertainty in 
background determination. We observe zero events in the data. Each photon is required to have an $E_T$$>$13~GeV. 
Distributions from the top left to the bottom right are: a) three-body invariant mass; b) invariant mass of two 
photons; c) invariant mass of each muon-photon pair; d) $E_T$ of each photon; e) $p_T$ of the muon; f) $H_T$, 
scalar sum of $\mett$ and $E_T$ of all identified photons, muons, and jets; g) $\mett$; and h) number of jets 
with $E_T$$>$15~GeV. } 	
\end{figure*}

Table~\ref{t:prgrandsignal} lists the expected and observed numbers of $\gamma\gamma e$ and $\gamma\gamma\mu$ 
events for $E_T^{\gamma}>13$~GeV. At this stage of event selection, we observe three $\gamma\gamma e$ events and 
zero $\gamma\gamma\mu$ events. The leading background in the $\gamma\gamma e$ channel is due to events where at 
least one of the photons is a misidentified electron. The leading background in the $\gamma\gamma\mu$ channel is 
the electroweak production of $Z\gamma\gamma$ events. Figures~\ref{fig:combinee}--\ref{fig:combinem} show several 
important kinematic distributions, including invariant mass, electron and photon $E_T$, $\mett$, jet multiplicity, 
and $H_T$ from data and the predicted backgrounds before applying the final selection, the silicon-track rejection 
(described next).
%---- Table
\begin{table}[!ht]
\begin{small}
\caption{Summary of the predicted and observed numbers of $\gamma\gamma e$ and $\gamma\gamma\mu$ events before 
applying silicon-track rejection. The systematic uncertainty includes uncertainty due to MC statistics, 
uncertainties in the data luminosity, predicted cross sections, and the misidentification probabilities. 
\label{t:prgrandsignal}}
\begin{center}
\renewcommand{\tabcolsep}{0.08in}
\begin{tabular}{cll}
\hline 
\hline 
 Source  &  \multicolumn{1}{c}{electron} &  \multicolumn{1}{c}{muon}  \\ \hline
 $Z\gamma\gamma$ & 0.90 $\pm$ 0.09 
	         & 0.55 $\pm$ 0.05  \\
 $W\gamma\gamma$ & 0.17 $\pm$ 0.02
 	         & 0.09 $\pm$ 0.01  \\
 $l\gamma+e$$\to$$\gamma$ & 5.14 $\pm$ 0.68
		          & 0.02 $\pm$ 0.02 \\
 $l\gamma+\mathrm{jet}$$\to$$\gamma$ & 0.48 $\pm$ 0.31
 	                             & 0.13 $\pm$ 0.09 \\ 
 Fake $l$+$\gamma\gamma$ & 0.13 $\pm$ 0.05
	                 & 0.004 $\pm$ 0.004 \\   \hline
 Total & 6.82 $\pm$ 0.75
       & 0.79 $\pm$ 0.11 \\ \hline
 Data & 3 & 0 \\ \hline \hline
 \end{tabular}
 \end{center}
 \end{small}
 \end{table}
%--
\begin{table}[!ht]
\begin{small}
\caption{Summary of the predicted and observed numbers of $\gamma\gamma e$ and $\gamma\gamma\mu$ events after 
applying silicon-track rejection. The systematic uncertainty includes uncertainty due to MC statistics, 
uncertainties in the data luminosity, predicted cross sections, and the misidentification probabilities. 
\label{t:prgrandsignal_fnl}}
\begin{center}
\renewcommand{\tabcolsep}{0.08in}
\begin{tabular}{cll}
\hline 
\hline 
 Source  &  \multicolumn{1}{c}{electron} &  \multicolumn{1}{c}{muon}  \\ \hline
 $Z\gamma\gamma$ & 0.82 $\pm$ 0.08
	         & 0.50 $\pm$ 0.05
  \\
 $W\gamma\gamma$ & 0.15 $\pm$ 0.02
 	         & 0.08 $\pm$ 0.01
  \\
 $l\gamma+e$$\to$$\gamma$ &  2.26 $\pm$ 0.46
		          &  0.004 $\pm$ 0.004 
 \\  
 $l\gamma+\mathrm{jet}$$\to$$\gamma$ & 0.44 $\pm$ 0.26
 	                             & 0.12 $\pm$ 0.08 \\ 
 Fake $l$+$\gamma\gamma$ &  0.12 $\pm$ 0.05
	                 &  0.004 $\pm$ 0.004 \\ \hline
 Total & 3.79 $\pm$ 0.54
       & 0.71 $\pm$ 0.10 \\ \hline
 Data & 1 & 0 \\ \hline
\hline
 \end{tabular}
 \end{center}
 \end{small}
 \end{table}
%__________________________________________________________________________
%
%  Figures of kinematic distributions for dipho+e/mu
%___________________________________________________________________________

The dominant source of background in the $\gamma\gamma e$ search are $\gamma+ee$ events where one of the 
electrons is misreconstructed as a photon. An electron may lose its track and be reconstructed as a photon 
because of catastrophic bremsstrahlung in the detector material in front of the COT. However, such an electron 
often leaves a few hits in the silicon detector and can be partially recovered by a special tracking algorithm 
(see Appendix~\ref{sec:photon} and Ref.~\cite{phoenix} for more details). We further compare the data and 
background prediction after removing events where at least one of the photons is matched to this type of 
electron track (silicon-track rejection). The silicon-track rejection suppresses $\sim$80$\%$ of fake photons 
from the electron bremsstrahlung for $E_T^{\gamma}$$>$45~GeV (see Fig.~\ref{fig:phxeff} from 
Appendix~\ref{sec:Jet2eleFkRt}) while it has only $\sim$1$\%$ inefficiency for real photons. Once this procedure 
is applied, the observed number of $\gamma\gamma e$ events is reduced to one. The final background predictions 
after the silicon-track rejection can be found in Table~\ref{t:prgrandsignal_fnl}.

%The robustness of our background estimation techniques is checked by using data from a different trigger and 
%independent methods for measuring the misidentification probabilities. Instead of counting electrons that satisfy 
%standard central-electron criteria, we use a number of electrons that satisfy photon-like electron criteria in the 
%$e$$\to$$\gamma$ fake rate denominator (see Appendix~\ref{sec:e2phoFkRt}). In addition, we cross-check the number 
%of $e\gamma + e$ events (to which the $e$$\to$$\gamma$ fake rate is applied) with $Z\gamma$ {\sc madgraph} MC. 
%These tests yield results that are consistent with the prediction described earlier. We also examine the background 
%estimate in larger samples, where either one photon and one electron ($e\gamma$) or one photon, one electron, and 
%one jet ($e\gamma j$) are required. The numbers of $e\gamma$ and $e\gamma j$ events in the data are in good 
%agreement with those from the background prediction. 

The robustness of our background estimation technique is validated in the following three ways. First, we use 
an independent method to measure the misidentification probabilities. Instead of counting electrons that satisfy 
standard central-electron criteria, we use the number of electrons that satisfy photon-like electron criteria in 
the $e$$\to$$\gamma$ fake rate denominator (see Appendix~\ref{sec:e2phoFkRt}). The difference in the prediction 
of fake photons from $ee\gamma$ events is $\sim$4-11$\%$. Second, we cross-check if the $ee\gamma$ data 
(to which the $e$$\to$$\gamma$ fake rate is applied) contain significant number of fake electrons. We fit 
data with a combined likelihood of multiple electron identification variables, where the signal shapes are
obtained from electrons in $Z$ decays and the background shapes are obtained from the sample enriched with fake 
electrons. The purity of electrons is estimated to be 97$\pm$2$\%$. In addition, we compare the yields of 
$ee\gamma$ and $\mu\mu\gamma$ events in data and those predicted by $Z\gamma$ {\sc madgraph} MC. We divide the 
ratio of data to MC yields in the muon channel by the same ratio in the electron channel. If the $ee\gamma$ 
events contain significant amount of fake electrons, the double ratio will be inconsistent with unity. The double 
ratio is found to be 1.10$\pm$0.15. Third, we examine the background estimate in larger samples, where either one 
photon and one electron ($e\gamma$) or one photon, one electron, and one jet ($e\gamma j$) events are required. The 
numbers of $e\gamma$ and $e\gamma j$ events in data are consistent with those from the background predictions 
within one standard deviation.

To summarize, we do not observe any evidence for anomalous production of $\gamma\gamma e$ and $\gamma\gamma\mu$ 
events. 

%%__________ gg+tau ___________________________________________________________________
\subsection{The $\gamma\gamma+\tau$ Final State}
\label{sec:dipho+tau}

\begin{figure}[htb]
\begin{centering}
\includegraphics[width=0.9\linewidth]{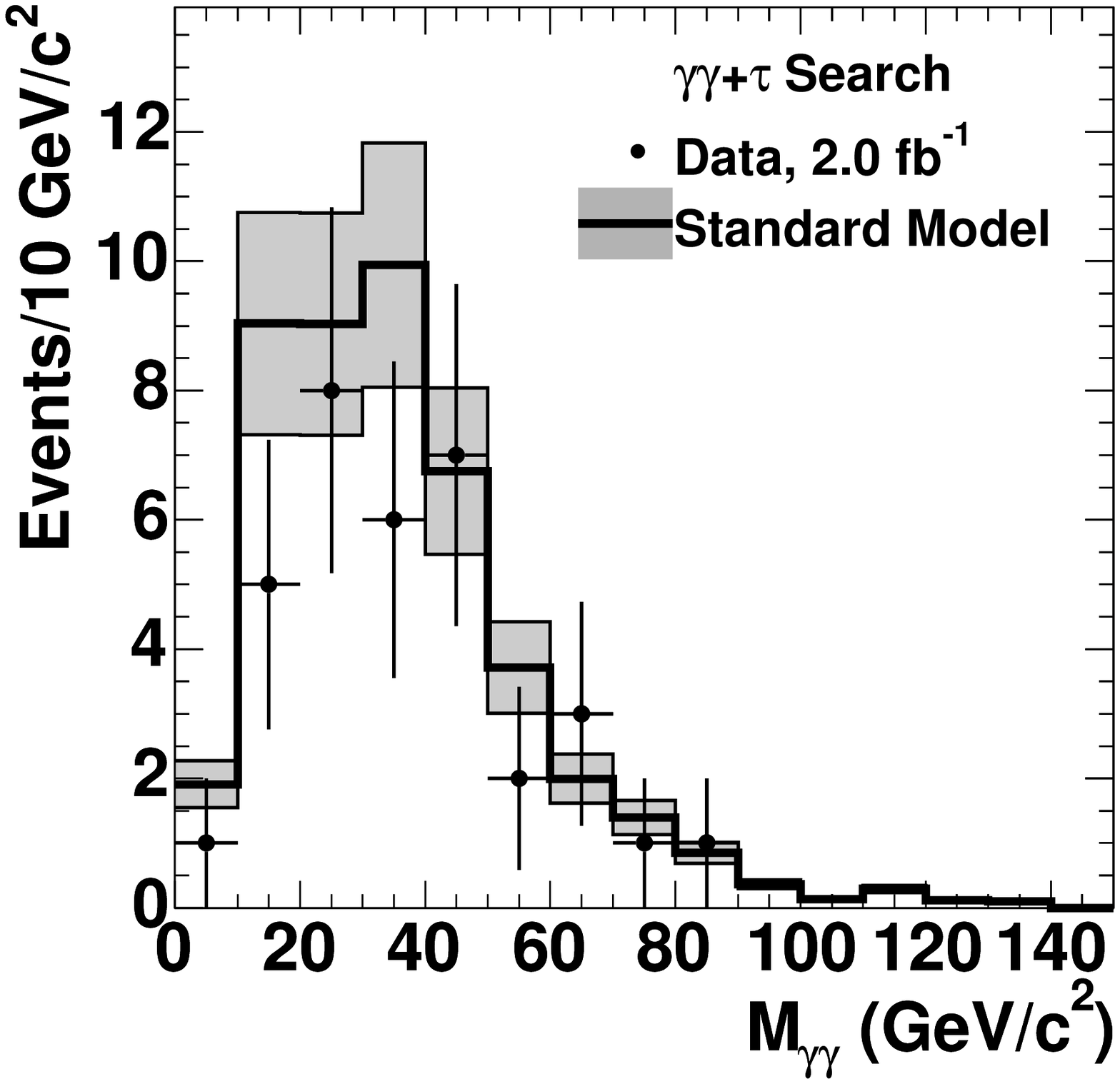}
\caption{
\label{fig:ggtau_fig1}
The mass of the two photons in $\gamma\gamma+\tau$ candidate events (marker) and the SM backgrounds (histogram). 
The gray boxes indicate the uncertainty in background determination.
}
\end{centering}
\end{figure}

We search for events with two photons and a hadronically-decaying $\tau$ lepton in data corresponding to 
2.0~fb$^{-1}$ of integrated luminosity. These events are a subset of the baseline diphoton sample (see 
Section~\ref{sec:data}) with at least one $\tau$ lepton candidate identified using the tight requirements 
and passing $E_T>$15~GeV (see Appendix~\ref{sec:tau}). We select 34 $\gamma\gamma+\tau$ candidate events. 

We consider two sources of backgrounds: the SM production of $W$$\to$$\tau\nu$ or $Z$$\to$$\tau\tau$ 
with photons and $\gamma\gamma$ events with jets misidentified as $\tau$ leptons. Other backgrounds are
negligible.

The electroweak backgrounds are estimated from $W\gamma$ and $Z\gamma$ {\sc madgraph}~\cite{madgraph} MC
simulation. The LO order predictions are multiplied by the appropriate next--to--leading--order $K$-factors
described in Section~\ref{sec:dipho+e/mu} and Ref.~\cite{cdf6601}. We find that these electroweak events with 
real leptons are dominated by events with at least one real photon, so we do not consider the case where both 
photons are misidentified jets. The simulation predicts the background from the cases of two real photons or 
one real with one fake photon to be 2.2$\pm$0.8 events, where the uncertainty comes from MC statistics.

The dominant background in this search is from events with two reconstructed photons (which may be real 
or fake), and jets, where one of the jets is misidentified as a $\tau$ lepton. To estimate this background, 
we select events with two photons and a jet identified as ``loose'' $\tau$ lepton candidate (see Appendix~\ref{sec:tau}) 
and apply the $jet$$\to$$\tau$ misidentification probability (see Appendix~\ref{sec:Jet2tauFkRt}). Since the 
misidentification probability is different for jets originated by quarks or by gluons, and the ratio of quark 
jets to gluon jets here may be different from the one in the sample used to derive the $jet$$\to$$\tau$ 
misidentification probability, we investigate a correction for the different types of jets in our sample. 
The probability for a quark jet to fake a $\tau$ lepton is three times larger than the probability for a 
gluon jet. The process becomes more complex because a photon candidate may also be a misidentified jet, and 
the probability for a quark jet to fake a photon is ten times larger than for a gluon jet. We use {\sc pythia} 
MC samples of diphotons, inclusive single photons and dijets to investigate the quark and gluon content of our 
data sample. Previous studies~\cite{run2diphoMET} have determined that the baseline diphoton sample has 
approximately 30$\%$ real diphoton events, 45$\%$ events with a real photon and a jet misidentifed as a photon, 
and 25$\%$ events with two jets misidentified as photons. The simulations indicate that the quark--to--gluon 
ratio is significantly higher in the case of one real photon and one fake photon (80$\%$ quarks) than either 
of the other cases (approximately 30$\%$ in dijet events and 40$\%$ in events with two real photons) and needs 
to be corrected for. We account for this effect by using two methods. In the first method, we simply apply the 
$jet$$\to$$\tau$ misidentification rate and then make a correction for the difference in the average 
quark--to--gluon ratio in the sample. In the second method, we allow for the possibility that quark jets will 
preferentially become misidentified photons, leaving the remaining jets to become misidentified taus. This 
method yields our reported central result, and the variation between the methods indicates a 13$\%$ systematic 
uncertainty which is added to the 20$\%$ systematic uncertainty in the misidentification probability.

\begin{figure}[hbt]
\centering
\includegraphics[width=1.0\linewidth]{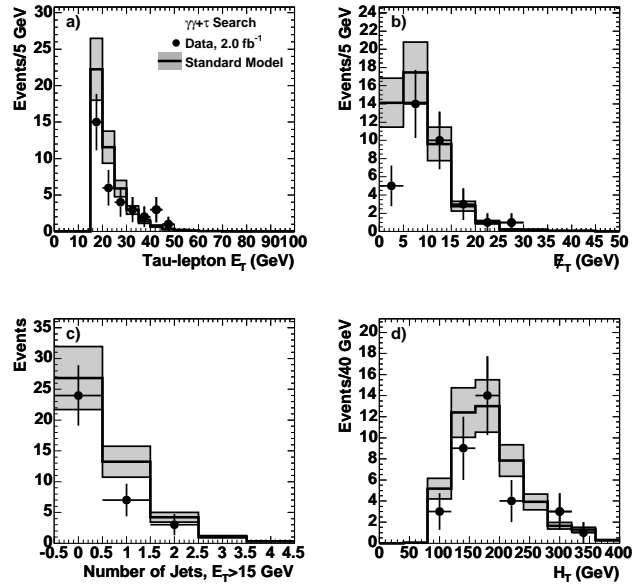}
\caption{The kinematic distributions in $\gamma\gamma+\tau$ candidate events (marker) and the SM backgrounds 
(histogram): a) $E_T$ of $\tau$ lepton candidate; b) $\mett$; c) number of jets with $E_T$$>$15~GeV; d) $H_T$, 
scalar sum of the transverse energies of photons, $\tau$ lepton candidate, jets, and $\mett$. The gray boxes 
indicate the uncertainty in background determination.}
\label{fig:ggtau_fig2}
\end{figure}

The misidentified $\tau$ background is 44$\pm$10 events and the total background estimate is 46$\pm$10 events, 
consistent with the 34 observed $\gamma\gamma+\tau$ candidate events. We perform three checks of the methodology 
by predicting the size of $\gamma\gamma+\tau$ sample where the two photons are selected with the relaxed criteria 
($\gamma\gamma$ control sample described in  Section~\ref{sec:data}) or with one of the photons in the forward 
region (1.1$<$$|\eta|$$<$2.0). The predictions for all control samples are consistent with the observations.
Figures~\ref{fig:ggtau_fig1}-\ref{fig:ggtau_fig2} show several important kinematic distributions for the selected 
$\gamma\gamma+\tau$ candidate events and the predicted SM background. These distributions include the diphoton 
invariant mass, $E_T$ of a $\tau$ lepton candidate, $\mett$, jet multiplicity, and $H_T$. No excess is found 
above the SM background.

In summary, we do not observe any evidence for the anomalous production of $\gamma\gamma+\tau$ events.

%%__________ gg+met _____________________
\subsection{The $\gamma\gamma+\mett$ Final State}
\label{sec:dipho+met}
%---------------- Metall -----------------------------------------------------------------
\begin{figure}[ht]
\centering
\includegraphics[width=0.9\linewidth]{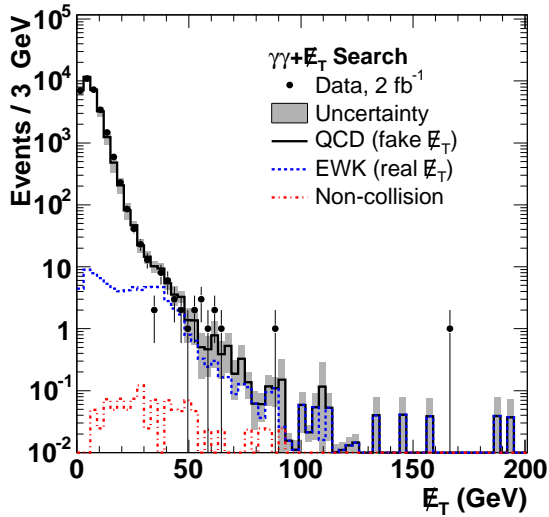}
\caption{The $\mett$ distribution in all $\gamma\gamma$ candidate events from the baseline sample. The data 
(marker) is compared with the total background prediction (solid line with the gray band representing the 
total uncertainty). The total background prediction is a sum (shown by the stacked histograms) of the QCD, 
electroweak (dashed line), and non-collision (dash-dotted line) backgrounds.
}
\label{fig:ggmet_fig1}
\end{figure}

\begin{figure}[htb]
\centering
\includegraphics[width=0.9\linewidth]{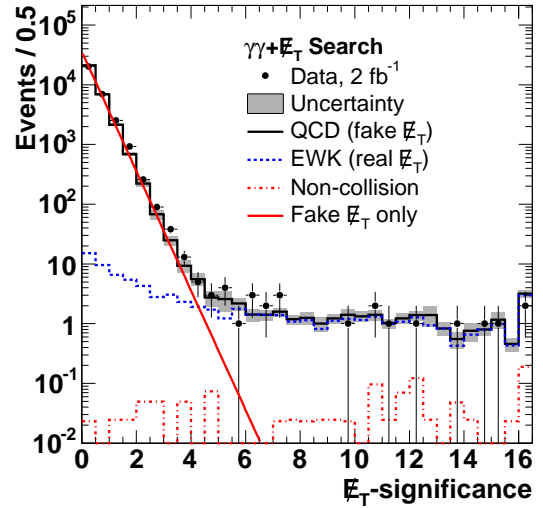}
\caption{The \mett-significance distribution in all $\gamma\gamma$ candidate events from the baseline sample. 
The data (marker) is compared with the total background prediction (solid line with the gray band representing 
the total uncertainty). The total background prediction is a sum (shown by the stacked histograms) of the QCD, 
electroweak (dashed line), and non-collision (dash-dotted line) backgrounds. The straight solid line represents 
the expected \mett-significance distribution if all $\gamma\gamma$ candidate events were to have fake $\mett$ 
due to the measurement fluctuations in the calorimeter (see Appendix~\ref{sec:MetModel} for more details).
}
\label{fig:ggmet_fig2}
\end{figure}

\begin{figure*}[!ht]
\centering
\includegraphics[width=0.45\linewidth]{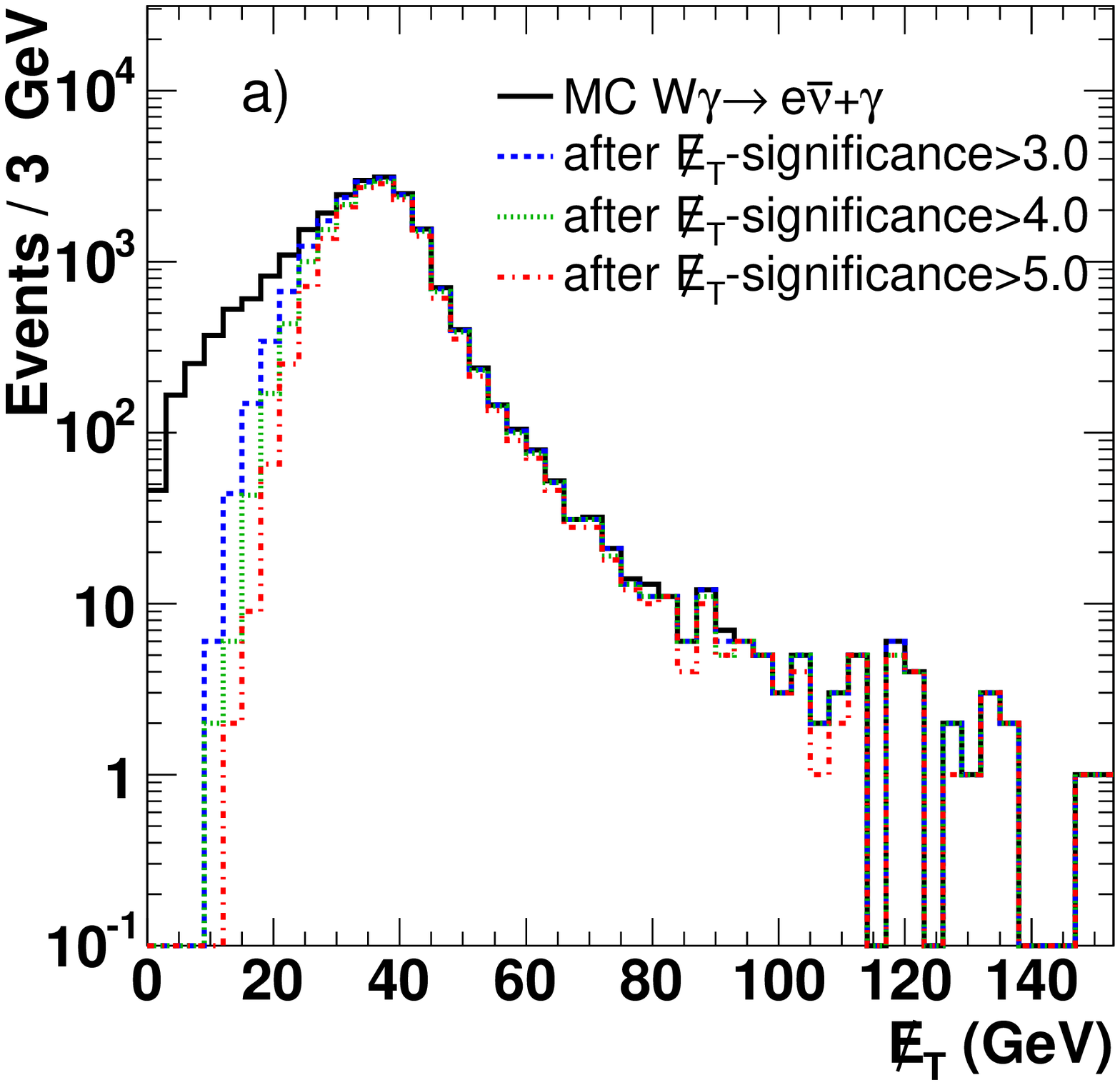}
\includegraphics[width=0.45\linewidth]{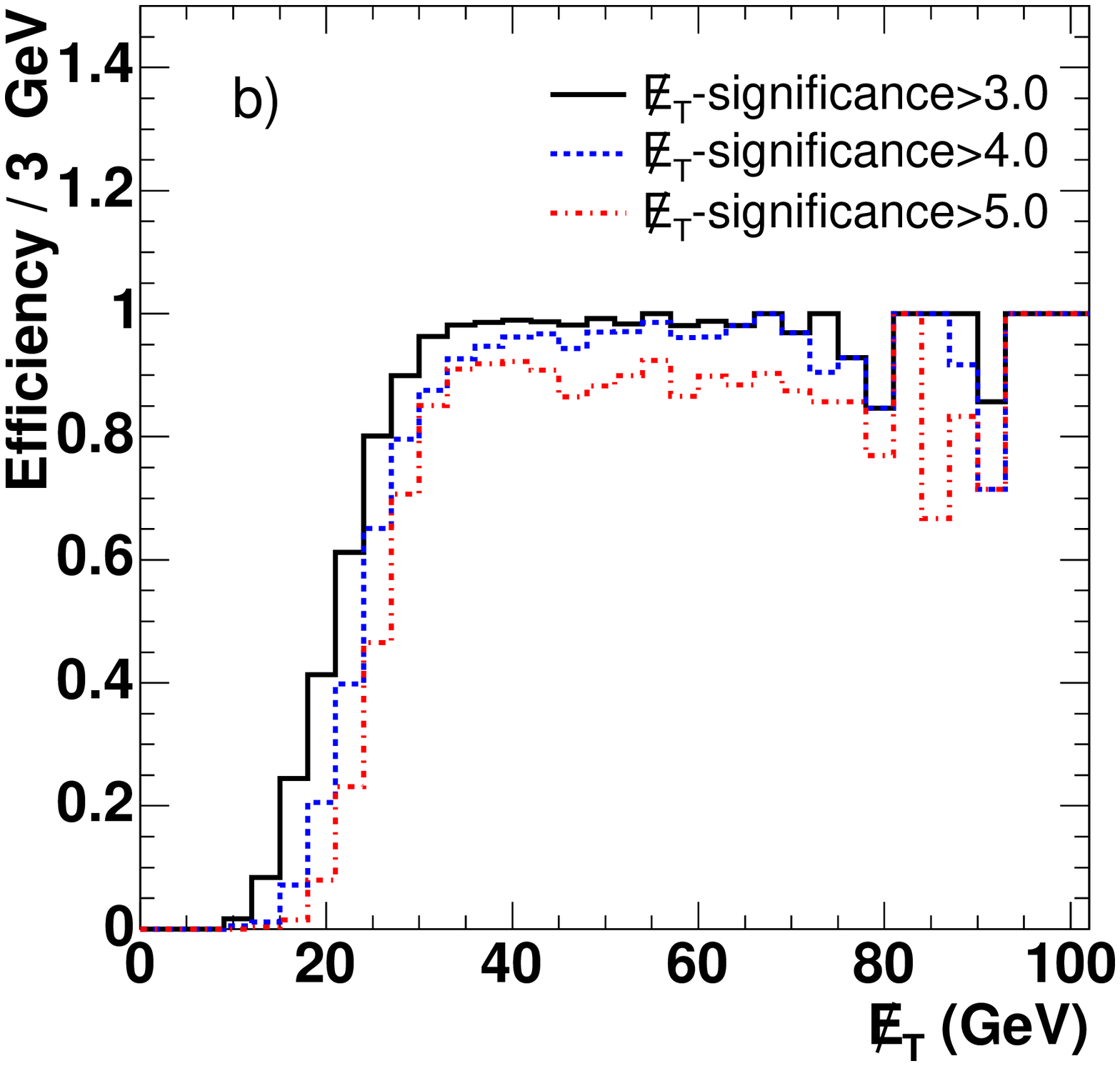}
\caption{Figure $a)$ demonstrates the $\mett$ distributions in MC $W\gamma$$\to$$e\nu$$+$$\gamma$ events 
before and after the \mett-significance cuts. Figure $b)$ illustrates the efficiency of the three 
\mett-significance cuts for these events.}
\label{fig:ggmet_metsigEff}
\end{figure*}

We search for the anomalous production of two photons and large missing transverse energy ($\mett$) in 
data corresponding to 2.0~fb$^{-1}$ of integrated luminosity. The subsample of $\gamma\gamma+\mett$ events 
is derived from the baseline diphoton sample described previously in Section~\ref{sec:data} and in 
Appendix~\ref{sec:photon}. The missing transverse energy is defined as an energy imbalance in the calorimeter 
(see detailed description of $\mett$ in Appendix~\ref{sec:MET}) and it is an experimental signature of 
neutrinos or new particles that do not interact electromagnetically or strongly with the detector material. 
The $\mett$, however, can be mimicked by a simple energy misreconstruction in SM events. Fluctuations in jet 
energy measurements are the most common source of such fake $\mett$. Figure~\ref{fig:ggmet_fig1} shows the 
$\mett$ distribution in the $\gamma\gamma$ baseline sample. This figure illustrates that events with fake 
$\mett$ are not only the dominant background in the region up to $\mett$$\sim$40 GeV, but they also have a 
significant contribution even to the tail of $\mett$ distribution. A better separation between events with 
real and fake $\mett$ can be achieved if a significance of the measured $\mett$ is considered rather than its 
absolute value. The $\mett$-significance is a dimensionless quantity based on the energy resolution of jets 
and soft unclustered particles. It also takes into account the event topology as shown in 
Appendix~\ref{sec:MetModel}. As it is demonstrated in Fig.~\ref{fig:ggmet_fig2}, the $\mett$-significance 
distributions have very different shapes in events with fake and real $\mett$: exponentially falling (solid 
line) and almost flat shapes, respectively. Thus, the \mett-significance is an efficient tool in separating 
such events. We study $\gamma\gamma+\mett$ events which pass three {\it a priori} \mett-significance requirements: 
\mett-significance$>$3,~4,~and~5. This choice of cut values has a straightforward motivation. If the 
$\gamma\gamma$ sample were only composed of events with fake $\mett$ due to energy misreconstruction in the 
calorimeter, then we would select 0.1$\%$, 0.01$\%$, and 0.001$\%$ of the total number of events by requiring 
\mett-significance$>$3,~4,~and~5, respectively. On the other hand, studies with MC $W\gamma$$\to$$e\nu+\gamma$ 
sample indicate that the \mett-significance$>$3 (\mett-significance$>$5) cut is $\sim$100$\%$ ($\sim$90$\%$) 
efficient for events with real $\mett$$>$35 GeV (see Fig.~\ref{fig:ggmet_metsigEff}).
%---------------- Metsig -----------------------------------------------------------------
%---------------- Metsig Efficiency -----------------------------------------------------------------

We consider three major sources of background for the $\gamma\gamma+\mett$ signature: QCD ($\gamma\gamma$, 
$j\gamma$, and $jj$ where $j$=$jet$$\to$$\gamma_{fake}$) events with large fake $\mett$ due to energy loss 
or mismeasurement in the calorimeter, electroweak (EWK) processes with real $\mett$ from neutrinos, and 
non-collision events with fake photons and $\mett$. Each of these sources is discussed below in the order 
of their importance. All of the background estimation techniques are tested on a control sample of ``loose'' 
diphoton events described in Section~\ref{sec:data} and in Appendix~\ref{sec:photon}.
  
%--------- QCD backgrounds
Significant losses or fluctuations in energy measurements in the inclusive $\gamma\gamma$ sample, which 
can lead to considerable values of fake $\mett$, happen only in a small fraction of events. However, the 
large production cross sections of QCD processes make them one of the largest backgrounds. We distinguish 
three types of QCD backgrounds: events with energy mismeasurement due to calorimeter energy resolution 
effects (QCD type-1); $\gamma\gamma$ candidate events with a wrong choice of the primary interaction vertex 
(QCD type-2); and $\gamma\gamma\gamma$ events where one of the photon candidates is lost in the calorimeter 
cracks (QCD type-3). 

%--------- QCD type-1
The QCD type-1 background estimate is based on a $\mett$-resolution model ({\sc metmodel}) described 
in Appendix~\ref{sec:MetModel}. For each $\gamma\gamma$ data event, we generate ten pseudoexperiments to 
simulate fake $\mett$ and calculate its significance given the event kinematics. In each pseudo-experiment, 
we smear the energies of jets and soft unclustered particles using appropriate resolutions. The difference 
between the smeared and measured energy of the object is taken as its individual contribution to the total 
fake $\mett$. We predict the QCD background due to energy mismeasurements by counting the number of 
pseudoexperiments that pass our \mett-significance requirements divided by the number of pseudoexperiments 
per event. Kinematic distributions from these pseudoexperiments are then used as QCD background templates 
for data. The systematic uncertainty (23$\%$ for \mett-significance$>$3, 47$\%$ for \mett-significance$>$4, 
and 130$\%$ for \mett-significance$>$5) is evaluated by comparing the {\sc metmodel} expectations obtained 
with the default parameters to predictions obtained by varying each parameter by one standard deviation 
($\pm 1~\sigma$). These parameters and associated sources of systematic uncertainties are discussed in  
Appendix~\ref{sec:MetModel}. The statistical and systematic uncertainties are added in quadrature to obtain 
the total uncertainty. The predictions for the QCD type-1 background and their associated uncertainties can 
be found in Table~\ref{tab:ggmetResults}.

%--------- QCD type-2
The background contribution due to $\gamma\gamma$ candidate events with mis-assigned primary vertex (QCD 
type-2) cannot be directly estimated by the {\sc metmodel} because the energy resolution parameterization 
does not include this effect. The vertex misassignment occurs when a $\gamma\gamma$ pair~\cite{wrongVx} is 
produced by a hard scattering interaction that overlaps with another interaction producing a vertex with 
the highest $\sum p_T$ of tracks. As a consequence of the wrong vertex choice, the $E_T$ of both photon 
candidates are incorrectly calculated, thus leading to fake $\mett$. Although the effect is small, it can 
occasionally result in a large fake $\mett$, for example, when two vertices are far apart and the photons 
are sufficiently energetic. We correct for these mismeasurements by recalculating the $E_T$ of photons with 
respect to the vertex which gives the smallest value of $\mett$. This procedure is verified to be valid for 
events with no intrinsic $\mett$. It is also tested in simulated $W\gamma$$\to$$e\nu\gamma$ events~\cite{sim} 
and data $e\gamma$ events with $\mett$$>$20~GeV. The selection of $e\gamma$ events is discussed in 
Appendix~\ref{sec:ElePho}. The effect is found to be small: after the procedure is applied, the number of 
simulated and data events with $\mett>20$ GeV is reduced by 1$\%$ and 2$\%$, respectively. In some fraction 
of events, however, the hard interaction completely fails to produce a reconstructed vertex and the vertex 
re-assignment cannot fix fake $\mett$. Since the {\sc metmodel} cannot account for this contribution, we 
employ a method based on a combination of data and Monte Carlo simulation to obtain the predictions. For 
this purpose, we use {\sc pythia} $\gamma\gamma$ events~\cite{pythia} passed through the detector 
simulation~\cite{geant}. These MC events also include additional interactions in the same bunch crossing 
that are modeled according to the luminosity profile in data. We select only events where the hard scattering 
interaction resulting in a $\gamma\gamma$ pair does not produce a reconstructed vertex, and the primary 
vertex is created by tracks from an overlapping additional interaction. We will refer to such events as 
``no vertex'' $\gamma\gamma$ events. The MC sample of ``no vertex'' events is normalized to the number of 
such events in real data (4.8$\pm$0.4$\%$ of the baseline $\gamma\gamma$ events). We then apply the standard 
analysis procedure to the sample and obtain the fraction of ``no vertex'' events in MC passing our 
\mett-significance cuts. The systematic uncertainties on the QCD type-2 background contribution include the MC 
statistical uncertainty (12$\%$-24$\%$), the uncertainty on the normalization factor (10$\%$), the uncertainty 
due to the jet energy scale (7-8$\%$), and the MC-data differences in the {\sc metmodel} parameterization 
(40$\%$). The predictions for the QCD type-2 background and their associated uncertainties can be found in 
Table~\ref{tab:ggmetResults}.   

%--------- QCD type-3
The $\gamma\gamma\gamma$ events are produced at a very low rate compared to that of $\gamma\gamma$ events. 
However a probability of losing a photon in calorimeter cracks is $\sim$10$\%$, so that the probability of 
losing one of the candidate photons in a {\it tri-photon} event is as large as $\sim$30$\%$. These events 
(QCD type-3) could reconstruct as $\gamma\gamma+\mett$ events. To reduce this background, we reject 
events if the $\mett$ vector points along the direction (within $|\Delta\phi|$$<$0.3) of a narrow jet~\cite{narrowjet} 
located close to the calorimeter cracks at $\eta$$\sim$0 and $|\eta|$$\sim$1.1. The remaining contribution 
of the QCD type-3 events is estimated using a large inclusive {\sc pythia} $\gamma\gamma$ MC sample. We select 
reconstructed tri-photon events ($E_T^{\gamma 1,2}$$>$13 GeV and $E_T^{\gamma 3}$$>$7 GeV) in MC and data. 
The numbers of reconstructed $\gamma\gamma\gamma$ candidates give us the MC-to-data normalization factor. 
To obtain an estimate of the remaining QCD type-3 background, we select {\sc pythia} tri-photon events at the 
generator level (before detector simulation), apply the standard analysis procedure to these events, and 
multiply the result by the normalization factor described above. The systematic uncertainties for this 
background prediction is due to the following sources: 1) MC statistical uncertainty (24$\%$-33$\%$); 
2) uncertainty on the normalization factor (19$\%$); 3) uncertainty due to MC-data differences in the 
{\sc metmodel} parameterization (10$\%$-44$\%$); 4) jet energy scale uncertainty (10$\%$-11$\%$). 
The predictions for all sources of QCD backgrounds and their associated uncertainties can be found in 
Table~\ref{tab:ggmetResults}.
%-----------------------------------------------------------------------------------------
\begin{table*}[tbp]
\begin{center}
\caption{The results of the search for anomalous production of $\gamma\gamma+\mett$ events. The data is
compared to the background predictions for three values of the \mett-significance cut. The quoted 
uncertainties include the effect of limited MC statistics as well as systematic uncertainties.
}
\label{tab:ggmetResults}
\begin{tabular}{cccc}
\hline
\hline
%-------------------------------------------------------------------------------------------
 & \mett-significance$>$3.0 & \mett-significance$>$4.0 & \mett-significance$>$5.0 \\ \hline 
%-------------------------------------------------------------------------------------------
EWK & 35.4$\pm$2.2 & 29.9$\pm$2.0 & 25.9$\pm$1.9 \\ \hline
%-------------------------------------------------------------------------------------------
QCD type-1 & 28.1$\pm$6.8 & 3.6$\pm$1.8 & 0.6$\pm$0.8 \\
%-------------------------------------------------------------------------------------------
QCD type-2 & 4.4$\pm$2.0 & 2.5$\pm$1.0 & 1.5$\pm$0.7 \\ 
%-------------------------------------------------------------------------------------------
QCD type-3 & 2.9$\pm$1.0 & 2.2$\pm$1.0 & 1.6$\pm$1.0 \\ \hline
%-------------------------------------------------------------------------------------------
Non-Collision & 0.9$\pm$0.3 & 0.8$\pm$0.3 & 0.8$\pm$0.3 \\ \hline
%-------------------------------------------------------------------------------------------
Total & 71.7$\pm$7.5 & 39.0$\pm$3.1 & 30.4$\pm$2.4 \\ \hline
%-------------------------------------------------------------------------------------------
Data & 82 & 31 & 23 \\ \hline\hline
%-------------------------------------------------------------------------------------------
\end{tabular}
\end{center}
\end{table*}

%---------------- met (sig>3,5) -----------------------------------------------------------------
\begin{figure*}[!ht]
\centering
\includegraphics[width=0.4\linewidth]{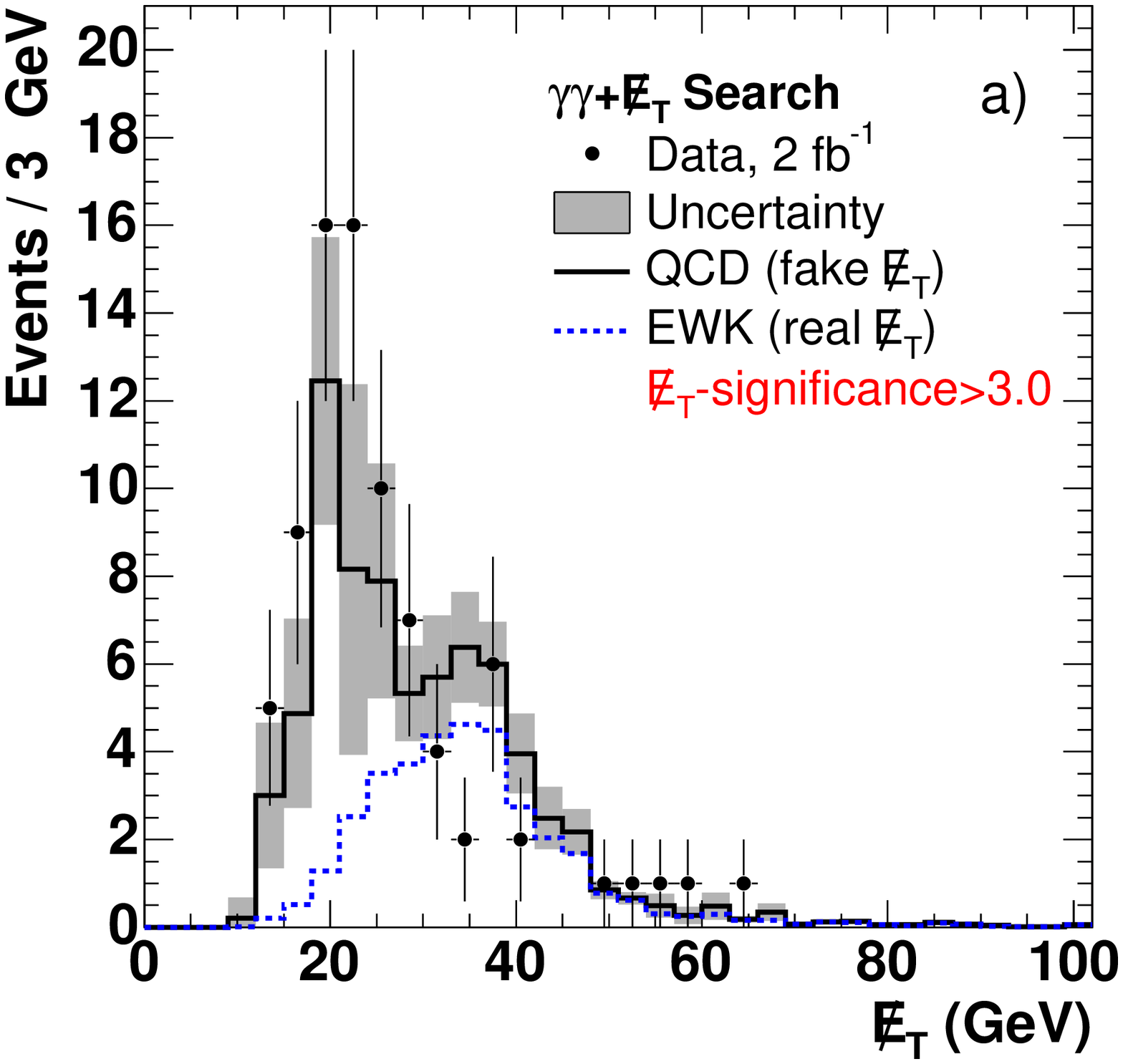}
\includegraphics[width=0.4\linewidth]{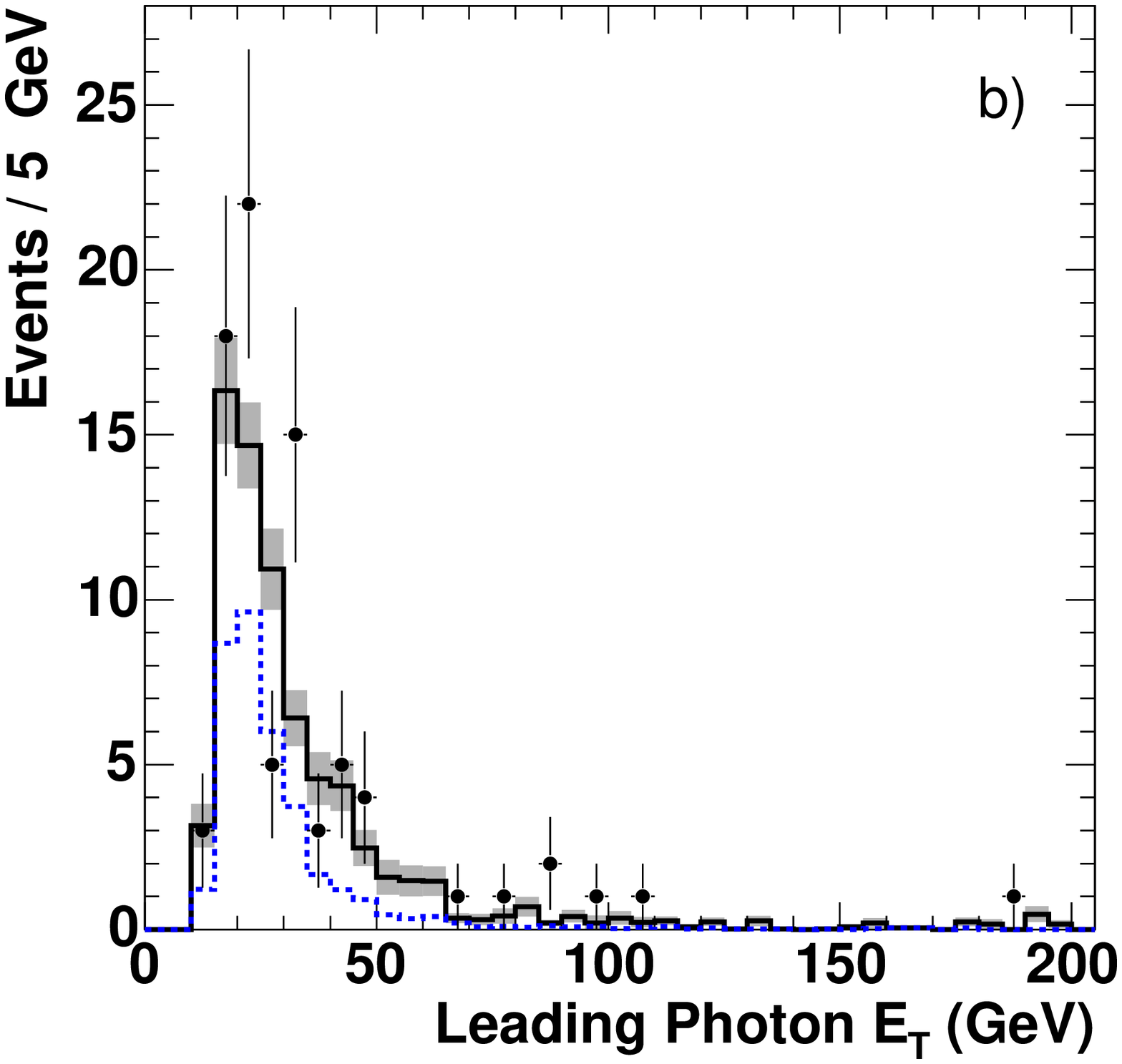}
\includegraphics[width=0.4\linewidth]{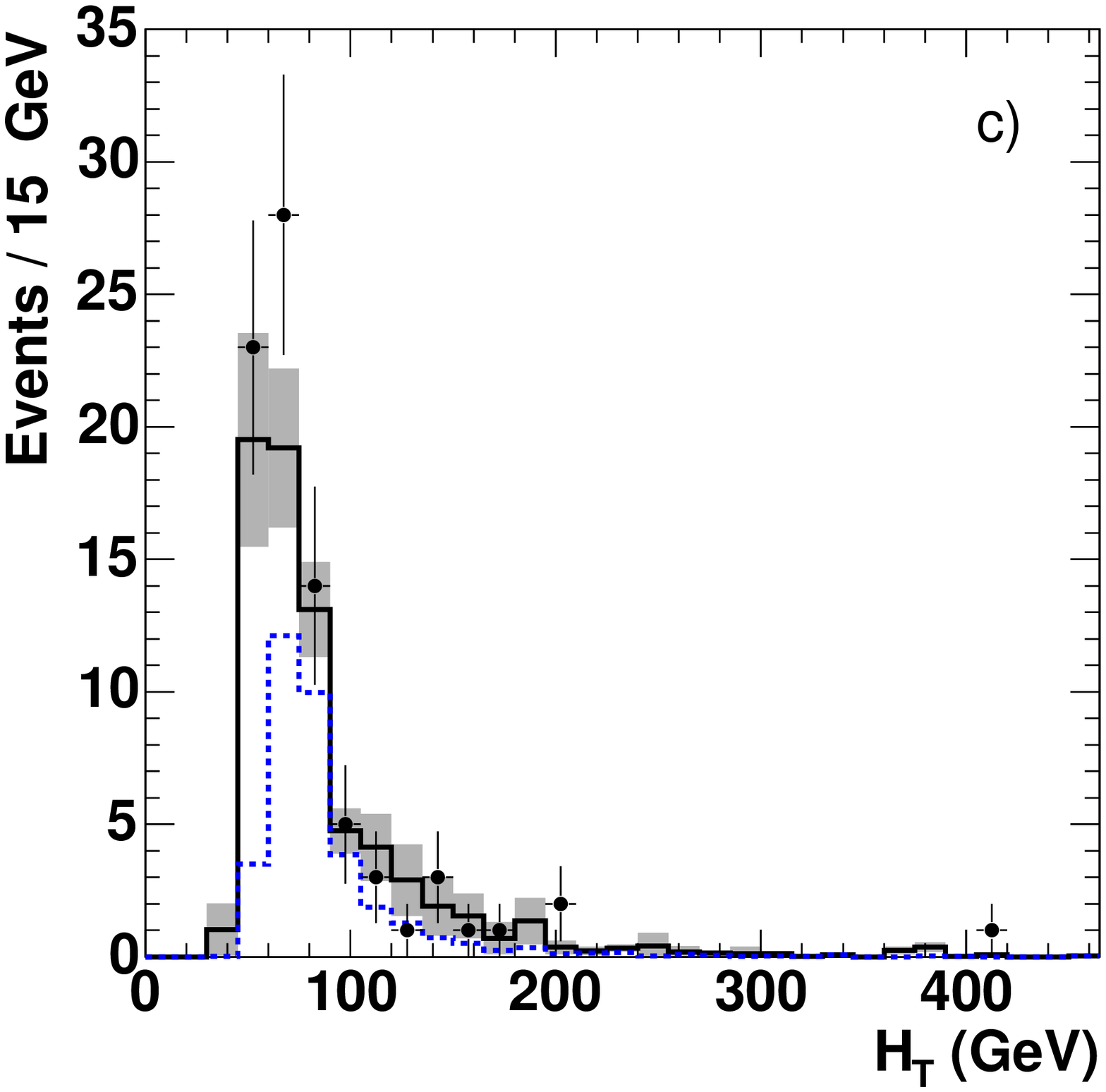}
\includegraphics[width=0.4\linewidth]{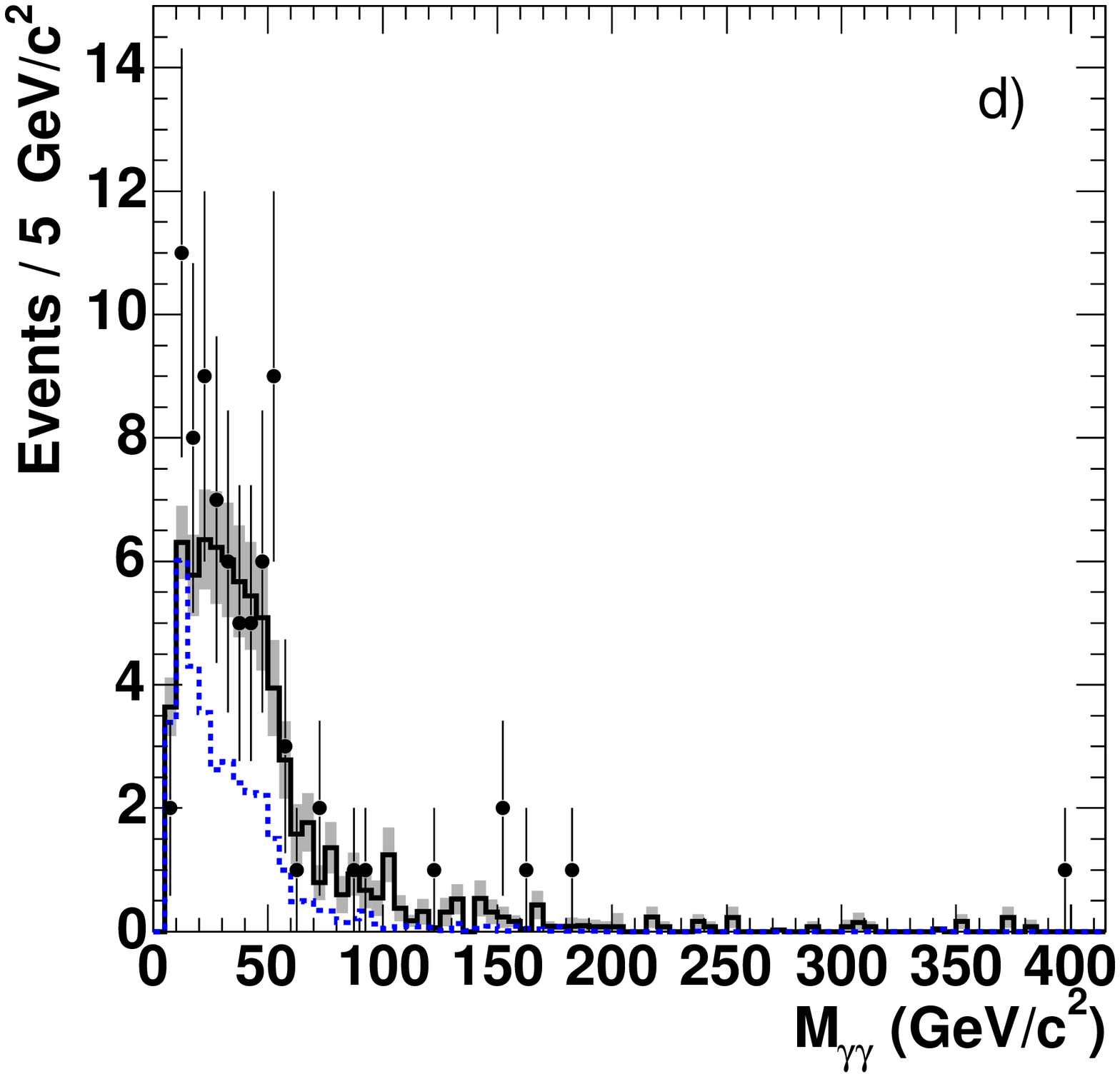}
\caption{The kinematic distributions for $\gamma\gamma+\mett$ candidate events with \mett-significance$>$3: 
a) $\mett$, the missing transverse energy; b) $E_T$ of leading photon candidate; c) $H_T$, the scalar sum 
of the transverse energies of photons, jets, and $\mett$; and d) invariant mass, $M_{\gamma\gamma}$, of two 
photons . In all figures, the data (marker) is compared with the total background predictions (solid line 
with the gray band representing the total uncertainty). The total background prediction is a sum (shown by 
the stacked histograms) of the QCD and electroweak (dashed line) backgrounds. The non-collision background 
is too small to be visible on a plot with linear scale.
}
\label{fig:ggmet_fig3}
\end{figure*}

\begin{figure*}[!ht]
\centering
\includegraphics[width=0.4\linewidth]{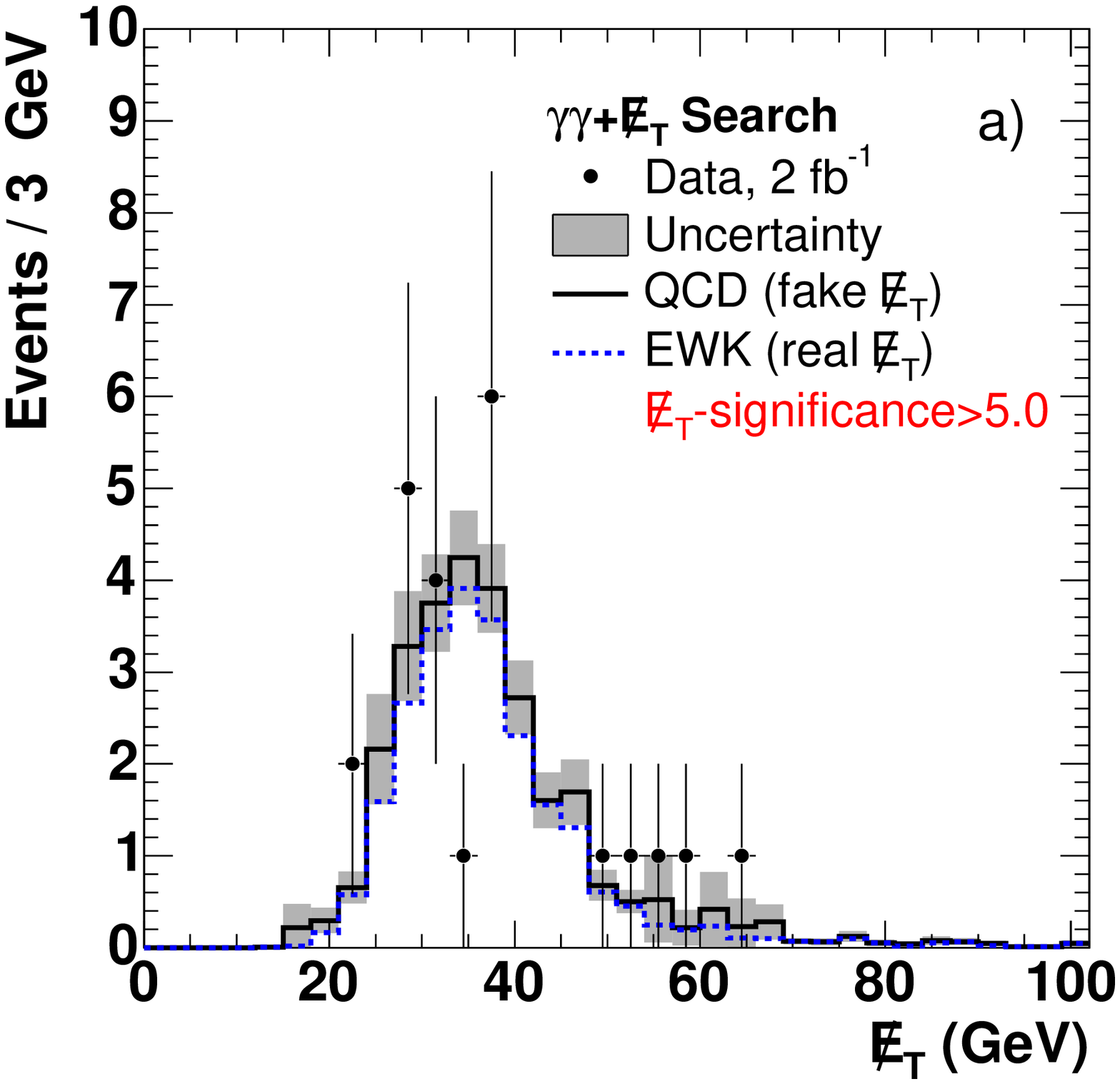}
\includegraphics[width=0.4\linewidth]{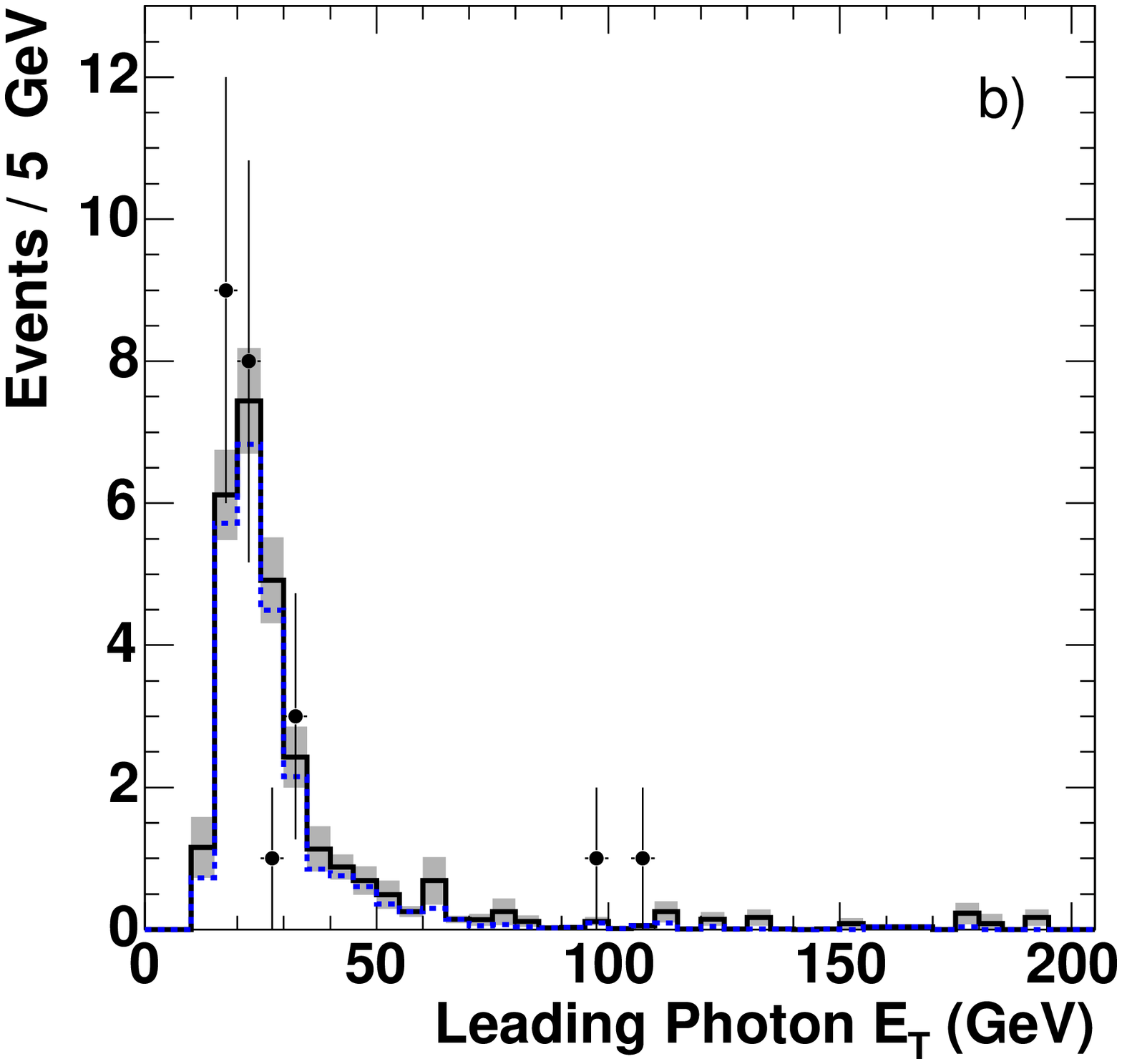}
\includegraphics[width=0.4\linewidth]{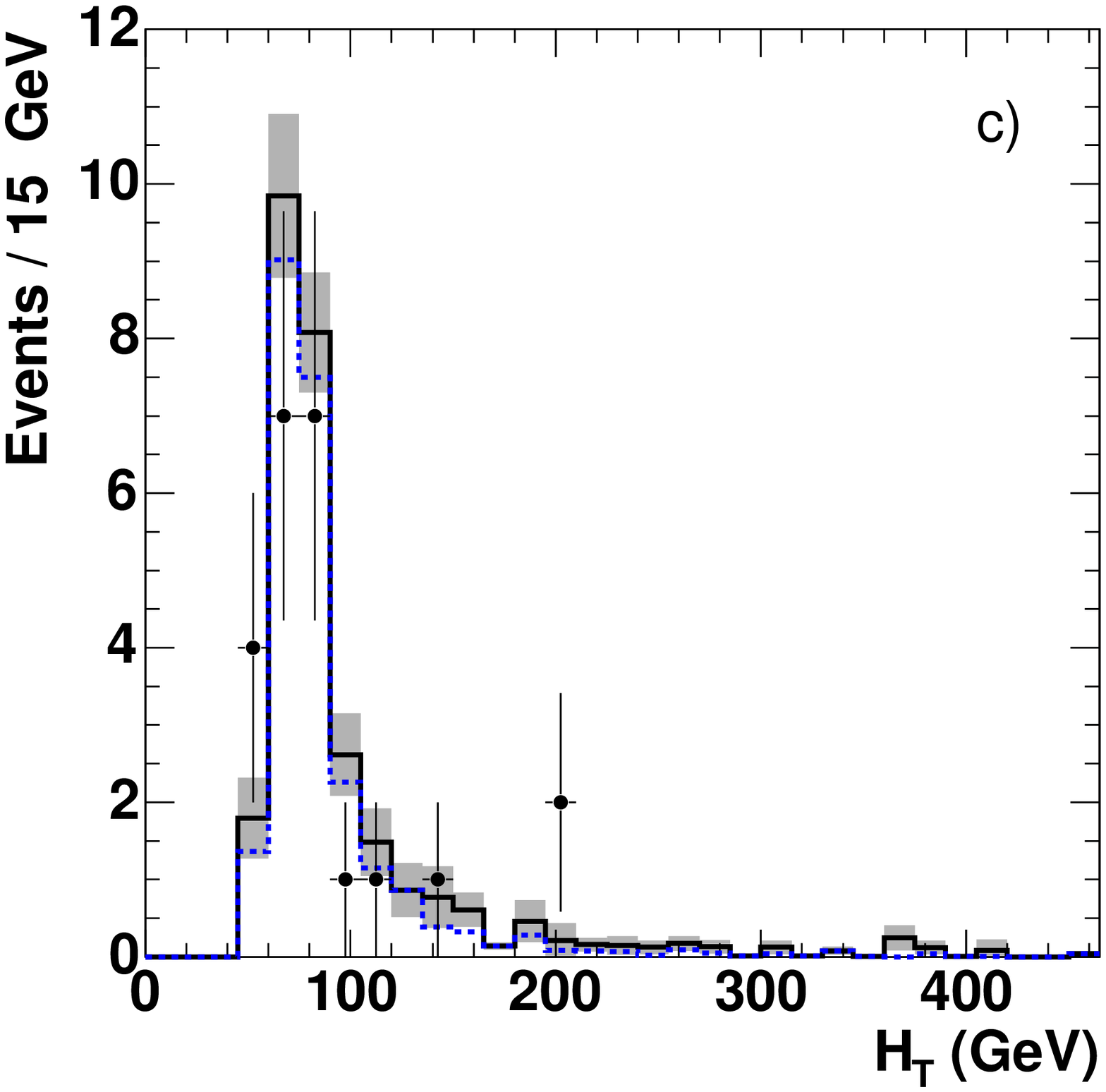}
\includegraphics[width=0.4\linewidth]{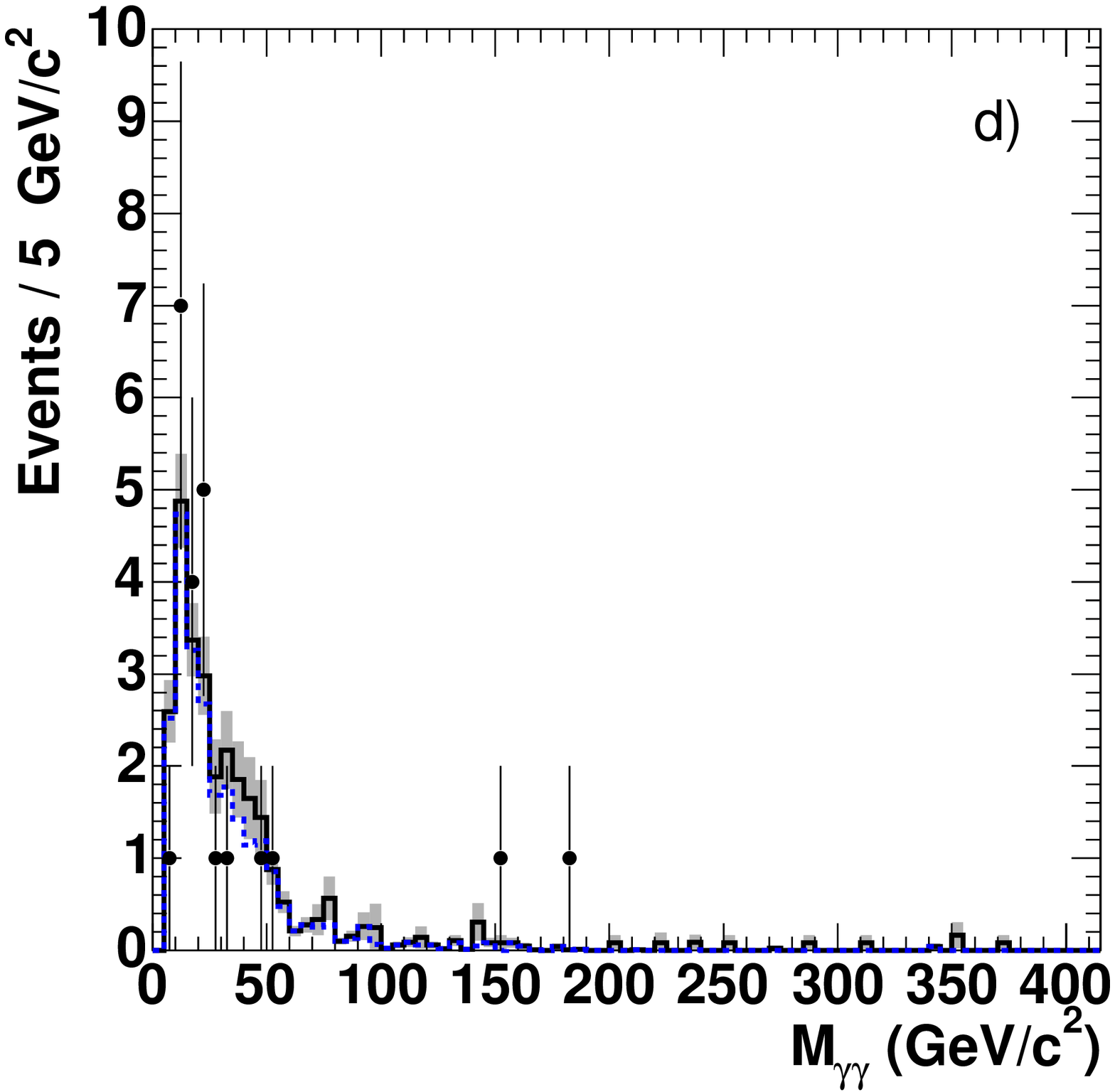}
\caption{The kinematic distributions for $\gamma\gamma+\mett$ candidate events with \mett-significance$>$5: 
a) $\mett$, the missing transverse energy; b) $E_T$ of leading photon candidate; c) $H_T$, the scalar sum of 
the transverse energies of photons, jets, and $\mett$; and d) invariant mass, $M_{\gamma\gamma}$, of two 
photons. In all figures, the data (marker) is compared with the total background predictions (solid line 
with the gray band representing the total uncertainty). The total background prediction is a sum (shown by 
the stacked histograms) of the QCD and electroweak (dashed line) backgrounds. The non-collision background 
is too small to be visible on a plot with linear scale.
}
\label{fig:ggmet_fig4}
\end{figure*}

\begin{figure*}[!ht]
\centering
\includegraphics[width=0.4\linewidth]{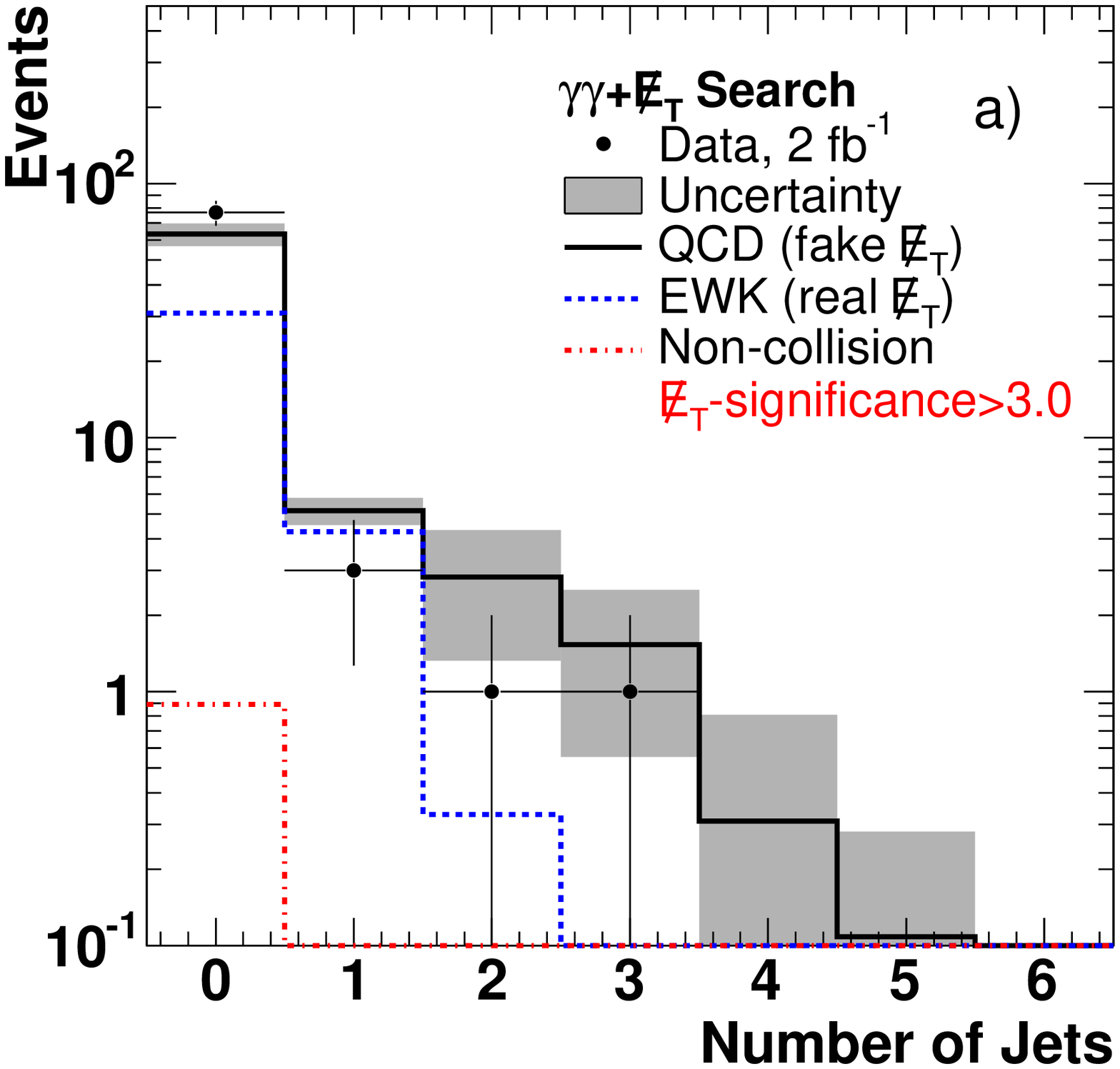}
\includegraphics[width=0.4\linewidth]{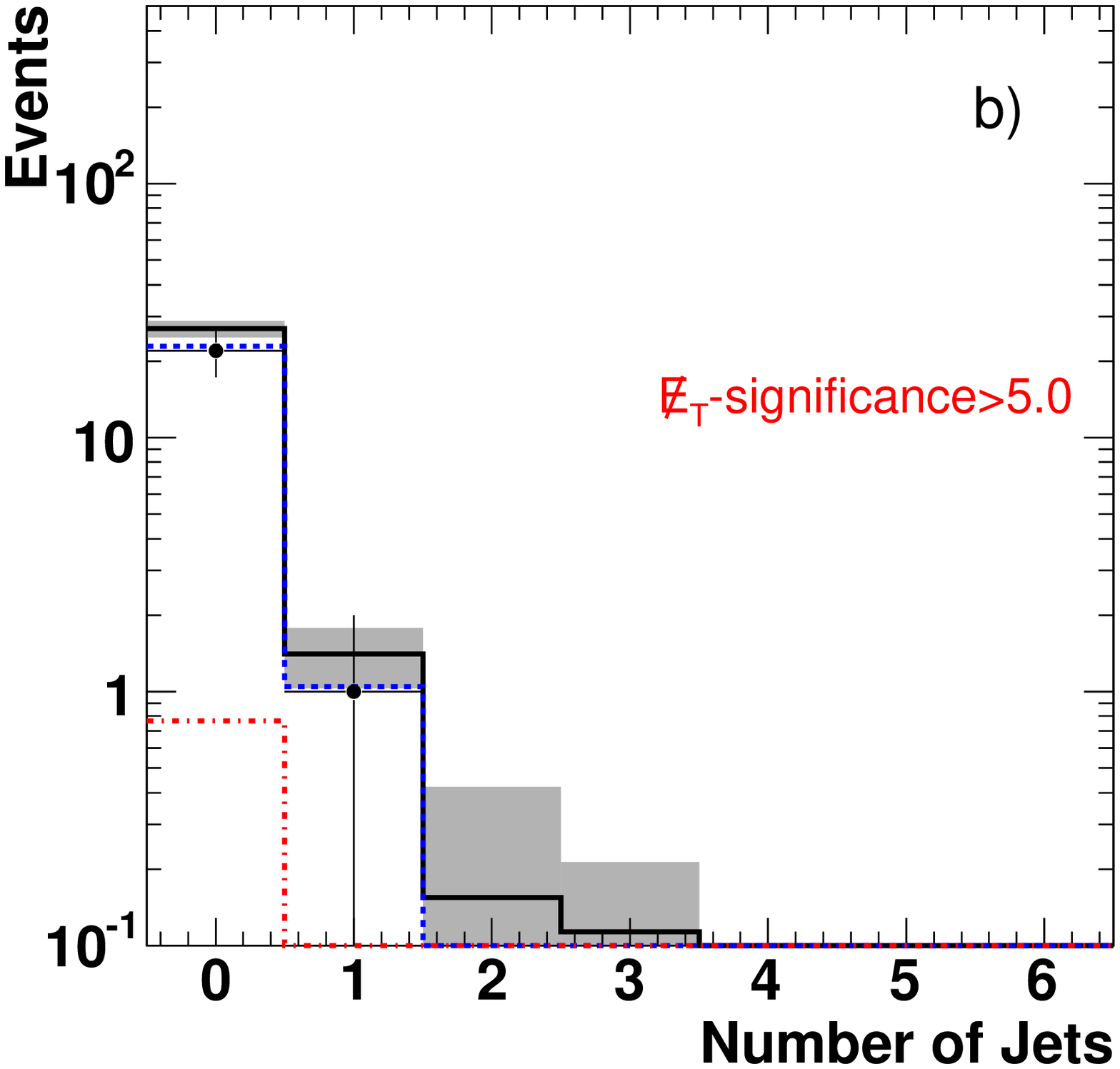}
\includegraphics[width=0.4\linewidth]{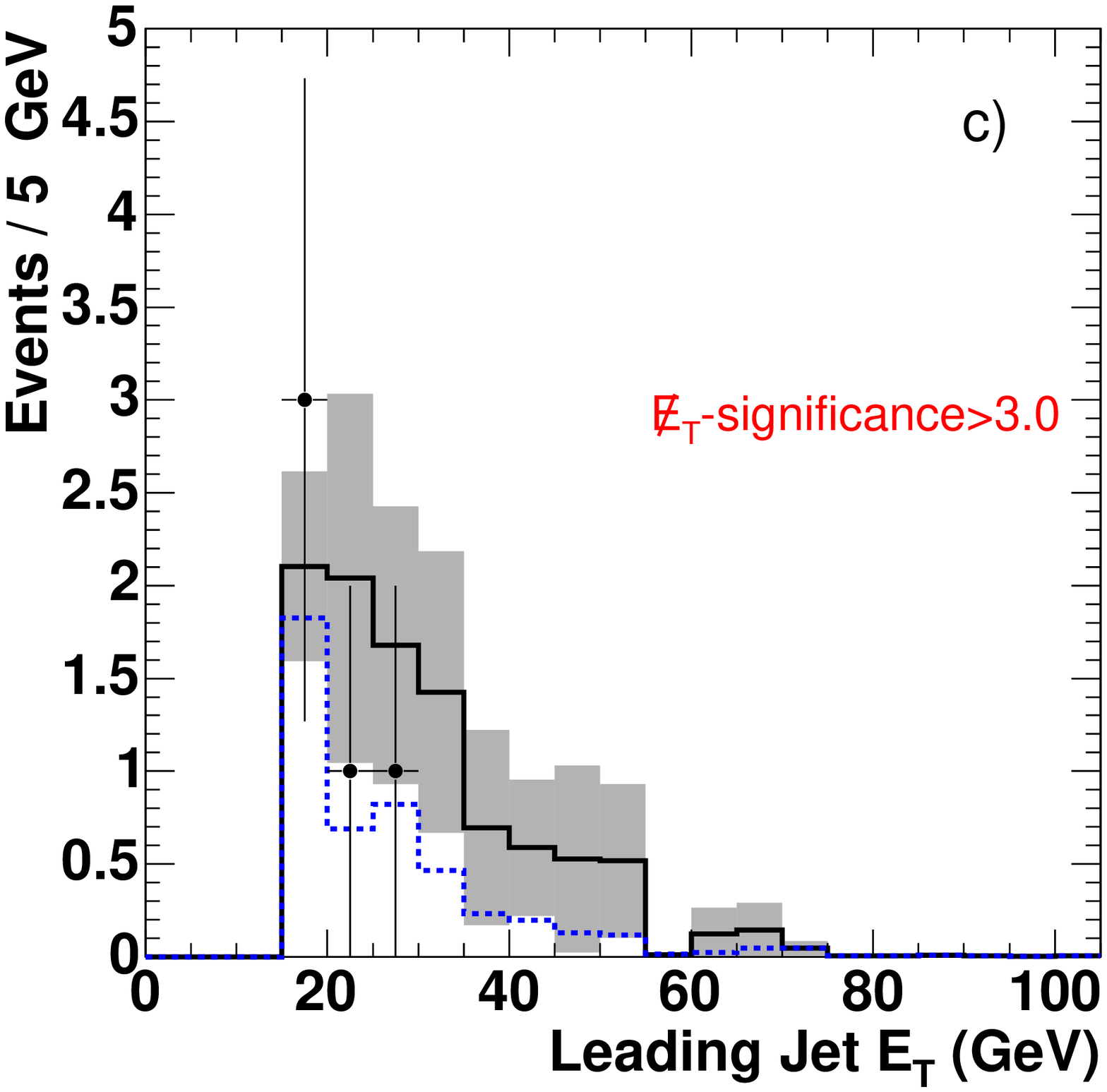}
\includegraphics[width=0.4\linewidth]{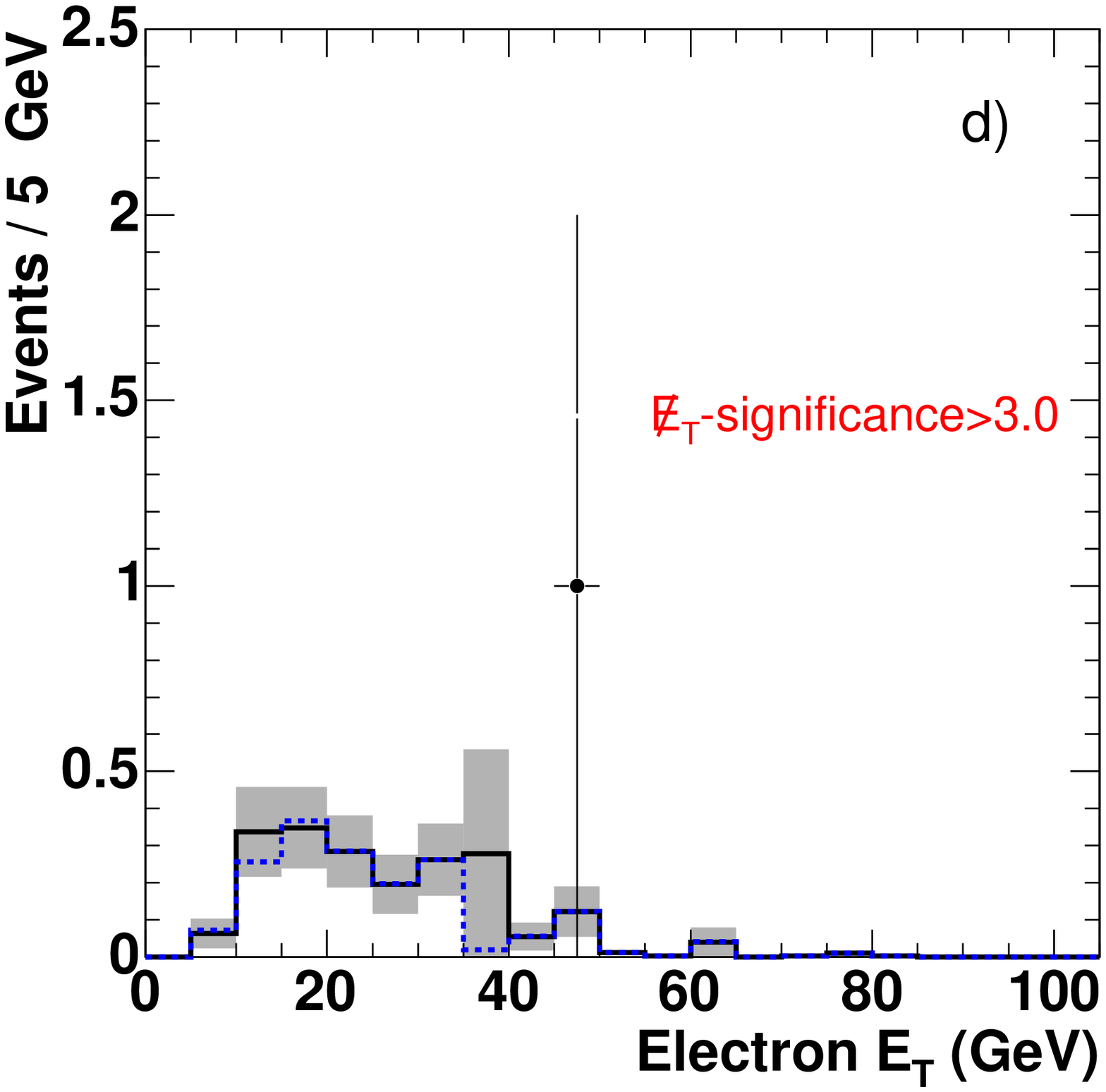}
\caption{Properties of additional objects in $\gamma\gamma+\mett$ candidate events: a) the multiplicity of 
jets with $E_T$$>$15~GeV for events with \mett-significance$>$3; b) the multiplicity of jets with $E_T$$>$15~GeV 
for events with \mett-significance$>$5; c) the leading jet $E_T$ for events with \mett-significance$>$3; and d) 
the electron $E_T$ for events with \mett-significance$>$3. In all figures, the data (marker) is compared with 
the total background prediction (solid line with the gray band representing the total uncertainty). The total 
background prediction is a sum (shown by the stacked histograms) of the QCD, electroweak (dashed line), and 
non-collision (dash-dotted line on figures a) and b) only) backgrounds.
}
\label{fig:ggmet_fig5}
\end{figure*}

%-------------------- EWK backgrounds
Electroweak processes involving $W$$\to$$l\bar\nu$ and $Z$$\to$$\nu\bar{\nu}/\tau^+\tau^-$ are the most common 
source of large real $\mett$ in $p\bar p$ collisions. There are three ways these processes can produce a 
$\gamma\gamma+\mett$ signature (listed in the order of importance): {\it 1)} $W\gamma$ and $Z\gamma$ events 
with one real and one fake photon; {\it 2)} $W\gamma\gamma$ and $Z\gamma\gamma$ events where both photons are 
real; {\it 3)} $W+jet$, $Z$$\to$$\tau^+\tau^-$, and $Z+jet$ events where both photon candidates are fake photons. 
We estimate the EWK backgrounds by using $W/Z+\gamma$~\cite{sim} (for {\it 1)} and {\it 2)}) and inclusive 
$W/Z$~\cite{pyth_sample} (for {\it 3)}) Monte Carlo events passed through the detector simulation. We consider 
all three leptonic decay modes of $W$ and $Z$ bosons. To avoid an overlap between $W/Z+\gamma$ and $W/Z$ samples, 
we remove {\sc pythia} $W/Z$ events where reconstructed photons are matched to generated photons originating 
from initial/final state radiation (ISR/FSR) of quarks or leptons. The MC-based predictions for the EWK 
backgrounds are then multiplied by a scale factor that diminishes possible data-MC differences and cancels out 
many of the systematic uncertainties (e.g., trigger efficiencies, acceptance and photon ID efficiencies, 
$K$-factors, modeling of ISR/FSR in MC, uncertainties in parton distribution functions, jet energy scale 
uncertainty, and luminosity uncertainty). This scale factor is obtained by comparing $e\gamma+\mett$ events 
(see Appendix~\ref{sec:ElePho}) in data and MC. It is defined as the ratio of numbers of data and MC $e\gamma$ 
events satisfying all analysis requirements. The resulting EWK background predictions and the corresponding 
uncertainties can be found in Table~\ref{tab:ggmetResults}. The total uncertainties includes the MC statistical 
uncertainties (3.5-4.4$\%$) and the MC-to-data normalization factor uncertainties (5.4-6.1$\%$). The last 
uncertainty includes statistical uncertainties from data and MC $e\gamma+\mett$ samples and systematic 
uncertainties associated with the purity of the $e+\gamma$ data sample and difference between the $E/p$-cut 
efficiency (see Appendix~\ref{sec:ElePho}) in data and MC. From Table~\ref{tab:ggmetResults}, one can see that 
the EWK processes are the dominant source of background when $\mett$-significance$>$4. We find that 59-63$\%$ 
(30-40$\%$) of the total EWK background for the $\gamma\gamma+\mett$ signature comes from the electron ($\tau$ 
lepton) decay channels of $W$ and $Z$ bosons. Note that the \mett-significance cuts are rather efficient for 
events with large real $\mett$: for example, 84$\%$ and 68$\%$ of $W$$+$$\gamma$$\to$$e\nu$$+$$\gamma$ events 
pass the \mett-significance$>$3 and \mett-significance$>$5 requirements, respectively.

%---------------------- Non-collision backgrounds
The last remaining source of background is non-collision events where both photons and $\mett$ are fake. These 
events may either be caused by cosmic rays (CR) or beam halo (BH) muons depositing energy in the calorimeter. 
CR events are suppressed by requiring the EM timing of both photon candidates ($T_1$ and $T_2$) to be consistent 
with the collision time: $|T_{1,2}|$$<$6.7~ns and $|T_1-T_2|$$<$4.1~ns (more details are given in Appendix~\ref{sec:photon}). 
BH events are removed by the topological cuts based on the distinct energy deposition pattern of BH muons traveling 
along the beam pipe. More details about CR and BH rejection cuts can be found in Appendix~\ref{sec:photon}. The 
number of remaining BH events is estimated from the number of identified BH candidates and known rejection power 
of the BH cuts. The background contribution due to CR events is estimated based on the number of these events in 
the 30~ns$<T_{1,2}<$120~ns EM timing window and known efficiency of the cosmic rejection cuts (see Appendix~\ref{sec:photon} 
and Ref.~\cite{delayedPhoPRD}). The total prediction for non-collision backgrounds can be found in 
Table~\ref{tab:ggmetResults}. The uncertainty for this estimate is dominated by the statistics in the samples of 
identified BH and CR events.   

%---------------------- Summary of gg+MET analysis

The results of the search are presented in Table~\ref{tab:ggmetResults}. The total expected SM background for
three \mett-significance cuts ($\mett$-significance$>$3,~4, and~5) is 71.7$\pm$7.5, 39.0$\pm$3.1, and 30.4$\pm$2.4   
events, respectively. These predictions agree well with the observed numbers of data events: 82, 31, and 23. 
We also examine various kinematic distributions in data and SM backgrounds for \mett-significance$>$3 and 
\mett-significance$>$5. Figures~\ref{fig:ggmet_fig3} and \ref{fig:ggmet_fig4} show the $\mett$, leading photon $E_T$, 
$H_T$, and $M_{\gamma\gamma}$ distributions for the selected $\gamma\gamma+\mett$ candidate events and the SM 
background predictions. Finally, Fig.~\ref{fig:ggmet_fig5} illustrates multiplicities and $E_T$ distributions 
of extra jets and electrons in selected events. We observe good agreement between data and predicted background 
shapes for all studied kinematic distributions that are expected to be sensitive to production of new particles. 

In summary, we have searched for anomalous production of $\gamma\gamma+\mett$ events in data corresponding to 
2.0~fb$^{-1}$ of integrated luminosity. No significant deviations from the SM background predictions are observed.

%%____________________________________________________________________________________
%%------------------------ Summary and Conclusion -------------------------
%%____________________________________________________________________________________
\section{Conclusions}
\label{sec:summary}

We performed a model-independent search for anomalous production of two photons with an electron, muon,
$\tau$ lepton, or large missing transverse energy. The analysis of a $\gamma\gamma$$+$$e/\mu$ signature was 
performed using data corresponding to 1.1~fb$^{-1}$ of integrated luminosity. After final selection, we 
observed one $\gamma\gamma$$+$$e$ candidate event and zero $\gamma\gamma$$+$$\mu$ events, in agreement 
with the expected background of 3.79$\pm$0.54 and 0.71$\pm$0.10 events, respectively. The kinematic properties 
of the $\gamma\gamma$$+$$e$ event were consistent with the SM predictions. The silicon-track rejection 
technique applied in this search allows for more than 60$\%$ reduction in the bremsstrahlung background (the 
dominant background in the electron channel) and has a promising potential for future searches with the 
$\gamma$$+$$e$$+$$X$ signature.    

The search for new physics in $\gamma\gamma$$+$$\tau$ was based on data corresponding to 2.0~fb$^{-1}$ of 
integrated luminosity. We observed 34 data events, in good agreement with the expected background of 
46$\pm$10 events. The kinematic distributions of the selected events did not reveal any deviations from the 
SM predictions. 
 
The study of the $\gamma\gamma$$+$$\mett$ signature was performed using data from 2.0~fb$^{-1}$ of integrated 
luminosity. The events of interest were selected based on the \mett-significance, rather than a fixed \mett-cut. 
This method proved to be very effective in rejecting events with fake \mett, while remaining sensitive to new 
physics processes even with moderate values of $\mett$ (\mett$\sim$20-40 GeV). We selected 82, 31, and 23 data 
events with the \mett-significance greater than 3,~4, and~5, respectively. These results are consistent with 
the expected SM background of 71.7$\pm$7.5, 39.0$\pm$3.1, and 30.4$\pm$2.4 events, respectively. The examined
kinematic distributions for the observed events with \mett-significance greater than three and five are in a 
good agreement with the predicted background shapes. The {\sc metmodel} developed as part of the 
$\gamma\gamma$$+$$\mett$ search was also successfully applied to suppress multijet background with fake $\mett$
in the first observation of vector boson pairs in a final state with two jets and $\mett$ at the 
Tevatron~\cite{diboson}. Finally, the reported in this paper model-independent analysis was later used as a 
basis for a search for supersymmetry with gauge-mediated breaking in $\gamma\gamma$$+$$\mett$ events~\cite{gmsbPRL}.   
The data samples used in these two analyses have a 60$\%$ overlap. 

In summary, no significant deviations from the standard model were observed in the numbers of recorded events 
and their kinematic properties in signatures with two photons and an additional electron, muon, $\tau$ lepton, 
or large $\mett$. We also did not observe any new $ee\gamma\gamma$$+$$\mett$ candidate events, similar to the 
one reported in Ref.~\cite{run1evnt}. With improved analysis techniques and up to 20 times more data compared 
to the previous searches~\cite{run1evnt}-\cite{run2d0}, this model-independent search is substantially more 
sensitive to new physics. 

%%____________________________________________________________________________________
%%------------------------ Acknowledgments -------------------------
%%____________________________________________________________________________________
\section{Acknowledgments}

We thank the Fermilab staff and the technical staffs of the participating institutions 
for their vital contributions. This work was supported by the U.S. Department of Energy 
and National Science Foundation; the Italian Istituto Nazionale di Fisica Nucleare; the 
Ministry of Education, Culture, Sports, Science and Technology of Japan; the Natural 
Sciences and Engineering Research Council of Canada; the National Science Council of the 
Republic of China; the Swiss National Science Foundation; the A.P. Sloan Foundation; the 
Bundesministerium f\"ur Bildung und Forschung, Germany; the World Class University Program, 
the National Research Foundation of Korea; the Science and Technology Facilities Council 
and the Royal Society, UK; the Institut National de Physique Nucleaire et Physique des 
Particules/CNRS; the Russian Foundation for Basic Research; the Ministerio de Ciencia e 
Innovaci\'{o}n, and Programa Consolider-Ingenio 2010, Spain; the Slovak R\&D Agency; and 
the Academy of Finland. 
%%____________________________________________________________________________________
%%------------------------ Appendix -------------------------
%%____________________________________________________________________________________

%%__________ Trigger _____________________
\appendix

\section{Definitions}

\subsection{Diphoton Triggers}
\label{sec:trigger}

%------------------------------- Table-1 ------------------------------------
\begin{table*}[tbp]
\begin{center}
\caption{Summary of the diphoton trigger requirements.}
\label{tab:trig}
\begin{tabular}{lccc}
\hline\hline
%------------------------------------------------------
Trigger Level & Diphoton-12 & Diphoton-18 \\  \hline
%------------------------------------------------------
        & EM $E_T>8$ GeV & same \\
Level-1 & $E^{HAD}/E^{EM}<0.125$ & same \\
        & $N_{cluster}=2$ & same \\ \hline
%------------------------------------------------------
        & EM $E_T>10$ GeV & EM $E_T>16$ GeV \\
Level-2 & $E^{HAD}/E^{EM}<0.125$ & same \\
        & $E_T^{ISO}<3$ GeV or $E_T^{ISO}/E_T<0.15$ & N/A \\
        & $N_{cluster}=2$ & same \\ \hline
%------------------------------------------------------
        & EM $E_T>12$ GeV & EM $E_T>18$ GeV \\
Level-3 & $E^{HAD}/E^{EM}<0.055+0.00045$$\times$$E$/GeV if $E<200$ GeV & same \\
        & $E_T^{ISO}<2$ GeV or $E_T^{ISO}/E_T<0.1$ & N/A \\
	& shower profile: $\chi^2_{CES}<20$  & N/A \\ 
        & $N_{cluster}=2$ & same \\
%------------------------------------------------------
\hline\hline
\end{tabular}
\end{center}
\end{table*}

There are two diphoton paths in the CDF three-level trigger: the first path requires two isolated 
electromagnetic (EM) clusters with $E_{T}$$>$12~GeV (diphoton-12), and the second path requires two 
electromagnetic clusters with $E_{T}$$>$18~GeV and has no isolation requirement (diphoton-18). The 
transverse energy of clusters is calculated with respect to the nominal center of the detector at 
$z$=0~cm. The trigger requirements at each level are briefly described below. 

At Level-1, events with two towers with EM $E_T$$>$8~GeV each are required. For each trigger tower, 
the amount of energy in the hadronic compartment of the calorimeter ($E^{HAD}$) has to be consistent 
with that of an electromagnetic object. A trigger tower consists of two adjacent towers in the same 
calorimeter wedge, so that the granularity is approximately $\Delta\eta\times\Delta\phi\simeq 0.2\times 15^o$. 

The Level-2 requirements are different for the two triggers. The diphoton-12 trigger selects events if 
there are two isolated clusters (seeds) with EM $E_T$$>$10~GeV each. The isolation (ISO) energy is 
calculated as a sum of the transverse energy in nine towers surrounding the seed tower according to five 
preset patterns. The ISO energy in each of the patterns has to be less than 3~GeV or 15$\%$ of the seed 
energy, whichever is larger. The diphoton-18 trigger requires two towers with EM $E_T$$>$16~GeV each at 
Level-2. 

The events are fully reconstructed at Level-3. At this level, for both triggers, the energy profile at the 
shower maximum ($\chi^2_{CES}$) of each photon candidate has to be consistent with that of a single photon. 
The diphoton-12 trigger selects events with two isolated photon candidates with $E_{T}$$>$12~GeV. The 
isolation energy at the level-3 is calculated as the sum of $E_T$ in all towers (except for photon towers) 
within the cone of $\Delta R=\sqrt{\Delta\eta^2 + \Delta\phi^2}<0.4$ centered around the photon candidate. 
This ISO energy has to be less than 2~GeV or 10$\%$ of the photon energy, whichever is larger. The 
diphoton-18 trigger has no isolation requirement and accepts events with two photon candidates with 
$E_{T}$$>$18~GeV. Table~\ref{tab:trig} gives a summary of all trigger requirements for events with EM 
objects in the central calorimeter.   

%%__________ Photons _____________________
\subsection{Photon Identification}
\label{sec:photon}

%------------------------------- Table-2 ------------------------------------
\begin{table*}[tbp]
\begin{center}
\caption{Summary of the standard (``tight'') and relaxed (``loose'') photon identification requirements 
for the signal and control $\gamma\gamma$ samples, respectively.}
\label{tab:phoID}
\begin{tabular}{lcc}
\hline\hline
%------------------------------------------------------
Cuts & ``Tight'' photon ID & ``Loose'' photon ID \\ \hline \hline 
%------------------------------------------------------
$E_T^{\gamma}$ & $\geq 13$ GeV & same \\ \hline
%------------------------------------------------------
Shower profile in CES: $\chi^{2}$ & $\leq$20 & same \\ \hline
%------------------------------------------------------
$E^{HAD}/E^{EM}$ & $\leq$0.055+0.00045$\times$$E$/GeV & $\leq$0.125 \\ \hline
%------------------------------------------------------
cal-ISO & $\leq$0.1$\times$$E_{T}$ if $E_{T}$$<$20 GeV or 
		    & $\leq$0.15$\times$$E_{T}$ if $E_{T}$$<$20 GeV or \\
                    & $\leq$2.0 GeV+0.02$\times$$(E_{T}-20$ GeV$)$ 
                    & $\leq$3.0 GeV+0.02$\times$$(E_{T}-20$ GeV$)$ \\ \hline 
%------------------------------------------------------
track-ISO & $\leq$ 2.0 GeV$+0.005\times E_{T}$ & $\leq 5$ GeV \\ \hline
%------------------------------------------------------
$N_{tracks}$ in cluster & $\leq 1$ & same \\
%------------------------------------------------------
track $p_T$ if $N_{tracks}=1$ & $\leq$1.0 GeV+0.005$\times$$E_{T}$ & $\leq$0.25$\times$$E_{T}$ \\ \hline
%------------------------------------------------------
$E_{T}$ of $2^{nd}$ CES  & $\leq$0.14$\times$$E_{T}$ if $E_{T}$$<$18 GeV & no cut \\
cluster & $\leq$2.4 GeV+0.01$\times$$E_{T}$ if $E_{T}$$\geq$18 GeV & \\ \hline\hline
%------------------------------------------------------
\end{tabular}
\end{center}
\end{table*}

Photon candidates have to satisfy strict (also referred to as ``tight'') photon identification requirements. 
The EM cluster has to be located inside the well--instrumented region of the CES chamber, away from the 
$\phi$-boundary of a calorimeter tower~\cite{CESfid}. The energy deposition pattern in both transverse 
profiles at CES has to be consistent with that of an electromagnetic object. The ratio of the energy measured 
in the hadron (HAD) calorimeter to the EM energy, $E^\mathrm{HAD}/E^\mathrm{EM}$, has to satisfy 
$E^{HAD}/E^{EM}$$<$0.055+0.00045$\times$$E^{\gamma}$ requirement. To distinguish photons from electrons, no 
high-$p_T$ charged track should point into the cluster ($N_{track}$$\leq$1 with track 
$p_T$$<$1.0+0.005$\times$$E_{T}$). The main sources of "fake" photons are $\pi^{0}$ and $\eta^{0}$ produced in 
jets. These mesons are usually produced in association with other particles. To reduce this contamination from 
jets, the photon candidate must be isolated in the calorimeter and tracking chamber. To calculate the calorimeter 
isolation (cal-ISO), the $E_T$ deposited in the calorimeter towers within the cone of $\Delta R<0.4$ around the 
EM cluster is summed, and the $E_T$ due to the EM cluster is subtracted. The cal-ISO is then corrected for the 
photon's energy leakage into towers in the neighboring wedge and for the contribution from multiple interactions 
in the same bunch crossing. The track isolation (track-ISO) is calculated as $\sum{p_T}$ of tracks inside a cone 
$\Delta$$R$$<$0.4 and satisfying $|z_{vertex}$-$z_{track}|$$<$5~cm. Both cal-ISO and track-ISO must be consistent 
with the amount of energy expected from the underlying event. In addition to calorimeter and tracking isolation, 
there should be no other significant energy ($E_T$ of 2$^{nd}$ CES cluster) deposited in the CES chamber 
containing the photon candidate. Table~\ref{tab:phoID} provides a summary of the photon identification 
requirements described above. 

We obtain the $\gamma\gamma$ control sample by selecting events where two photon candidates pass relaxed 
(``loose'') photon identification requirements, but at least one of them fails the ``tight'' cuts. The 
main difference between ``loose'' and ``tight'' photon requirements is in the amount of allowed isolation 
energy (see  Table~\ref{tab:phoID}). The resulting $\gamma\gamma$ control sample is dominated by $jet-\gamma$ 
and $jet-jet$ events where one or both photon candidates are faked by jets. The fraction of real 
$\gamma\gamma$ events in the control sample is only 5$\%$.      

In addition to the photon identification requirements described above, we also apply cuts to remove 
contamination from non-collision sources. Muons produced in the beam halo are known to fake a photon 
signature~\cite{delayedPhoPRD}. These energetic muons travel parallel to the beam pipe and deposit energy 
in many towers of one wedge, consistent with a minimum ionizing particle. When the muon undergoes energetic 
bremsstrahlung, it may also create one or two fake photon candidates. Probability for a single photon BH event 
to overlap with a collision event with a photon candidate is very low. Therefore, events with both fake photons 
from one muon are a dominant source of the BH background. We use this fact to suppress such BH events.     
We reject events if $\Delta\phi_{\gamma\gamma}$$<$0.524~rad and if there are more than two hadronic and four 
central electromagnetic towers above 0.1~GeV threshold~\cite{BHcuts}. The efficiency of these requirements for 
collision events is estimated with data $Z$$\to$$e^{+}e^{-}$ events and found to be $\sim$100$\%$. The rejection 
power of the cuts for beam halo events is found to be 90.4$\%$$\pm$0.2$\%$, as estimated using a very pure sample 
of beam halo events with two photon candidates located in the same calorimeter wedge. The criteria used to select 
this sample are discussed in detail in Ref.~\cite{delayedPhoPRD}.      

Muons from cosmic rays may also bremsstrahlung in the calorimeter and create fake photon candidates. To 
suppress these events, we use different approaches for data collected before and after the timing system in 
the EM calorimeter~\cite{EMtiming} was installed. In the first 0.44~fb$^{-1}$ of integrated luminosity for which 
the EM timing is not available, we reject events if there is a segment of hits (``stub'') in the muon drift 
chambers within a cone of 30$^{o}$ around the direction of any of the photon candidates that is not linked to a 
track in the COT (trackless muon stub). This requirement rejects approximately 85$\%$ of cosmic rays and is 
approximately 98$\%$ efficient for $\gamma\gamma$ events. In data from the later 1.6~fb$^{-1}$ of integrated 
luminosity for which the EM timing is available, we reject events if one of the photon candidates has arrival 
time $|T_{1,2}|$$>$4$\sigma_{T}$ or two photons have $|\Delta T|=|T_{1}-T_{2}|$$>$4$\sigma_{\Delta T}$, where 
$\sigma_{T}$$=$1.67~ns and $\sigma_{\Delta T}$$=$1.02~ns are the timing resolutions obtained by studying the EM 
timing of electrons from $Z$$\to$$e^{+}e^{-}$ events~\cite{EMres}. These EM timing requirements reject 99.4$\%$ 
of cosmic rays, while they are 99.9$\%$ efficient for prompt $\gamma\gamma$ events. 
  
Another source of fake photons is electrons from electroweak processes which are misreconstructed as 
prompt photons. This occurs when either an electron undergoes a catastrophic bremsstrahlung in the detector 
material in front of the COT or when its track does not get reconstructed. In both cases, electrons usually 
leave a few hits in the silicon detectors and their tracks can be partially recovered by a special tracking 
algorithm~\cite{phoenix}. This algorithm looks for silicon hits along two helix curves connecting vertex 
and EM cluster positions. The helix curvature is uniquely defined by the EM cluster $E_T$, and two curves 
correspond to a positive and negative charge hypotheses. If any of the photon candidates is matched to such 
a track, we reject the event. This technique is used in the $\gamma\gamma+e/\mu$ and $\gamma\gamma+\mett$ 
searches and it is referred to as the ``silicon track rejection'' in the main text.

%%__________ Electrons _____________________
\subsection{Electron Identification}
\label{sec:electron}

We select electrons using the CDF standard criteria. An electron is characterized by a narrow shower in 
the electromagnetic calorimeter and a matching track (either in the COT or silicon detector) originating 
from the primary vertex. The transverse EM energy, $E_T^\mathrm{EM}$, must be greater than 20~GeV. 
The $E^\mathrm{HAD}/E^\mathrm{EM}$ ratio has to be less than 0.055. The lateral energy distribution of 
the shower must be consistent with that for an electron. Candidates are required to be isolated in the 
calorimeter and to contain at least 90\%\ of the total transverse energy within a cone of $\Delta R$$=$0.4. 
For an electron detected in the central region ($\left|\eta\right|$$<$1.05), the matching track is required 
to be well reconstructed by the COT and have $p_T$$>$10~GeV/$c$. The ratio of the electron energy to track 
momentum, $E/p$, must be less than 2.0. Electrons from photon conversions are suppressed by rejecting the 
candidates which have an oppositely-charged track with a small separation in the $xy$-plane and a minimal 
difference in the polar angle. For an electron detected in the forward region (1.2$<$$\left|\eta\right|$$<$2.0), 
the matching track is required to have a minimum of three hits measured in the silicon detector. We do 
not apply further requirements on the forward matching tracks because fewer measurements per track are 
available and the momentum measured in the forward region is not as reliable as that measured in the 
central region. More details of electron identification can be found in Ref.~\cite{LeptonID}.

%%__________ Muons _____________________
\subsection{Muon Identification}
\label{sec:muon}

We select muons using the CDF standard criteria. A muon is characterized by a well-reconstructed COT track 
which is matched to track segments (stubs) in the central muon detectors, and an energy deposition in the EM 
and HAD calorimeters consistent with a minimum ionizing particle. The $p_T$, measured either with the COT only, 
or with the COT and the silicon detector if the silicon hits are available, must be greater than 20~GeV/$c$. 
Two types of muons are selected: CMUP ($\left|\eta\right|$$<$0.6) and CMX (0.6$<$$\left|\eta\right|$$<$1.0). 
The CMUP muon candidate requires a match between the track and the stubs in the CMU and CMP detectors. The CMX 
muon candidate requires a match to a muon stub in the CMX detector. In order to reduce the background from 
cosmic rays or hadrons which decay in flight, we require the track to be consistent with originating along the 
beamline. Cosmic muons are further suppressed via their back-to-back track topology and asynchronous timing 
measured in the COT. More details on the muon identification can be found in Ref.~\cite{LeptonID}.

%%__________ Tau _____________________
\subsection{Tau-Lepton Identification}
\label{sec:tau}

The $\tau$ lepton has a $\sim$18$\%$ branching fraction for decays into an electron or muon, with neutrinos. 
When this occurs, the event would be categorized in the $e/\mu$ final state and addressed in that study. In 
the $\tau$ lepton search we address only the hadronic decay modes.

To identify the hadronic decays of $\tau$ leptons~\cite{tauID}, we require a narrow cluster of one or 
three tracks and calorimeter energy. This cluster must be consistent with a $\tau$ lepton in several ways, 
inconsistent with an electron, and isolated from other nearby calorimeter energy.

The clustering begins with a single tower with $E_T$$>$6~GeV. Up to five more towers may be added to the 
cluster if they are adjacent and have $E_T$$>$1~GeV. At least one high-quality track with $p_T$$>$6~GeV$/c$ 
must be associated with the cluster.  This track defines the origin point of the $\tau$ lepton. The cone 
subtending an angle of $\theta_{sig}$ from the track direction defines the signal region where the 
$\tau$ lepton decay products are expected. This angle is fixed to be 0.17 at low $\tau$ lepton $E_T$ and 
is smaller for $E_T$$>$30~GeV, shrinking to 0.05 at $E_T$$=$100~GeV, allowing for greater rejection as the 
$\tau$ lepton decay products become highly collimated. A second cone given by $\theta$$<$0.52 defines an 
isolation annulus. The calorimeter and the shower maximum detector are used to define $\pi^0$ candidates 
in the $\tau$ lepton signal cone and the isolation annulus. To reject electrons some hadronic energy, 
consistent with the observed signal-cone tracks, is required.

The $\tau$ lepton four-vector is defined by the total four-vector of the tracks and $\pi^0$ candidates in 
the signal cone. If the calorimeter cluster energy is significantly greater than this sum, the calorimeter 
cluster energy is used instead. The ``visible'' mass of the $\tau$ lepton is found as the magnitude of this 
total four-vector.

We define two levels of $\tau$ lepton identification: a ``loose'' identification (used in studies and 
background techniques) and the standard or ``tight'' identification, used for the signal region search.

Apart from the selection included in the reconstruction as described above, the loose identification 
requires only $E_T$$>$15~GeV. The tight selection also requires visible $\tau$ lepton mass less than 1.8~GeV, 
total track $p_T$ in the isolation cone less than 1.0~GeV, $\pi^0$ $E_T$ in the isolation cone less than 
0.6~GeV, and one or three tracks in the signal cone, with total charge of $\pm$1.

%%__________ Jets _____________________
\subsection{Jets}
\label{sec:jets}

We reconstruct jets by using the cone clustering algorithm~\cite{jetclu} with a cone radius 
$R=\sqrt{\Delta\phi^2+\Delta\eta^2}=$0.4. Identified photons and electrons are removed from a list of 
jets. The jet energy is corrected for a non-linearity of the detector response and for contributions 
due to underlying event and multiple interactions in the same bunch crossing~\cite{jes}. Unless otherwise 
stated, we only consider jets with $E_T$$>$15~GeV and $|\eta|$$<$3.0.  

%%__________ MET _____________________
\subsection{Missing Transverse Energy and $H_T$}
\label{sec:MET}

The missing transverse energy, $\mett$, is defined as an energy imbalance in the calorimeter and it 
is a signature of neutrinos or new particles that do not interact with the detector material. The $\mett$ 
is calculated from all calorimeter towers with $E_T$$>$0.1~GeV in the region $|\eta|$$<$3.6 according 
to $\vec{\mett}=-\sum_{i}{E}_T^i{\vec{n}_i}$ where ${\vec{n}_i}$ is a unit vector that points from the 
interaction vertex to the $i^{th}$ calorimeter tower in the transverse plane. To improve resolution 
and reduce the number of events with large fake $\mett$, we apply corrections to the $\mett$ to 
account for a non-linearity of the detector response for jets with $E_T$$>$15~GeV and for presence of 
reconstructed muons, which do not deposit their total energy in the calorimeter.  

One of the global kinematic characteristics of any hard scattering process is the total transverse 
energy of final products, $H_T$. We define $H_T$ for each event as a sum of the transverse energies 
of all identified objects: photons, electrons, muons, visible energy of $\tau$ leptons, jets, and 
$\mett$. According to many theoretical models, new physics is expected to appear at large energy scales 
and may reveal itself in an anomalous rate of events with large values of $H_T$. 
 
%%__________ soft _____________________
\subsection{Unclustered Energy}
\label{sec:soft}

The activity due to the underlying event and additional interactions in the same bunch crossing is 
characterized by the soft unclustered energy, $\displaystyle\sum^{soft}{E_T}$. We calculate 
$\displaystyle\sum^{soft}{E_T}$ for each event by taking the difference between the total transverse 
energy in the event and transverse energies of all reconstructed photons, electrons, and jets: 
$\displaystyle\sum^{soft}{E_T}=\sum^{all}{E_T}-\sum{E_T^{jet}}-\sum{E_T^{\gamma}}-\sum{E_T^{ele}}$.

%%__________ e+gamma events in gg+MET _____________________
\subsection{The $e\gamma$ Events in $\gamma\gamma+\mett$ Analysis}
\label{sec:ElePho}

We use inclusive $e\gamma$ events to obtain a data/MC normalization factor for the MC-based estimate of
the EWK backgrounds in the search for anomalous production of $\gamma\gamma+\mett$ events. To minimize 
differences between $e\gamma$ and $\gamma\gamma$ samples, we obtain the $e\gamma$ events in data and MC 
by using the same diphoton triggers (for data) and analysis selection procedures as used to derive our 
$\gamma\gamma$ baseline sample. In this selection, we treat an electron as a photon (i.e., we apply the 
same cuts as in Table~\ref{tab:phoID}) with only one exception that we also require the presence of a 
track pointing to an EM cluster. This track must satisfy the 0.8$<$$E/p$$<$1.2 requirement where $E$ is 
the energy of the EM cluster and $p$ is the track momentum. All additional tracks must pass the cuts 
listed in Table~\ref{tab:phoID}.

%%------------------------------------------------------------------------------------
%%  Appendix: Fake Rates
%%____________________________________________________________________________________
\section{Fake Rates}

%%__________ e->pho fake rate ________________________________________________________
\subsection{The $e\to\gamma$ Fake Rate}
\label{sec:e2phoFkRt}

 \begin{figure*}[!ht]
     \begin{center}
        \includegraphics[width=0.45\linewidth]
	{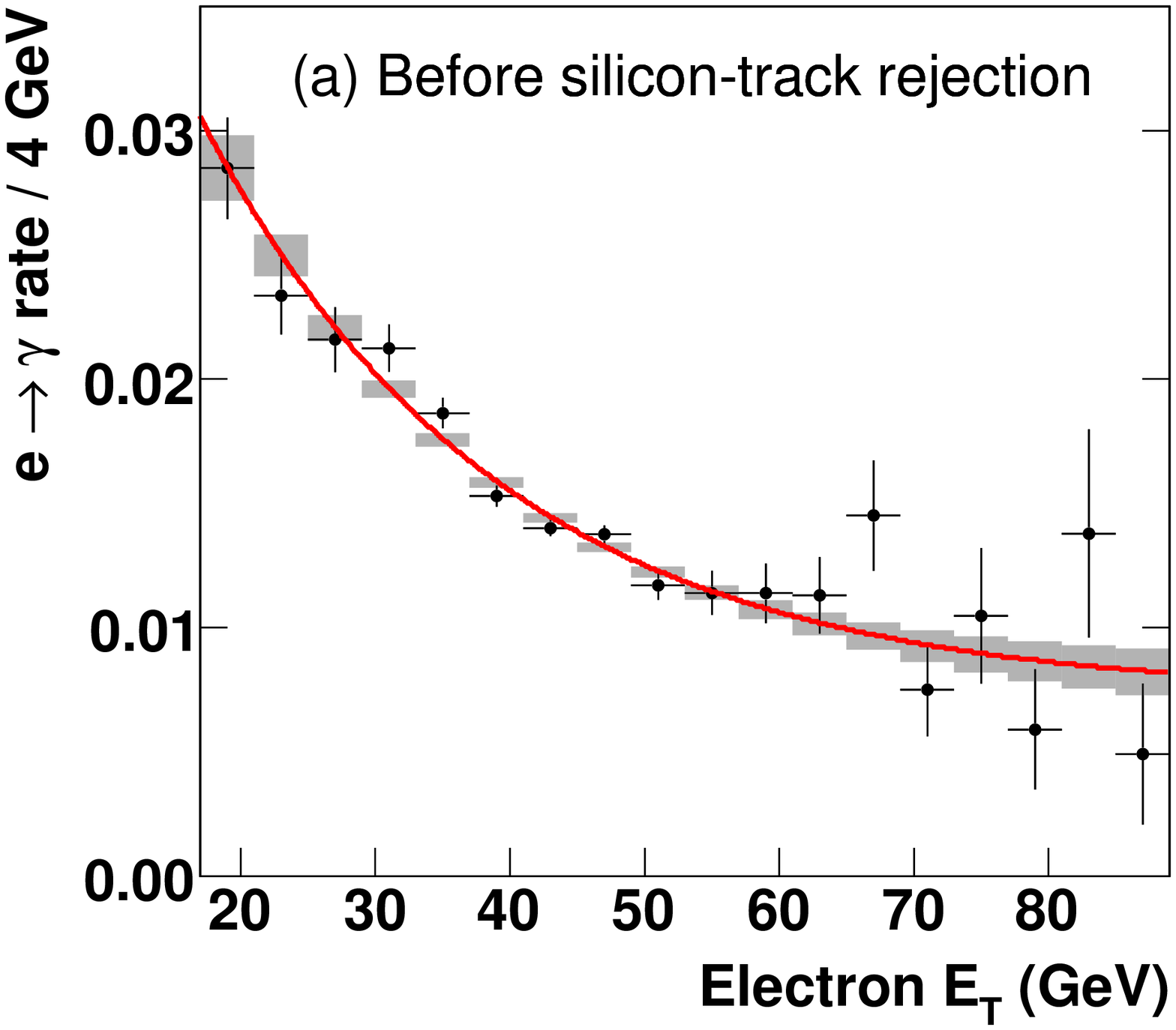}
        \includegraphics[width=0.45\linewidth]
	{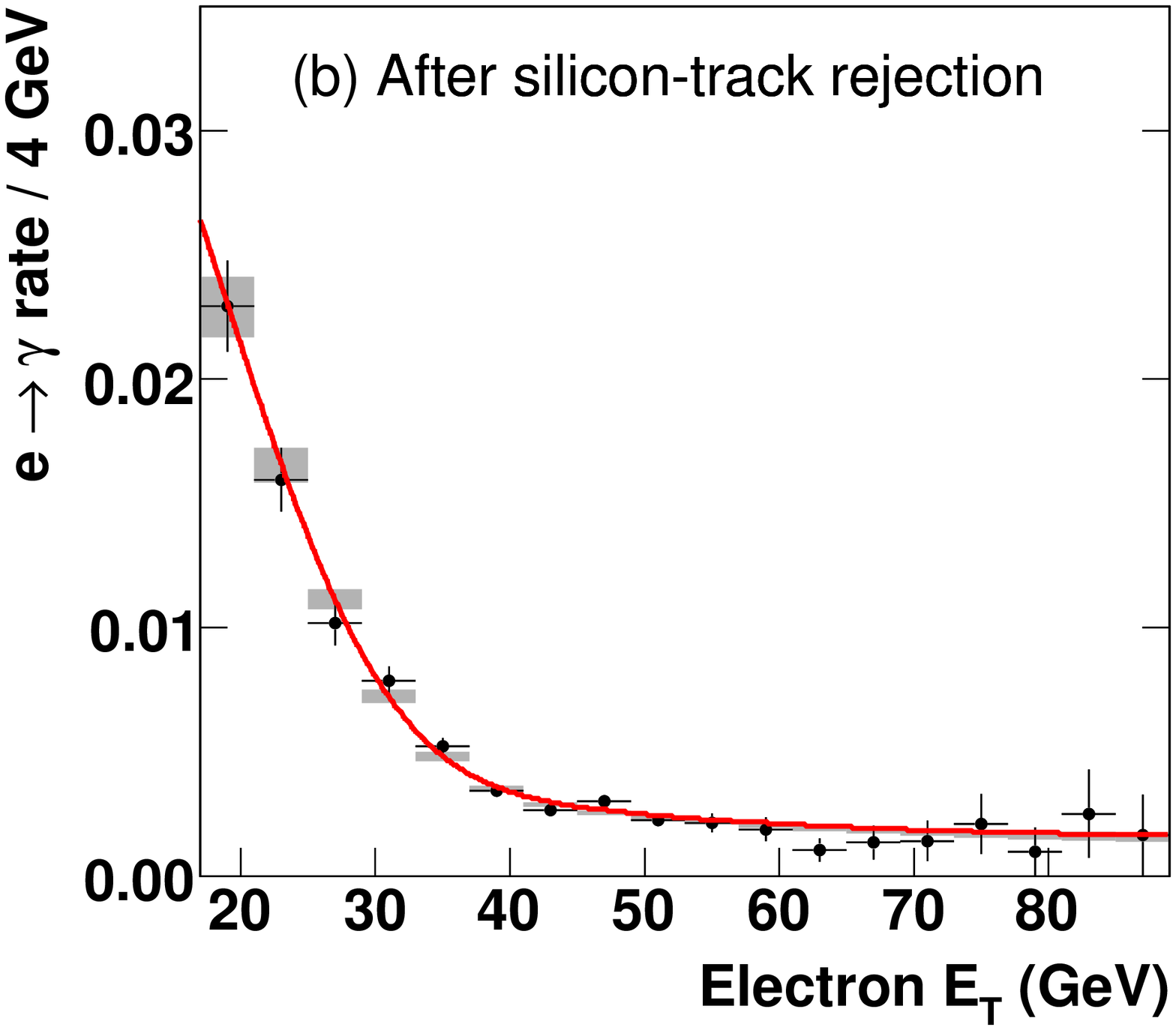}
     \end{center}
        \caption  		 
	{\label{fig:fakeratee1}
	Probability for a CDF standard central electron to be misidentified 
	as a standard central photon as measured in the Drell-Yan MC, 
	before (a) and after (b) applying the silicon-track rejection. 
	The misidentification probabilities (points) are parameterized as a 
	function of electron $E_T$ at the parton level. The gray boxes 
        indicate the systematic uncertainties in each $E_T$ bin due to the 
        uncertainties on fit parameters. 
	} 	
  \end{figure*}

 \begin{figure}[!ht]
     \begin{center}
        \includegraphics[width=1.0\linewidth]
	{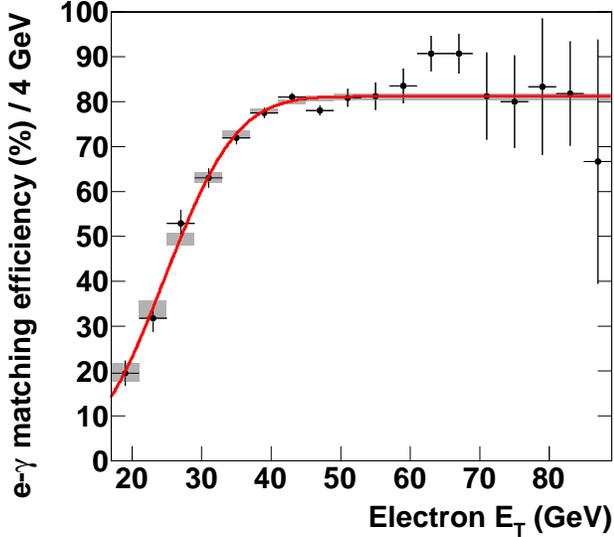}
     \end{center}
        \caption  		 
	{\label{fig:phxeff}
	Efficiency for an electron reconstructed as a CDF standard central photon to be matched to 
	the silicon-electron track, as a function of electron $E_T$ at the parton level, measured 
        in the Drell-Yan MC. The gray boxes indicate the systematic uncertainties in each $E_T$ bin 
        due to the uncertainties on fit parameters. 	
	} 	
  \end{figure}

Electrons may be misidentified as signal photons due to hard bremsstrahlung in the detector material, 
inefficiency of track reconstruction, or collinear final state radiation (FSR). We measure the 
misidentification probability, ${\cal P}(e$$\rightarrow$$\gamma)$, using Drell-Yan $Z/\gamma^*$ events.  
The ${\cal P}(e$$\rightarrow$$\gamma)$ is defined as the ratio of the number of reconstructed 
$Z/\gamma^*$$\rightarrow$$e\gamma$ events to the number of reconstructed $Z/\gamma^*$$\rightarrow$$ee$ 
events. The $E_T$ dependence of ${\cal P}(e$$\rightarrow$$\gamma)$ is obtained from the simulation. The 
overall normalization is scaled by the ratio of data--to--MC probabilities measured at $E_T$$=$40-50~GeV, 
around the $Z$ peak. The electrons have been selected using two types of identification: 1) standard 
central-electron criteria in Section~\ref{sec:electron}, 2) photon-like electron criteria in 
Section~\ref{sec:ElePho}. The result from identification-1 has been used in the $\gamma\gamma e$ and 
$\gamma\gamma\mu$ searches, while the result from identification-2 has been used in the cross-checks of 
$\gamma\gamma e$, $\gamma\gamma\mu$, and $\gamma\gamma\mett$ searches. The ${\cal P}(e$$\to$$\gamma)$ for 
identification-1, before (${\cal P}(e$$\to$$\gamma)^{B}$) and after (${\cal P}(e$$\to$$\gamma)^{A}$) 
applying silicon-track rejection, is measured with the data and MC Drell-Yan samples and parameterized 
as a function of the electron $E_T$ (in GeV):
\begin{eqnarray}
 {\cal P}(e\rightarrow \gamma)^\mathrm{B} = S^\mathrm{B} \cdot 
 ( e^{-2.991-0.045\cdot E_T} + 0.007), \\ 
 {\cal P}(e\rightarrow \gamma)^\mathrm{A} = S^\mathrm{A} \cdot 
    {\cal P}(e\rightarrow \gamma)^\mathrm{B} \cdot 
    (1-\epsilon(E_T)), 
\end{eqnarray}
 where $S^\mathrm{B}$ and $S^\mathrm{A}$ are the data to MC scaling factors: 
\begin{eqnarray*}
   S^\mathrm{B} & = & 1.08\pm0.09, \\
   S^\mathrm{A}  & = & 1.32\pm0.20, 
 \end{eqnarray*}
and the $\epsilon(E_T)$ is the efficiency of silicon-track rejection measured with the Drell-Yan MC:
\begin{eqnarray}
     \epsilon(E_T) = 0.405 \cdot \frac{2}{\sqrt{\pi}}\int^{\infty}_{z}e^{-t^2}dt \\
\nonumber z=0.086 \cdot ( 24.711 - E_T)
\end{eqnarray}
The electron $E_T$ at the parton level is further translated to the photon $E_T$ at the reconstruction 
level using a simulation study. Figure~\ref{fig:fakeratee1} shows the ${\cal P}(e$$\to$$\gamma)^\mathrm{B}$ 
and ${\cal P}(e$$\to$$\gamma)^\mathrm{A}$ measured in the Drell-Yan MC, without data--to--MC scaling factors 
applied. Figure~\ref{fig:phxeff} shows the $\epsilon(E_T)$. The average ${\cal P}(e$$\to$$\gamma)^\mathrm{B}$ 
is about 1.5$\%$ with an 11$\%$ fractional uncertainty, and the average ${\cal P}(e$$\to$$\gamma)^\mathrm{A}$ 
is about 0.4$\%$ with a 17$\%$ fractional uncertainty. The uncertainties come from the limited size of $Z$ 
data sample which determines the data--to--MC scaling factor, the variation of fitting methods which determines 
the number of $Z$ candidates, and the difference between results measured in the diphoton and inclusive electron 
triggers.

%%__________ jet->pho fake rate ______________________________________________________
\subsection{The $jet\to\gamma$ Fake Rate}
\label{sec:Jet2phoFkRt}

Hadrons in jets, such as $\pi^0$, $\eta^0$, and $K_s^0$, may decay into multiple photons. The segmentation 
of the CDF EM calorimeter is not sufficiently small to separate these photons and the standard reconstruction 
algorithm will reconstruct these hadron daughters as a single photon candidate. The probability to misidentify 
a jet as a signal photon, ${\cal P}(jet$$\to$$\gamma)$, has been measured in Ref.~\cite{cdf6601}, using data 
collected with inclusive jet triggers. The ${\cal P}(jet$$\to$$\gamma)$ is defined as the number of identified 
photon candidates times the fake-photon fraction ($F_\mathrm{QCD}$) and divided by the number of jets. The 
fraction $F_\mathrm{QCD}$ is required because the identified photon candidates in the jet data will contain 
real photons not relevant to the fake rate. Ref.~\cite{cdf6601} has determined $F_\mathrm{QCD}$ statistically 
by combining the following information: (a) the lateral shower shape measured in the wire and strip chamber 
(CES), (b) the extra energy in a cone of $\Delta R$$=$0.4 around the photon candidate (cal-ISO) measured in 
the calorimeters, and (c) the conversion rate measured in the central preshower detector. The 
${\cal P}(jet$$\to$$\gamma)$ is parameterized as a function of the $E_T$ of jet (in GeV) and found to be:
\begin{eqnarray}
{\cal P}(jet\rightarrow \gamma) =  
 10^{-3}\cdot(e^{2.397-0.153\cdot E_T(\mathrm{jet})}+0.404).
\end{eqnarray}
The fake photon $E_T$ is smaller than the original jet $E_T$ because the fake photon is often accompanied by 
other particles from that jet. The translation of the jet $E_T$ to the photon $E_T$ has been studied using 
simulations and is represented by a Gaussian distribution with a mean of 0.937 and a width of 0.048. The 
${\cal P}(jet$$\to$$\gamma)$ is about 0.2$\%$ at $E_T^{\gamma}=$13~GeV and 0.04$\%$ for $E_T^{\gamma}$$>$50~GeV, 
with a systematic uncertainty ranging from 50$\%$ to 200$\%$. The sources of systematic uncertainties include: 
the differences between the methods for determining $F_\mathrm{QCD}$, the differences of results when using a 
loose photon candidate as the fake denominator, variation of the mixture of quark jets and gluon jets, and 
variation of fragmentation model in the simulation which changes the Gaussian function of $E_T$ translation.

%%__________ jet->e/mu fake rate ________________________________________________________
\subsection{The $jet\to e/\mu$ Fake Rate}
\label{sec:Jet2eleFkRt}

Hadrons in jets may be misidentified as electrons due to inelastic charge exchange or the production
of an energetic conversion electron. The inelastic charge exchange in the EM calorimeter, $\pi^- p$$\to$$\pi^0 n$ 
or $\pi^+ n$$\to$$\pi^0 p$, results in a track in the COT due to the $\pi^{\pm}$ and an EM shower in the 
calorimeter due to the photons from $\pi^0$ decay. The combination of a charged track and an EM shower gives 
a fake electron candidate. Hadrons can also decay into muons before interacting with calorimeter (e.g., 
$K^+$$\to$$\mu^+\nu_{\mu}$) or pass through the calorimeter into the muon chamber (punch-through) with minimal 
interaction and give fake muon candidates. The probability to misidentify a jet as an electron or a muon, 
${\cal P}(jet$$\to$$e,\mu)$, has been measured in Ref.~\cite{canepa}, using data collected with inclusive jet 
triggers. ${\cal P}(jet$$\to$$e,\mu)$ is defined as the ratio of the number of identified electron/muon 
candidates to the number of ``fakeable'' objects (denominator). The fakeable object is a jet with uncorrected 
$E_T$$>$4~GeV for central electrons, a jet with uncorrected $E_T$$>$15~GeV for forward electrons, and an 
isolated track with $p_T$$>$4~GeV/$c$ and minimal extra energy in the cone of $\Delta R$$=$0.4 for muons. 
A track is considered to be isolated if the total $E_T$ of calorimeter towers within the cone of $\Delta R$$<$0.4 
around the track is less than 4 GeV or less than 10$\%$ of track's momentum. 
The misidentification probabilities, parameterized as a function of the jet $E_T$ (in GeV), for the electrons are: 
\begin{eqnarray}
  {\cal P}(jet\rightarrow e^\mathrm{central}) =  
0.00013 + e^{-7.940-0.194\cdot E_T(\mathrm{jet})},\\
  {\cal P}(jet\rightarrow e^\mathrm{forward}) =  
0.00032 + 0.000012 \cdot E_T(\mathrm{jet}). 
\end{eqnarray}
The translation of the jet $E_T$ to the electron $E_T$ has been studied using simulations. The ratio of electron 
$E_T$ to jet $E_T$ is represented by a Gaussian distribution with a mean of 0.89 and a width of 0.06. The 
misidentification probabilities, parameterized as a function of the track $p_T$ (in GeV/$c$), for the muons are: 
\begin{eqnarray}
  {\cal P}(track\rightarrow \mu^\mathrm{CMUP}) =  
0.00086 + 0.00017 \cdot p_T(\mu), \\
  {\cal P}(track\rightarrow \mu^\mathrm{CMX}) =  
0.00082 + 0.00020 \cdot p_T(\mu).
\end{eqnarray}
Since the misidentification probabilities were measured up to $E_T$=50~GeV and $p_T$=50~GeV/$c$ and the 
misidentification probabilities are expected to reach a plateau, we assign a constant value to all 
misidentification probabilities for $E_T$$\geq$50~GeV and $p_T$$\geq$50~GeV/$c$. The ${\cal P}(jet$$\to$$e,\mu)$ 
averages $\approx$0.01$\%$ for central electrons, $\approx$0.04$\%$ for forward electrons, and $\approx$1.0$\%$ 
for central muons, with a 50$\%$ systematic uncertainty estimate provided by the variation of results 
measured in different jet triggers.

%%__________ jet->tau fake rate ______________________________________________________
\subsection{The $jet\to\tau$ Fake Rate}
\label{sec:Jet2tauFkRt}

\begin{figure}
\begin{centering}
\includegraphics[width=1.0\linewidth]{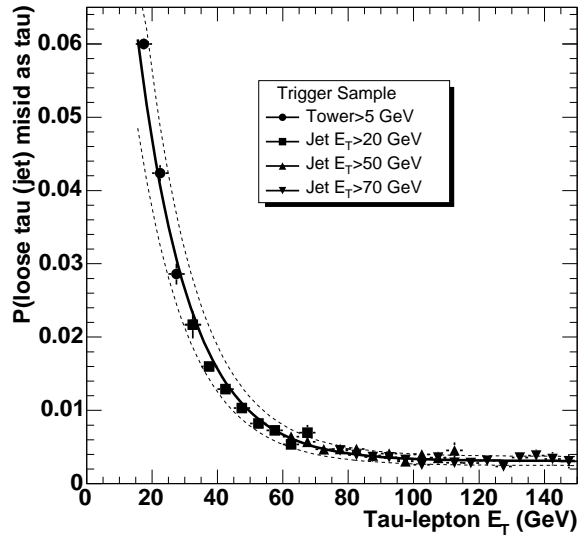}
\caption{
\label{fig:stitch}
The solid line represents the probability for objects passing loose $\tau$ lepton ID cuts to also pass 
tight $\tau$ lepton ID cuts ($jet\to\tau$ fake rate) as a function of $\tau$ lepton $E_T$, with the 
overlapping regions removed. The dashed lines are the systematic uncertainties on the $jet\to\tau$ fake 
rate ($\pm$1 standard deviation).}
\end{centering}
\end{figure}

The probability of a quark or gluon jet to be misreconstructed as an hadronically--decaying $\tau$ lepton is 
measured and then applied to a sample of jets to estimate the number of fake $\tau$ leptons we expect in that 
sample.

We measure this misidentification rate in a sample of inclusive jet triggers~\cite{jettriggers}, using only 
the energy clusters for which the trigger is fully efficient. This jet sample has a negligible fraction of 
real $\tau$ leptons because the rates for $W/Z$$\to$$\tau$$+$$X$ and $c/b$$\to$$\tau$$+$$X$ processes are 
very small compared to the jet production rates. Therefore, the measurement of the misidentification rate 
is straightforward. We identify all loose and tight $\tau$ leptons (See Section~\ref{sec:tau}) and measure 
the rate by dividing the number of tight $\tau$ leptons by the number of loose $\tau$ leptons as a function 
of the $\tau$-candidate $E_T$. 

We check the misidentification rate by using it to predict various distributions in the jet samples. We compare 
the number of $\tau$ leptons observed and predicted as a function of the first, second and third jet $E_T$, the 
event total energy, the underlying event energy, the number of jets, the number of interactions in the event, 
and the distance the nearest jet. The only notable discrepancy is the case where the $\tau$ lepton is close to 
a second jet, where the jet's energy tends to spoil the $\tau$ lepton's isolation and reduce the fake rate. We 
include this effect in the application of the misidentification rate. The assigned systematic uncertainty of 
20$\%$ accounts for any other discrepancies. The resulting function is shown in Fig.~\ref{fig:stitch}.

Finally, we identify the primary source of each jet in inclusive {\sc pythia} MC jet samples by searching 
for the highest-$E_T$ parton consistent with the jet direction. We then measure the misidentification rate in 
quark and gluon jets separately. We find the rate for gluon jets is approximately three times smaller than the 
rate for quark jets, which have a higher probability to fragment to a few energetic particles.

%%__________ Met Model _____________________
\section{The $\mett$ Resolution Model}
\label{sec:MetModel}

A major sources of background in the $\gamma\gamma+\mett$ final state is diphoton candidate events with 
significant fake $\mett$ due to energy mismeasurement in the calorimeter. Given the large production rates for 
QCD processes ($\gamma\gamma$, $\gamma$-$jet$, and $jet$-$jet$), fluctuations in energy measurements can 
result in a considerable fraction of such events. We predict the shape of this fake $\mett$ and calculate 
its significance on an event--by--event basis by means of the $\mett$ resolution model denoted as {\sc metmodel}. 

The {\sc metmodel} is based on a simple assumption that fluctuations in energy measurements of jets, soft 
unclustered particles from the underlying event, and multiple interactions are the dominant sources of 
fake $\mett$. Therefore, the individual contributions of each of these components to fake $\mett$ can 
be modeled, on average, by smearing their energies according to the corresponding energy resolution 
functions. Jets are the dominant source of fake $\mett$ because they are collimated sprays of energetic 
particles in a certain direction and may have large measurement fluctuations in that direction. The 
unclustered energy, on the other hand, tends to be uniformly spread in the calorimeter. Therefore, the 
portion of $\mett$ due to this source is usually small and mostly results in a smearing of the jet 
component of fake $\mett$. Taking into account the above considerations and for reasons of simplicity, we 
model only the fake $\mett$ due to mismeasurements of jets and all soft unclustered energy (rather than 
individual unclustered particles).
%---------------------- Uncl parameterization ----------------------------------
\begin{figure}[tbp]
\centering
%------- temporary plot, to be replaced	
\includegraphics[width=1.0\linewidth]{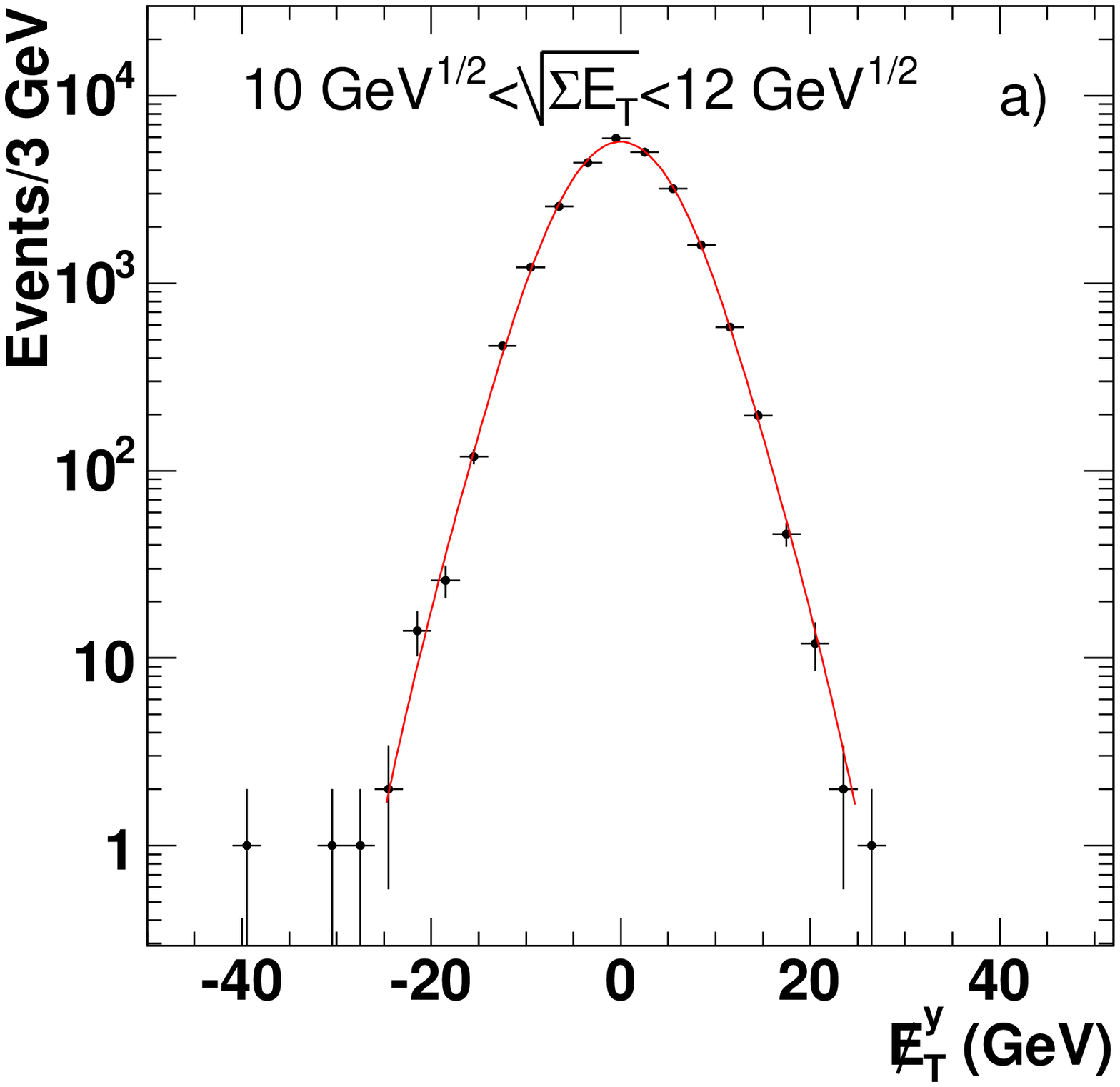}
\includegraphics[width=1.0\linewidth]{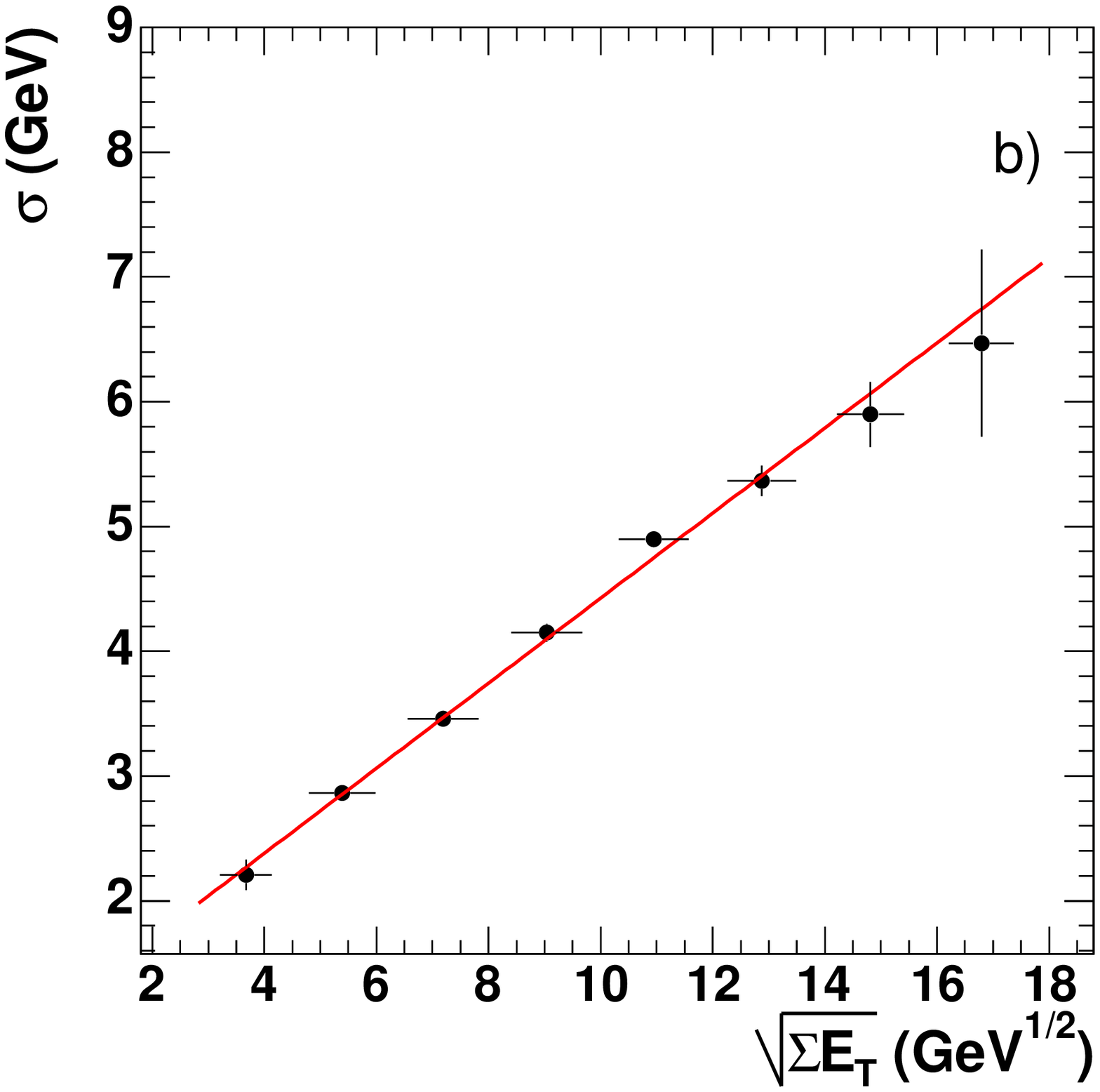}
\caption{Example of the fake $\mett$ parameterization due to unclustered energy. Figure {\it a)} shows a 
two-Gaussian fit of the ${\mett}^{y}$ distribution for {\sc pythia} $\gamma\gamma$ events from one of the bins 
in $\sqrt{{\sum}E_{T}}$. Figure {\it b)} demonstrates how a width, $\sigma$, of the leading Gaussian depends 
on the $\sqrt{{\sum}E_{T}}$. On both plots, points are {\sc pythia} data and curves are the fit functions. 
}
\label{fig:uncl_param}
\end{figure}

The $\mett$ resolution due to the soft unclustered energy is studied in the $\gamma\gamma$ control sample 
(see Appendix~\ref{sec:photon}) and $Z/\gamma^{*}$$\to$$e^{+}e^{-}$ events with 
85~GeV/c$^2$$<$$M_{ee}$$<$97.5~GeV/c$^2$. We fit distributions of $x$ and $y$ components of the $\mett$ for 
events without jets, $N_{jet}(E_{T}$$>$15~GeV)=0, in small bins of $\sqrt{\sum E_{T}}$, with a sum of two 
Gaussian distributions. We assume that both Gaussian distributions have the same mean, but different 
widths ($\sigma$ and scale$\times$$\sigma$, respectively). From the individual fits of $\mett^x$ and $\mett^y$ 
distributions, we obtain the mean, $\sigma$, scale, and relative normalization, $Norm$, of two Gaussians for 
each bin of $\sqrt{\sum E_{T}}$ (bin size is 2~GeV$^{\frac{1}{2}}$). The parameters are then fitted by simple 
polynomial functions of $z$=$\sqrt{\sum E_{T}}$: $p_0+p_1z$ for $\sigma$ and the scale, $p_0+p_1z^2$ for the 
mean, and $Norm=p_0$. These functions provide a parameterization of the unclustered energy contribution into 
the $x$ and $y$ components of the fake $\mett$ in the event. The default set of parameters is obtained from 
the $\gamma\gamma$ control sample. We also use the results of fits in the data $Z$$\to$$e^{+}e^{-}$ sample as 
an alternative set of parameters to study the associated systematic uncertainties. Figure~\ref{fig:uncl_param} 
demonstrates an example of the $\mett^y$ resolution parameterization due to the unclustered energy in 
{\sc pythia} $\gamma\gamma$ events. Distributions for both $x$ and $y$ components of $\mett$ look essentially 
identical to those shown in Fig.~\ref{fig:uncl_param}. We also do not observe any large difference in the 
parameterization of the $\mett$ resolution due to unclustered energy between $Z/\gamma^{*}$$\to$$e^{+}e^{-}$ 
and ``loose'' $\gamma\gamma$ events in data as well as between data and MC. 

\begin{figure}[!ht]
\centering
\includegraphics[width=1.0\linewidth]{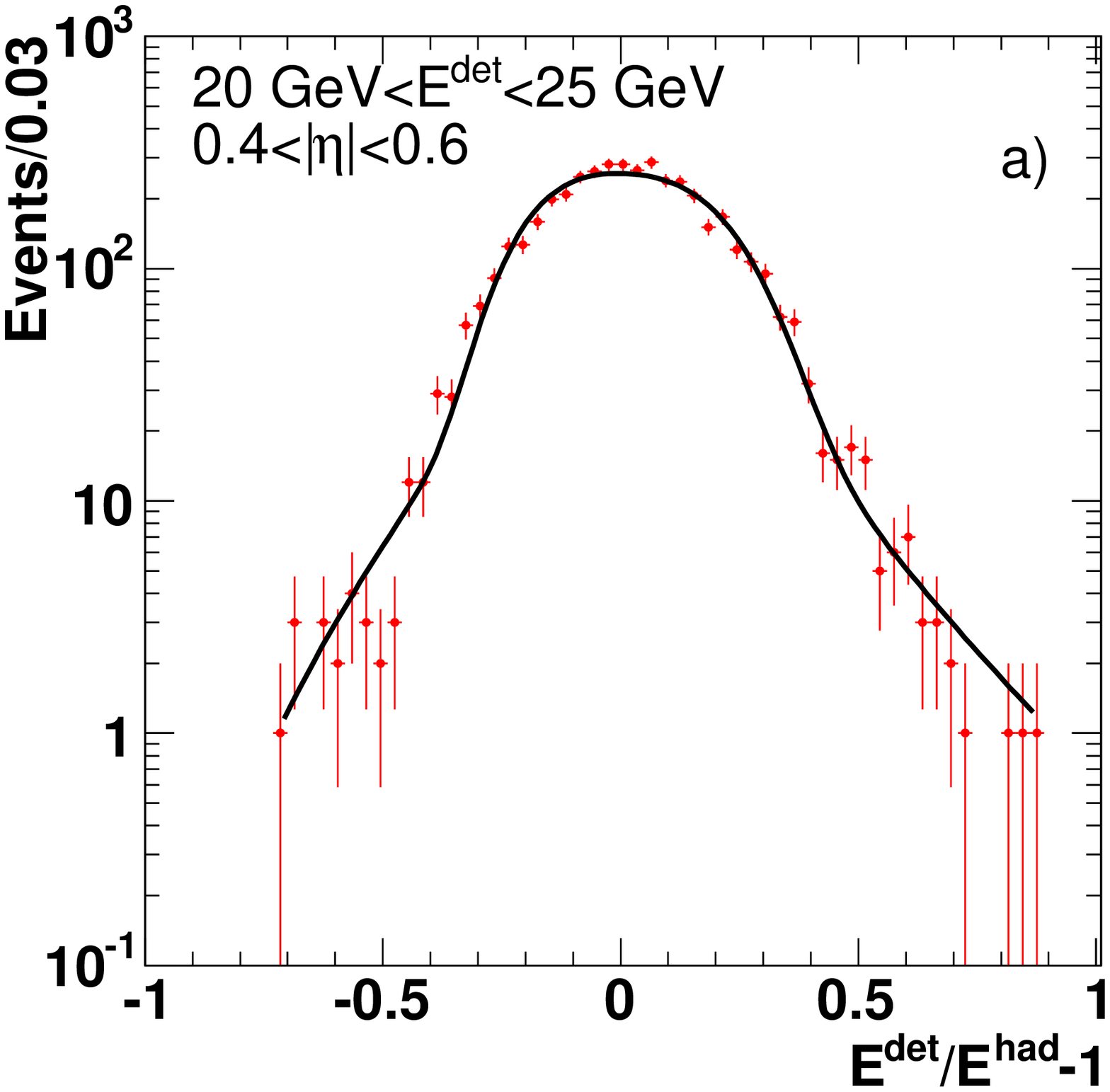}
\includegraphics[width=1.0\linewidth]{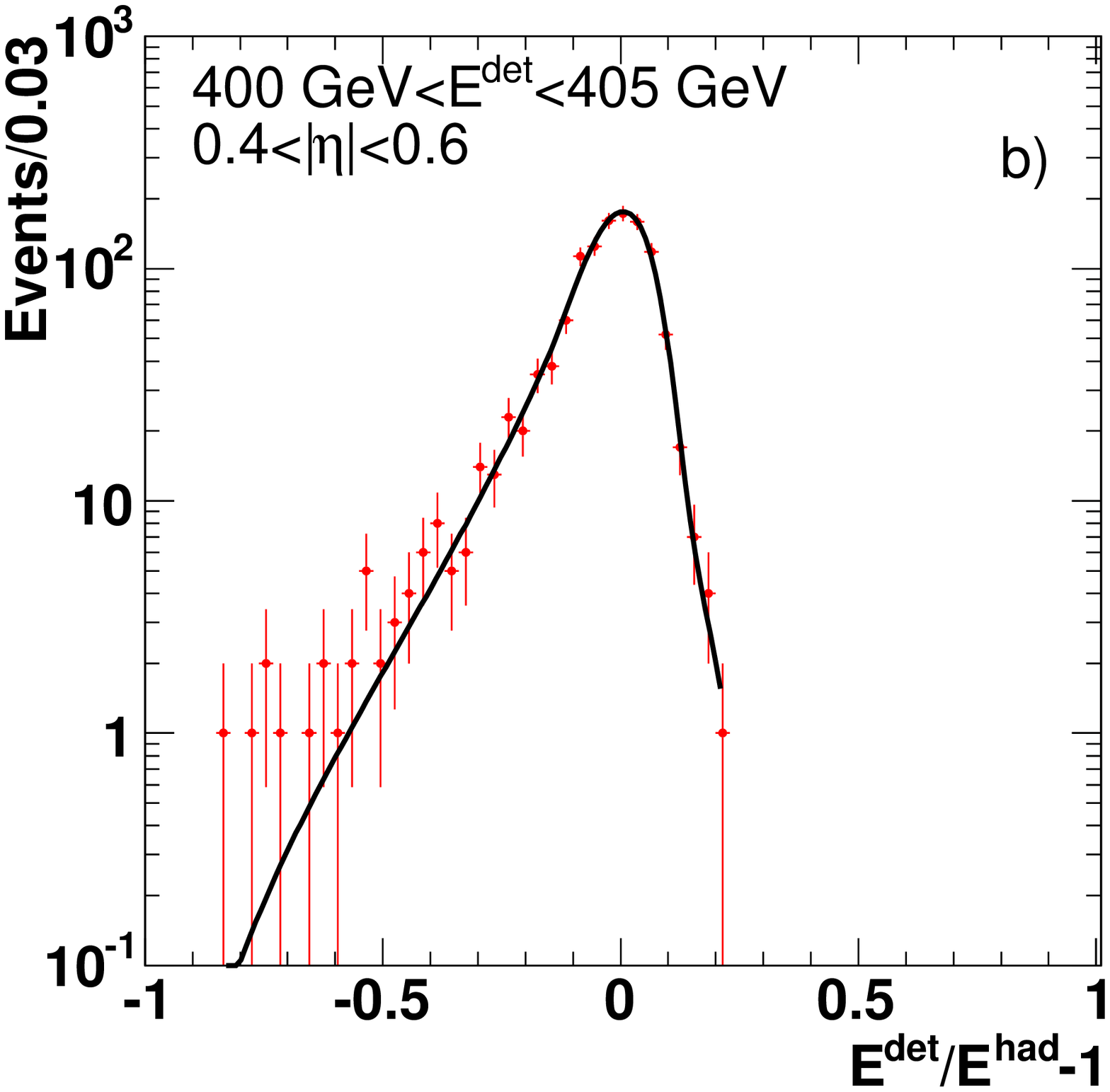}
\caption{Examples of jet energy resolution (JER) fits using a linear combination of Gaussian and Landau 
functions of $x$=$-JER/(1+JER)$ where $JER=E^{det}/E^{had}-1$ for two different jet energy bins: {\it a)} 
20~GeV$<$$E^{det}$$<$25~GeV and {\it b)} 400~GeV$<$$E^{det}$$<$405~GeV.}
\label{fig:JerFit}
\end{figure}

To account for contributions from jets into the fake $\mett$, we obtain the jet energy resolution, $JER$, as 
a function of jet energy and pseudorapidity, $E$ and $\eta$. For this purpose, we use {\sc pythia} samples 
of dijet and $Z$$-$jet events passed through the {\sc geant}-based detector simulation. In these events, we 
reconstruct jets before ({\it hadron} jet) and after ({\it detector} jet) the detector simulation by using the 
same cone clustering algorithm at both levels. The jet energy resolution is then defined as a ratio of the
detector ($E^{det}$) and hadron level ($E^{had}$) jet energies, $JER$=$E^{det}/E^{had}$-1, for hadron and 
detector jets with $p_T$$>$3~GeV/c that are matched within a cone of $R(\phi,\eta)$$<$0.1. Unlike the energy 
balance in dijet and $Z$-jet events, this definition of $JER$ is mostly sensitive to detector effects and 
allows us to significantly minimize the dependence of resolution on the effects of initial and final state 
radiation. However, we still compare the dijet and $Z$-jet balance in data and MC to make sure that the 
simulation adequately describes the resolution. We fill $JER$ histograms for jets in 5~GeV bins in jet energy 
and $\Delta\eta$=0.2 bins in pseudorapidity. We fit these histograms by a linear combination of Gaussian and 
Landau functions of $x$, where $x$=$-JER/(1+JER)$ ensures stable fits in the entire range of jet energies. 
Examples of fits for one particular $\eta$-bin can be found in Fig.~\ref{fig:JerFit}. These plots illustrate 
that our fit function successfully describes the jet energy resolution in a wide range of jet energies. It is 
also important to mention that the same fit function is used for all $\eta$-bins. From the individual fits 
for each ($E^{jet}$,$\eta$)-bin, we obtain a relative normalization, $C$, and parameters of a Gaussian (mean 
and $\sigma$) and Landau (mean and $\sigma$) fits. These parameters are plotted as a function of $E^{jet}$ 
for each $\eta$-bin, and fit with the following functions: $\sigma$=$\sqrt{p_{0}/E+p_{1}}$; 
$mean$=$p_{0}+p_{1}E+p_{2}/E$; and $C$=$(p_{0}+p_{1}\sqrt{E})/E+p_{2}$. This provides a smooth 
parameterization of $JER$ for all reconstructed jets with $E^{jet}$$>$3~GeV and $|\eta|$$<$3.6. 
%__________________________________________________________________
%---------------------- Fig.xx ----------------------------------

\begin{figure*}[htp]
\centering
\includegraphics[width=0.47\linewidth]{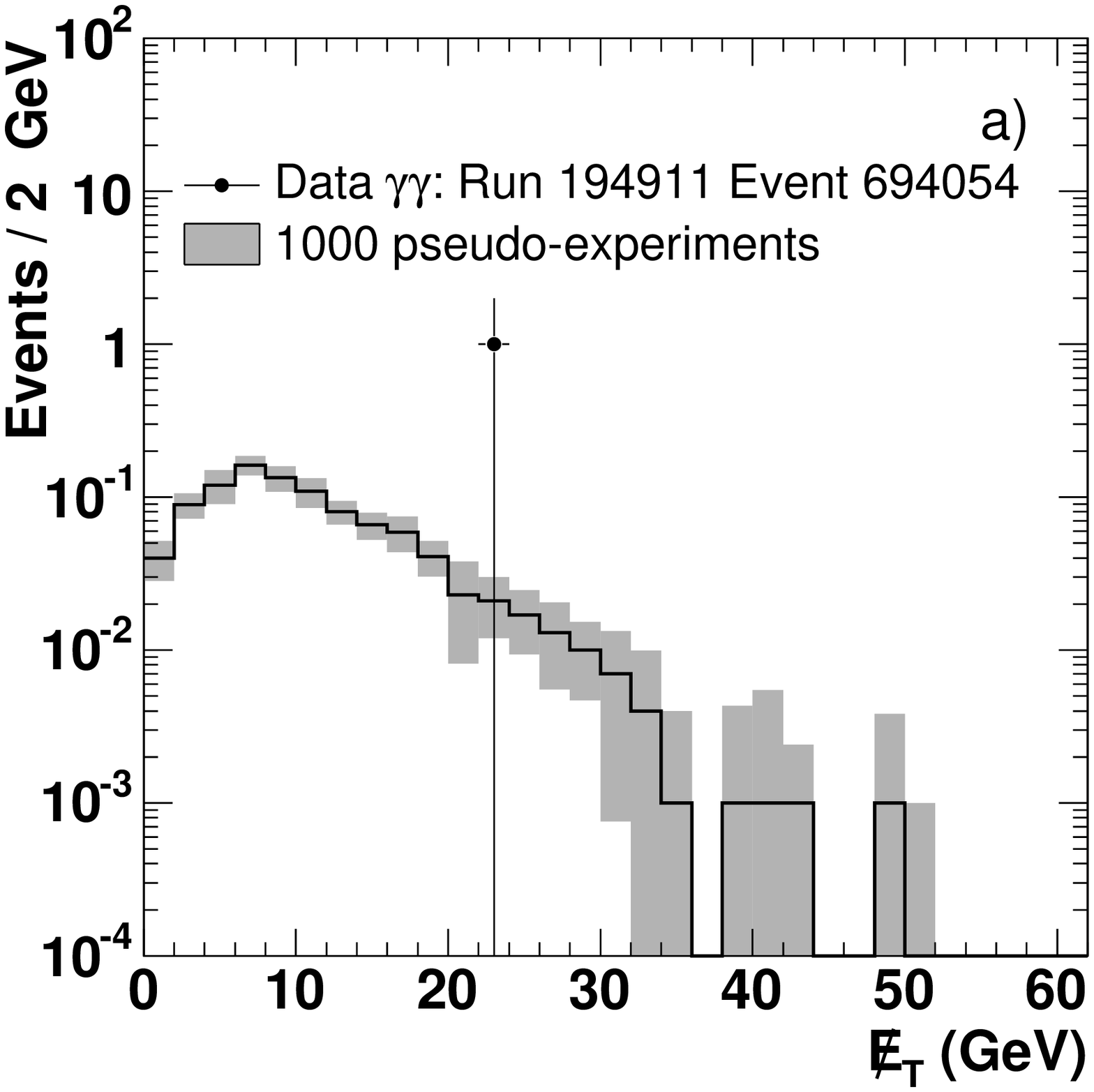}
\includegraphics[width=0.47\linewidth]{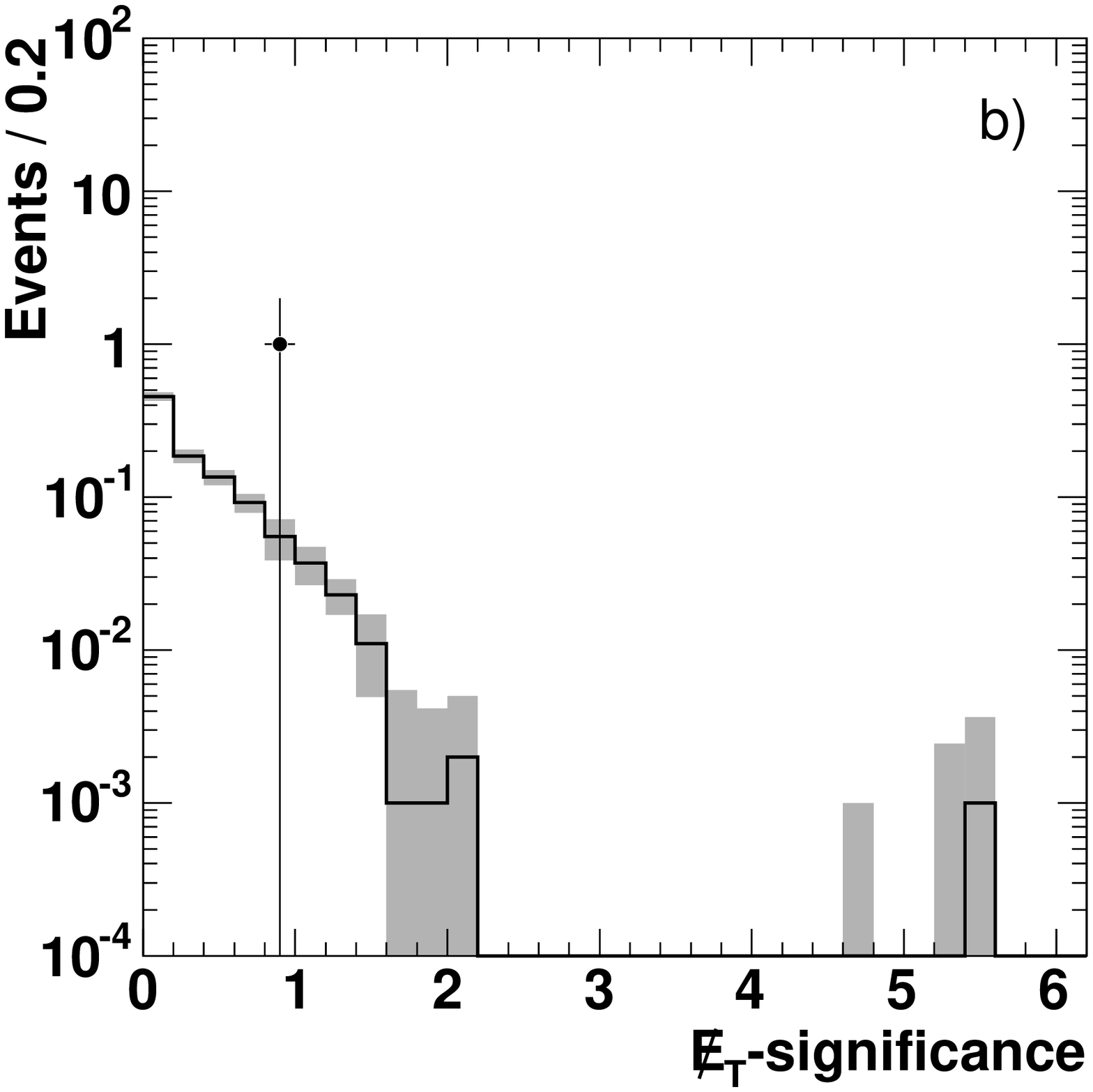}
\caption{Examples of the generated {\it a)} $\mathcal{P}(\mett)$ and {\it b)} \mett-significance distributions 
for one of the signal sample events.}
\label{fig:MetModel_example}
\end{figure*}

\begin{figure*}[htp]
\centering
\includegraphics[width=0.47\linewidth]{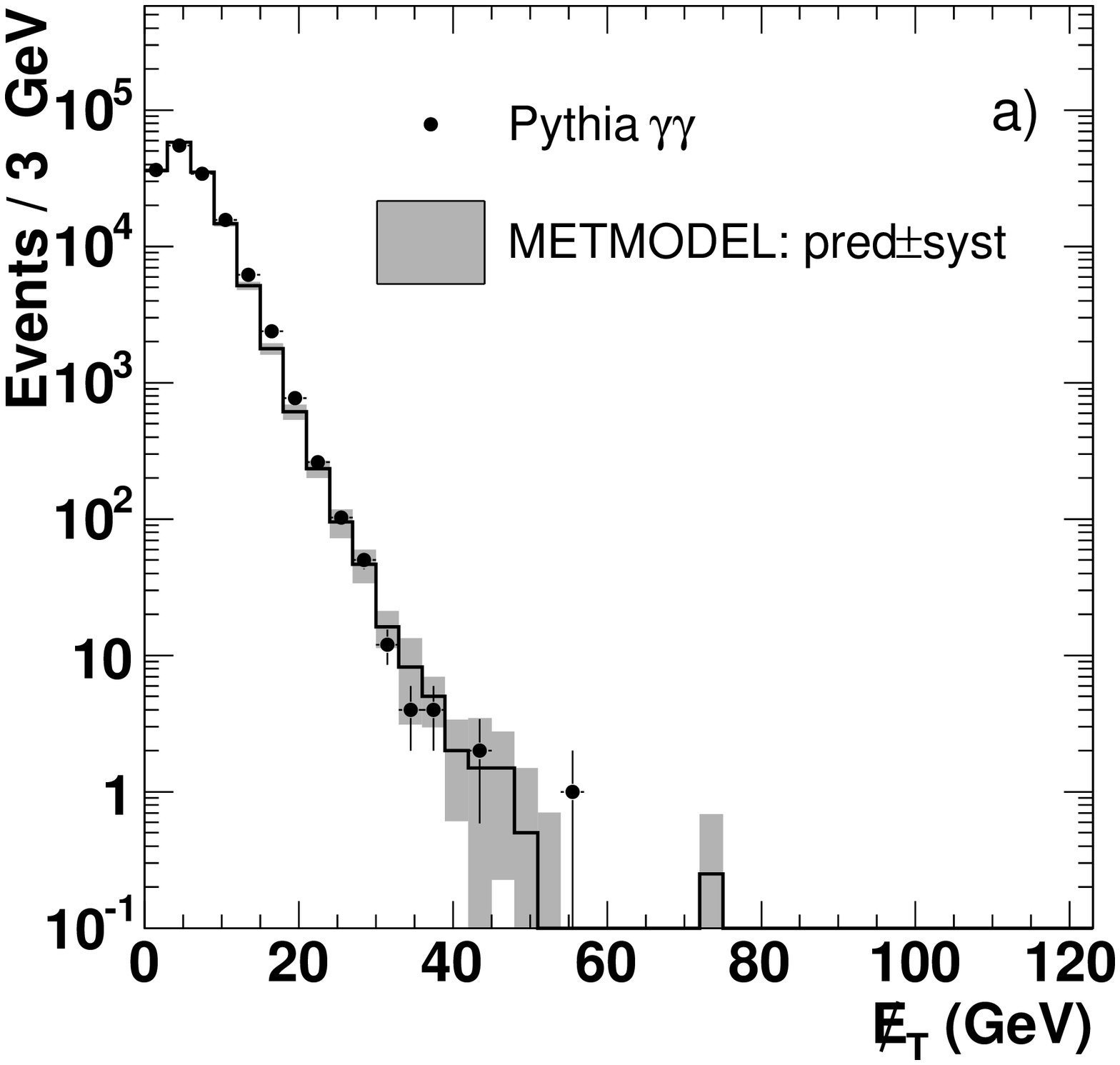}
\includegraphics[width=0.47\linewidth]{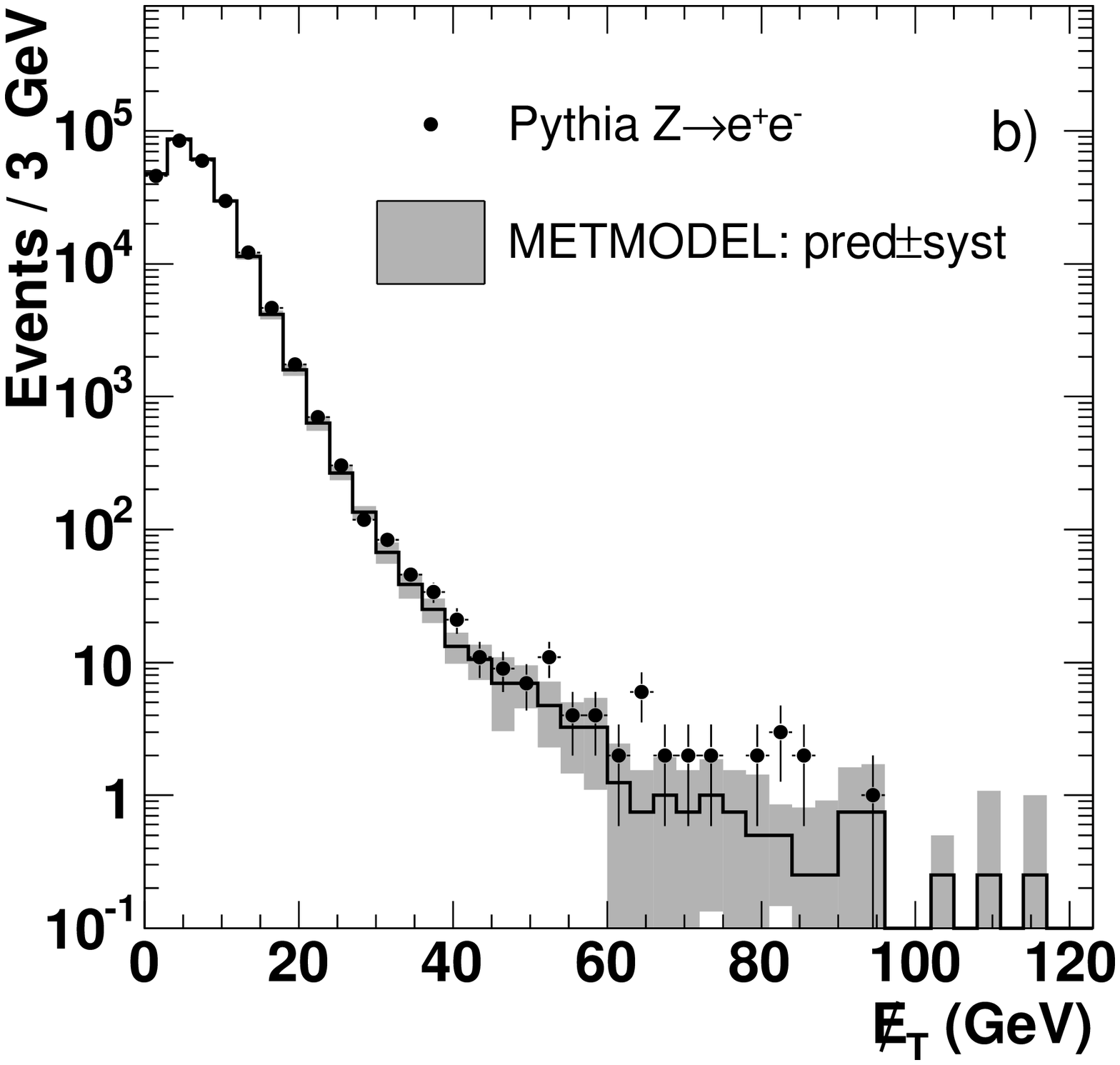}
\caption{Examples of the {\sc metmodel} predictions for $\mett$-distributions in {\sc pythia} {\it a)} 
$\gamma\gamma$ and {\it b)} $Z$$\to$$e^{+}e^{-}$ events. These events do not have the intrinsic $\mett$. 
However, fluctuations in energy measurements can result in the fake $\mett$ as large as 100~GeV. Both 
distributions are well described by the {\sc metmodel} predictions in the entire range of the observed 
$\mett$.}
\label{fig:Met_Pythia}
\end{figure*}

We predict the shape of fake $\mett$ based on the energy resolution functions described above. For each 
event, we produce a probability distribution function, $\mathcal{P}(\mett)$, of all possible values of the 
fake $\mett$ by smearing the energies of jets and unclustered energy according to these objects individual 
resolution functions in a large number of pseudoexperiments. Then, we sum up these individual 
$\mathcal{P}(\mett)$ distributions for all events to obtain a shape of the predicted fake $\mett$ due energy 
mismeasurements in our data sample. Technical details of how we generate $\mathcal{P}(\mett)$ are given 
below. An example of this $\mathcal{P}(\mett)$ distribution for one of the $\gamma\gamma$ baseline sample 
events can be found in Fig.~\ref{fig:MetModel_example}. The method is validated in MC samples with and 
without intrinsic $\mett$. Figure~\ref{fig:Met_Pythia} demonstrates that the {\sc metmodel} successfully 
predicts the shape of $\mett$-distributions in {\sc pythia} $\gamma\gamma$ and {\sc pythia} $Z$$\to$$e^{+}e^{-}$ 
events with fake $\mett$. The technique is also cross-checked by performing the entire analysis with the data 
$\gamma\gamma$ control sample and data $Z$$\to$$e^{+}e^{-}$ sample. 

%-----------------------------------------------
%-----------------------------------------------

The $\mathcal{P}(\mett)$ distribution for a given event can be obtained using a large number of 
pseudoexperiments. For each pseudo-experiment, we start by forming a list of all jets with $E_{T}$$>$3~GeV 
and $|\eta|$$<$3.0 in this event and then smear their energies according to $JER(E^{jet},\eta)$ described 
above. If the smeared jet energy, $E_{T}^{smear}$, is above the 15~GeV threshold, we calculate the contribution 
of that jet into the fake $\mett$: ${\vec{\mett}}^{jet,i}$=${\vec{E}}_{T} - {\vec{E}}_{T}^{smear}$. Therefore, 
the {\sc metmodel} should account for a correlation between the directions of $\mett$ and jets. Then, we 
re-calculate the unclustered energy based on $E_{T}^{smear}$ of all jets to avoid double-counting when one 
of the jets with $E_T$$<$15~GeV has $E_{T}^{smear}$$>$15~GeV. For the next step, we randomly generate the 
expected $x$ and $y$ components of the $\mett$ contribution due to the unclustered energy deposited in the 
calorimeter. This procedure also accounts, on average, for effects of energy resolution of photons and 
electrons as well as residual effects of the wrong vertex choice. Finally, we take a vector sum of all 
individual $\mett$ components due to the soft unclustered energy and each of the jets with 
$E_{T}^{smear}$$>$15~GeV to obtain the final prediction of the fake $\mett$. 

The {\sc metmodel} is not designed to predict the exact value of the fake $\mett$ in each event. Instead, it 
provides a two-dimensional probability density function, $\mathcal{P}(\vec{\mett})$, for values of the fake 
$\mett$ which could arise from energy mismeasurements in the calorimeter. This $\mathcal{P}(\vec{\mett})$ can 
be used to determine a significance of the observed $\mett$ in a given event according to the following formula:
\begin{eqnarray}
\text{\mett-significance}=-\log_{10}\left(1-\int_{0}^{\vec{w}}\mathcal{P}(\vec{z})d\vec{z}\right),
\label{eq:metsig_ideal}
\end{eqnarray}
\noindent where $\vec{z}$ is the generated fake $\vec{\mett}$ and $\vec{w}$ is the observed $\vec{\mett}$. The 
\mett-significance defined by Eq.~\ref{eq:metsig_ideal} correctly takes into account all of the correlations 
between jets and the observed $\mett$. However, the method has one significant drawback since it requires 
generating a large number of pseudoexperiments (e.g., $>$10$^6$ pseudoexperiments for \mett-significance=6). 
To overcome this problem, we take a simplified path of calculating an upper limit on the \mett-significance 
(``raw'' \mett-significance) according to the formula:
\begin{eqnarray}
\text{\mett-significance}=-\log_{10}(\widetilde{\mathcal{P}}_{jets}\widetilde{\mathcal{P}}_{uncl}),~~~
\label{eq:metsig_real} \\
\nonumber \widetilde{\mathcal{P}}_{uncl}=\prod_{i=x,y}\left(1-\int_{-u_i}^{u_i}\mathcal{P}^{i}_{uncl}(u)du\right),\\
\nonumber u_i=\mett^{x},\mett^{y},\\
%%\nonumber \widetilde{\mathcal{P}}_{jets}=\prod_{i}^{jets}\int_{-1}^{v_i}\mathcal{P}_{i}(v)dv,~\mathrm{if}~v_i<0,\\
%%\nonumber \mathrm{or}~\widetilde{\mathcal{P}}_{jets}=\prod_{i}^{jets}\left(1-\int_{-1}^{v_i}\mathcal{P}_{i}(v)dv\right),~\mathrm{if}~v_i>0,\\
\nonumber \widetilde{\mathcal{P}}_{jets}=\prod_{i,v_i>0}^{jets}\int_{-1}^{v_i}\mathcal{P}_{i}(v)dv \times \prod_{i,v_i<0}^{jets}\left(1-\int_{-1}^{v_i}\mathcal{P}_{i}(v)dv\right) \\
\nonumber v_i=\mett/(E_{T}^{i}cos\Delta\phi_{i}),
\end{eqnarray}
\noindent where $\mathcal{P}^{x,y}_{uncl}(u)$ is the probability density function for unclustered energy contribution 
to $\mett$ resolution (illustrated in Fig.\ref{fig:uncl_param}a), $\mathcal{P}_{i}(v)$ is the probability density 
function for jet energy resolution (shown in Fig.\ref{fig:JerFit}), $E_T^i$ is the transverse energy of the $i$-th 
jet, and $cos\Delta\phi_{i}$ is the azimuthal angle between that jet and measured $\mett$. The ``raw'' 
\mett-significance obtained from Eq.~\ref{eq:metsig_real} is then calibrated to have a simple shape defined 
{\it a priori}: $dN/dx$=$N_{evnt}$$\cdot$$\ln(10.0)$$\cdot$$10^{-x}$, where $x$ is the \mett-significance and $N_{evnt}$ 
is the number of events in a sample. The shape of the \mett-significance has one important property: if all events in 
a data sample were to have only fake $\mett$, then $N_{evnt}\cdot 10^{-cut}$ events would pass a requirement 
\mett-significance$>$$cut$. This property makes it very easy to calibrate the \mett-significance by means of 
pseudoexperiments. In each pseudo-experiment, we obtain a randomly generated value of $\mett$. Then we 
calculate the significance of this generated $\mett$ as if it were measured $\mett$. We repeat this procedure 
for all events in the data sample and obtain the significance distribution for pseudoexperiments. Finally, 
an adjustment factor is derived for each bin of the distribution so that the corrected \mett-significance satisfies 
the $N(\text{\mett-significance}$$>$$cut)$=$N_{evnt}\cdot10^{-cut}$ requirement. 
    
The systematic uncertainties associated with the {\sc metmodel} predictions are evaluated by comparing the 
results obtained with the default set of parameters to predictions obtained with the {\sc metmodel} parameters 
changed by one standard deviation ($\pm\sigma$). In total, ten sources of the systematic uncertainties are 
considered: 1) difference in the unclustered energy parameterization of the $\mett$ resolution for 
$\gamma\gamma$ control and $Z$$\to$$e^+e^-$ events; 2) uncertainties on four parameters of the unclustered 
energy parameterization; 3) uncertainties on five parameters of the {\it JER} parameterization. The correlations 
between these parameters are also taken into account. The statistical uncertainty that depends on the number 
of pseudoexperiments per event and the systematic uncertainty are added in quadrature to obtain the total 
uncertainty.


\begin{thebibliography}{99}

\bibitem{models} S.~Abe {\it et al.}, arXiv:hep-ph/0003154; R.L.~Culbertson {\it et al.}, arXiv:hep-ph/0008070;
B.~Allanach {\it et al.}, arXiv:hep-ph/9906224; S.~Ambrosanio {\it et al.}, arXiv:hep-ph/0006162; M.S.~Carena 
{\it et al.}, arXiv:hep-ph/0010338.
\bibitem{SM} S.L.~Glashow, Nucl. Phys. {\bf 22}, 579 (1961); S.~Weinberg, Phys. Rev. Lett. {\bf 19}, 1264 
(1967); A.~Salam, {\it Proc. 8th Nobel Symposium, Stockholm} (1979); D.J.~Gross and F.~Wilczek, Phys. Rev. D {\bf 8}, 3633 (1973).      
\bibitem{ggMETgmsb} S.~Dimopoulos, S.D.~Thomas and J.D.~Wells, Nucl. Phys. B {\bf 488}, 
39 (1997); S.~Ambrosanio, G.D.~Kribs and S.P.~Martin, Phys. Rev. D {\bf 56}, 1761 (1997); 
G.F.~Giudice and R.~Rattazzi, Phys. Rep. {\bf 322}, 419 (1999); S.~Ambrosanio, G.L.~Kane, 
G.D.~Kribs, S.P.~Martin and S.~Mrenna, Phys. Rev. D {\bf 55}, 1372 (1997).
\bibitem{Hsector} B.A.~Dobrescu, Phys. Rev {\bf D63}, 015004 (2000); G.~Landsberg and K.~Matchev, 
Phys. Rev {\bf D62}, 035004 (2000).
\bibitem{tRho} K.D.~Lane and S.~Mrenna, Phys. Rev {\bf D67}, 115011 (2003).
\bibitem{gen4} G.~Bhattacharyya and R.N.~Mohapatra , Phys. Rev. D {\bf 54}, 4204 (1996).
\bibitem{ledH} L.~Hall and C.~Kolda, Phys. Lett. B {\bf 459}, 213 (1999); M.C.~Kumar, Prakash~Mathews, V.~Ravindran, 
and Anurag~Tripathi, Phys. Lett. B {\bf 672}, 45 (2009). 
\bibitem{eta} The CDF~II detector uses a cylindrical coordinate system in which $\phi$ is the azimuthal 
angle, $\theta$ is the polar angle with respect to the proton beam, $r$ is the radius from the nominal 
beam line, and $z$ points in the proton beam direction, with the origin at the center of the detector. 
The transverse $\rphi$, or $x$-$y$ plane, is the plane perpendicular to the $z$ axis. The pseudorapidity, 
$\eta$, is defined as $-\ln(\tan(\theta/2))$. The transverse energy of a particle is $E_T$=$E\cdot sin(\theta)$.
The transverse momentum of a particle is defined as $p_T$=$p\cdot sin(\theta)$. 
\bibitem{run1evnt} F.~Abe {\it et al.} (CDF Collaboration), Phys. Rev. Lett. {\bf 81}, 1791 
(1998); Phys. Rev. D {\bf 59}, 092002 (1999).
\bibitem{run2cdf} D.~Acosta {\it et al.} (CDF Collaboration), Phys. Rev. D {\bf 71}, 031104 
(2005).
\bibitem{run2d0} V.M.~Abazov {\it et al.} (D0 Collaboration), Phys. Lett. B {\bf 659}, 
856 (2008).
\bibitem{lgmetPRD} A.~Abulencia {\it et al.} (CDF Collaboration), Phys. Rev. D {\bf 75}, 112001 (2007).
\bibitem{RSdipho} T.~Aaltonen {\it et al.} (CDF Collaboration), Phys. Rev. Lett. {\bf 99}, 
171801 (2007).
\bibitem{Hdipho} T.~Aaltonen {\it et al.} (CDF Collaboration), Phys. Rev. Lett. {\bf 103}, 061803 (2009).
\bibitem{vista} T.~Aaltonen {\it et al.} (CDF Collaboration), Phys. Rev. D {\bf 79}, 011101 (2009).
%%---------------- Detector references ------------------------------------------------------------
\bibitem{CDF} D.~Acosta {\it et al.} (CDF Collaboration), Phys.\ Rev.\ D {\bf 71}, 032001 (2005). 
\bibitem{L00} C.S.~Hill {\it et al.}, Nucl. Instrum. Methods Phys. Res., Sect. A {\bf 511}, 118 (2003).
\bibitem{SVX} A.~Sill {\it et al.}, Nucl. Instrum. Methods Phys. Res., Sect. A {\bf 447}, 1 (2000).
\bibitem{ISL} A.~Affolder {\it et al.}, Nucl. Instrum. Methods Phys. Res., Sect. A {\bf 453}, 84 (2000).
\bibitem{COT} A.~Affolder {\it et al.}, Nucl. Instrum. Methods Phys. Res., Sect. A {\bf 526}, 249 (2004).
\bibitem{cem} L.~Balka {\it et al.}, Nucl. Instrum. Methods Phys. Res., Sect. A {\bf 267}, 272 (1988); 
S.R.~Hahn {\it et al.}, {\it ibid.} {\bf 267}, 351 (1988). 
\bibitem{cwha} S.~Bertolucci {\it et al.}, Nucl. Instrum. Methods Phys. Res., Sect. A {\bf 267}, 301 (1988).
\bibitem{plug} R.~Oishi {\it et al.}, Nucl. Instrum. Methods Phys. Res., Sect. A {\bf 453}, 227 (2000); 
M.~Albrow {\it et al.}, {\it ibid.} {\bf 480}, 524 (2002).
\bibitem{EMtiming} M.~Goncharov {\it et al.}, Nucl. Instrum. Methods Phys. Res., Sect. A {\bf 565}, 543 
(2006).
\bibitem{muon_systems} G.~Ascoli {\it et al.}, Nucl.\ Instrum.\ Meth.\  A {\bf 268}, 33 (1988); A.~Artikov 
{\it et al.}, Nucl.\ Instrum.\ Meth.\  A {\bf 538}, 358 (2005).
\bibitem{CLC} D.~Acosta {\it et al.}, Nucl. Instrum. Methods Phys. Res., Sect. A {\bf 494}, 57 (2002).
\bibitem{trigger} F.~Abe {\it et al.} (CDF Collaboration), Nucl. Instrum. Methods Phys. Res., Sect. A 
{\bf 271}, 387 (1988).
%%---------------- Event Selection ------------------------------------------------------------
\bibitem{run2diphoMET} D.~Acosta {\it et al.} (CDF Collaboration), Phys. Rev. D {\bf 71}, 031104 (2005).
\bibitem{madgraph} F.~Maltoni and T.~Stelzer, Comput.\ Phys.\ Commum. {\bf 357}, 81 (1994).
\bibitem{pythia} T. Sjostrand, Comput. Phys. Commun. {\bf 82}, 74 (1994); S. Mrenna, Comput. Phys. 
Commun. {\bf 101}, 232 (1997).
\bibitem{geant} R. Brun {\it et al.}, CERN Report No. CERN-DD/EE/84-1, 1987.
\bibitem{cdf6601} D.~Acosta {\it et al.}  (CDF Collaboration), Phys.\ Rev.\ Lett.\ {\bf 94}, 041803 
(2005); H.~S.~Hayward, Ph.D. thesis, University of Liverpool, 2005; M.~H.~Kirby, Ph.D. thesis, Duke 
University, 2004.
\bibitem{pdf} H.L.~Lai {\it et al.}, Eur. Phys. J. C {\bf 12}, 375 (2000). 
\bibitem{canepa} T.~Aaltonen {\it et al.} (CDF Collaboration), Phys.\ Rev.\  D {\bf 77}, 052002 (2008);
M. Griffiths, Ph.D. thesis, University of Liverpool, 2007.
\bibitem{phoenix}  D.~Acosta {\it et al.} (CDF Collaboration), Phys. Rev. D {\bf 71}, 051104 (2005).
\bibitem{wrongVx} The vertex misassignment mostly happens in real $\gamma\gamma$ events. However, it 
also occurs in $\gamma\gamma_{fake}$ or $\gamma_{fake}\gamma_{fake}$ events, but at a much lower rate 
because these events tend to have more tracks associated with a vertex produced by the hard primary 
interaction vertex.
\bibitem{baur} U. Baur, T. Han, and J. Ohnemus, Phys. Rev. D {\bf 48}, 5140 (1993); J. Ohnemus, Phys. 
Rev. D {\bf 47}, 940 (1993).
\bibitem{sim} The $W\gamma$ and $Z\gamma$ events are simulated using the leading-order (LO) event 
generator~\cite{baur}. The initial state radiation (resulting in additional jets or photons), underlying 
event, and additional interactions in the same bunch crossing are modeled with the {\sc pythia}~\cite{pythia} 
Monte Carlo program. The generated events are processed through a realistic {\sc geant}-based~\cite{geant} 
detector simulation.
\bibitem{narrowjet} Sometimes a photon or an energetic $\pi^0/\eta^{0}$ from a jet can hit the area very 
close to the calorimeter cracks at $\eta$$\sim$0 and $|\eta|$$\sim$1.1. In such cases, these objects can get 
reconstructed as narrow jets with just a few towers ($<$10) and a few tracks ($<$5) pointing to them. Often, 
most of the energy ($\sim$85$\%$) of these narrow jets is deposited either in the electromagnetic or hadronic 
calorimeters.   
\bibitem{pyth_sample} The inclusive $W$ and $Z/\gamma^{*}$ events are simulated using the {\sc pythia} 
\cite{pythia} Monte Carlo program and processed through a realistic {\sc geant}-based~\cite{geant} detector 
simulation.
\bibitem{delayedPhoPRD} T.~Aaltonen {\it et al.} (CDF Collaboration), Phys. Rev. D {\bf 78}, 032015 (2008).
\bibitem{diboson} T.~Aaltonen {\it et al.} (CDF Collaboration), Phys. Rev. Lett. {\bf 103}, 091803 (2009).
\bibitem{gmsbPRL} T.~Aaltonen {\it et al.} (CDF Collaboration), Phys. Rev. Lett. {\bf 104}, 011801 (2010). 
\bibitem{CESfid} For the central calorimeter, the fiducial region covers $\sim$87$\%$ of the total area.
\bibitem{BHcuts} When counting central electromagnetic towers, we exclude towers associated with photon 
candidates. To reduce the rate at which real collision events are misidentified as beam halo, the total 
transverse energy of all contributing towers has to be less than 0.6 GeV.  
\bibitem{EMres} In this analysis, the EM timing is calculated with respect to the nominal center of the 
detector, $z=0$ cm. This leads to a larger resolution compared to the case when the actual interaction 
position, $z_{vertex}$, and time, $t_{vertex}$, are taken into account~\cite{EMtiming}.
\bibitem{LeptonID} A.~Abulencia {\it et al.} (CDF Collaboration), J.\ Phys.\ G {\bf 34}, 2457 (2007);
A.~Abulencia {\it et al.} (CDF Collaboration), Phys.\ Rev.\  D {\bf 74}, 072006 (2006).
\bibitem{tauID} A.~Abulencia {\it et al.} (CDF Collaboration), Phys. Rev. D {\bf 75}, 092004 (2007).
\bibitem{jetclu} F.~Abe {\it et al.} (CDF Collaboration), Phys.\ Rev.\  D {\bf 45}, 1448 (1992).
\bibitem{jes} A.~Bhatti {\it et al.}, Nucl. Instrum. Methods Phys. Res., Sect. A {\bf 566}, 375 (2006).
\bibitem{jettriggers} T.~Aaltonen {\it et al.} (CDF Collaboration), Phys. Rev D {\bf 78}, 052006 (2008). 
 
\end{thebibliography}
\end{document}